\documentclass[prd,twocolumn,showpacs,reprint,preprintnumbers,nofootinbib]{revtex4-1}
\input{header}

\begin{document}

%============================================================
\title{Extending the reach of FASER, MATHUSLA, and SHiP towards smaller lifetimes
\\
using secondary particle production}
%============================================================

\author{Krzysztof~Jod\l{}owski}
\email{krzysztof.jodlowski@ncbj.gov.pl}
\affiliation{National Centre for Nuclear Research, Pasteura 7, 02-093 Warsaw, Poland}

\author{Felix~Kling}
\email{felixk@slac.stanford.edu}
\affiliation{SLAC National Accelerator Laboratory, 2575 Sand Hill Road, Menlo Park, CA 94025, USA}

\author{Leszek~Roszkowski}
\email{leszek.roszkowski@ncbj.gov.pl}
\affiliation{Astrocent, Nicolaus Copernicus Astronomical Center Polish Academy of Sciences,ul. Bartycka 18, 00-716 Warsaw, Poland}
\affiliation{National Centre for Nuclear Research, Pasteura 7, 02-093 Warsaw, Poland}

\author{Sebastian~Trojanowski}
\email{s.trojanowski@sheffield.ac.uk}
\affiliation{Consortium for Fundamental Physics, School of Mathematics and  Statistics, University of Sheffield, Hounsfield Road, Sheffield, S3 7RH, UK}
\affiliation{National Centre for Nuclear Research, Pasteura 7, 02-093 Warsaw, Poland}

\begin{abstract}
Many existing or proposed intensity-frontier search experiments look for decay signatures of light long-lived particles (LLPs), highly displaced from the interaction point, in a distant detector that is well-shielded from SM background. This approach is, however, limited to new particles with decay lengths similar to or larger than the baseline of those experiments. In this study, we discuss how this basic constraint can be overcome in BSM models that go beyond the simplest scenarios. If more than one light new particle is present in the model, an additional secondary production of LLPs may take place right in front of the detector, opening this way a new lifetime regime to be probed. We illustrate the prospects of such searches in the future experiments FASER, MATHUSLA and SHiP, for representative models, emphasizing possible connections to dark matter or an anomalous magnetic moment of muon. We also analyze additional advantages from employing dedicated neutrino detectors placed in front of the main decay volume.
\end{abstract}

\renewcommand{\baselinestretch}{0.85}\normalsize
\maketitle
\tableofcontents
\renewcommand{\baselinestretch}{1.0}\normalsize

%============================================================
\section{\label{sec:intro}Introduction}
%============================================================

Motivated by a successful history of discoveries of new elementary particles, it has long been the dominant paradigm in experimental searches to look for heavier and heavier particles that could manifest their existence in increasingly more powerful colliders. This approach led to a well-established experimental program that is now being continued at the Large Hadron Collider (LHC). In parallel with these persevering efforts, however, there has also been a growing interest in recent years in exploring scenarios that might have been overlooked in the past.

Among them, a prime focus has been directed on light and very weakly-interacting beyond the Standard Model (BSM) particles that could have escaped detection in previous searches due to the lack of sufficient luminosity. The corresponding efforts are often referred to as \textsl{intensity frontier} searches for light long-lived particles (LLPs). This captures the fact that the relevant detection prospects depend upon the ability to study very rare events that should also be discriminated from \textsl{a priori} overwhelming Standard Model (SM) background (BG).

A variety of atypical experimental signatures have been proposed to search for LLPs~\cite{Beacham:2019nyx,Alimena:2019zri}. In particular, very clean searches for LLPs have been considered to employ their displaced decays in a distant detector that is physically separated and shielded from the primary interaction point (IP). This allows one to drastically reduce BG, often to negligible levels. However, an obvious limitation of this strategy is the lifetime of BSM particles that can be probed this way. If a LLP is too short-lived, it primarily decays before reaching a detector's fiducial volume in which case prospects for discovery are greatly reduced.

This is evident in a variety of simplified models that have been proposed as benchmark scenarios for intensity frontier studies~\cite{Beacham:2019nyx}. In particular, in models employing a single coupling of an LLP to the SM particles, the shape of current exclusion bounds and future sensitivity lines is often driven by the tension between increasing production rates and diminishing the lifetime of the LLP. This can be overcome in less simplified models employing e.g. non-universal couplings of the LLP to SM hadrons and leptons, or by the use of a compressed mass spectrum to increase the lifetime of decaying LLP. By tuning different types of couplings or a mass splitting between the particles present in the model, one can keep a relatively high production rate of LLPs, while simultaneously increasing their decay length.

On the other hand, in more complete models, if an LLP is present, it is often accompanied by other light BSM species. A notable example can be light dark matter (DM) with a mediator particle of similar mass that couples to the SM and yields correct thermal DM relic density~\cite{Boehm:2003hm,Feng:2008ya}, or various realizations of the twin Higgs model~\cite{Chacko:2005pe}, that have been advocated for in the context of \textsl{neutral naturalness} and often predict the existence of several light BSM particles in the mirror sector. 

%------------------------
\begin{figure*}[tb]
\centering
\includegraphics[width=0.8\textwidth]{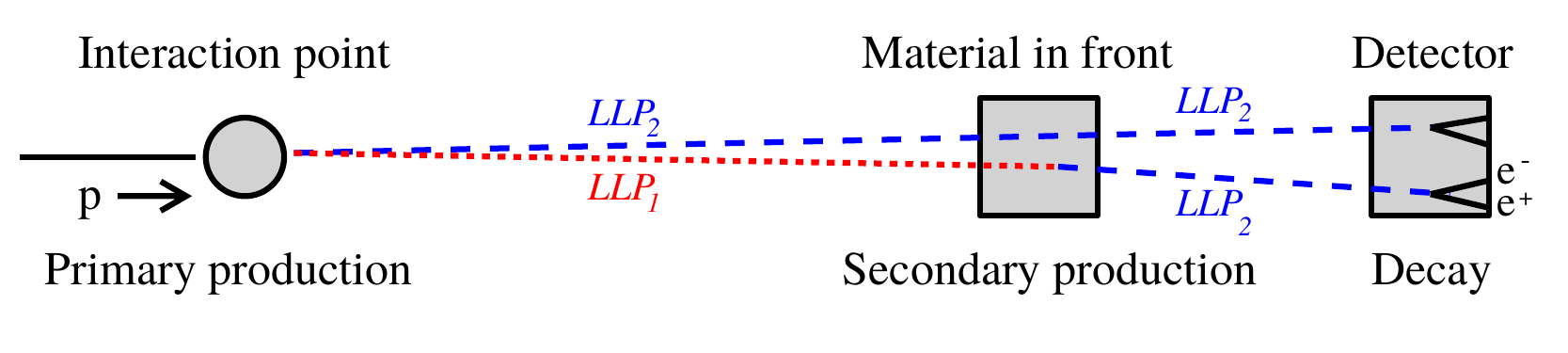}
\caption{Schematic illustration of primary and secondary production of the light long-lived particle $\textrm{LLP}_2$, shown with the dashed blue line, that subsequently decays inside the detector. In the case of secondary production, the intermediate role of other light species, $\textrm{LLP}_1$, shown with the dotted red line, allows for producing $\textrm{LLP}_2$ much closer to the detector. In models studied below, the $(\textrm{LLP}_1,\textrm{LLP}_2)$ pair can be identified with: $(\chi,A^\prime)$ for dark bremsstrahlung (see \secref{sec:darkbrem}), $(\chi_1,\chi_2)$ in the case of inelastic dark matter (see \secref{sec:iDM}), and $(S,A^\prime)$ for the model with a dark photon and a secluded dark Higgs boson (see \secref{sec:darkHiggs}).
}
\label{fig:idea}
\end{figure*}
%------------------------

In such scenarios, the aforementioned effective decoupling of production and decay of the LLPs can appear even more naturally, without requiring a tuning between the model parameters or introducing exotic non-universal couplings. In particular, if one of the LLPs can be effectively produced in interactions of the other, that happen in front of the detector or inside it, this can lead to secondary production of LLPs at a position much closer to a decay vessel. We illustrate this schematically in \cref{fig:idea}.

\begin{figure*}[tb]
\begin{subfigure}{0.32\textwidth}
\centering
\includegraphics[width=1.0\textwidth]{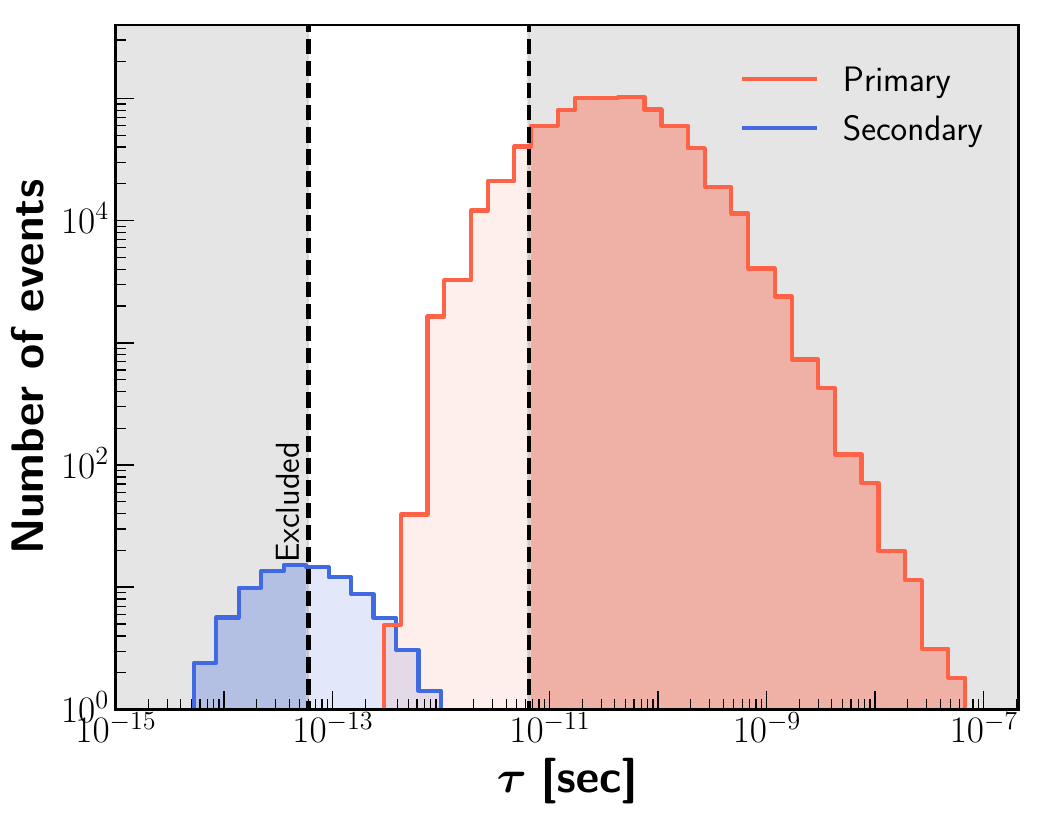}
\caption{FASER 2\\
($m_{A^\prime}\simeq 30~\mev$, model \cref{sec:darkHiggs})}\label{fig:histFASER2}
\end{subfigure}
\begin{subfigure}{0.32\textwidth}
\centering
\includegraphics[width=1.0\textwidth]{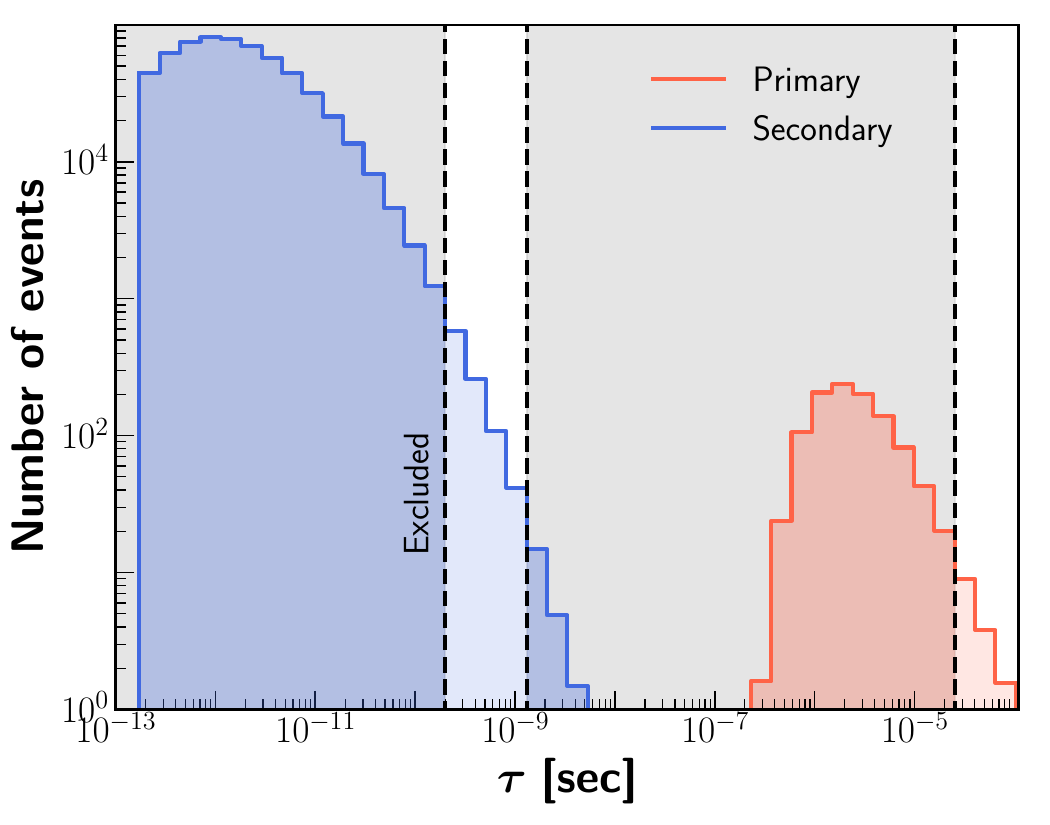}
\caption{MATHUSLA\\
($m_{A^\prime}\simeq 70~\mev$, model \cref{sec:iDM})}\label{fig:histMATHUSLA}
\end{subfigure}
\begin{subfigure}{0.32\textwidth}
\centering
\includegraphics[width=1.0\textwidth]{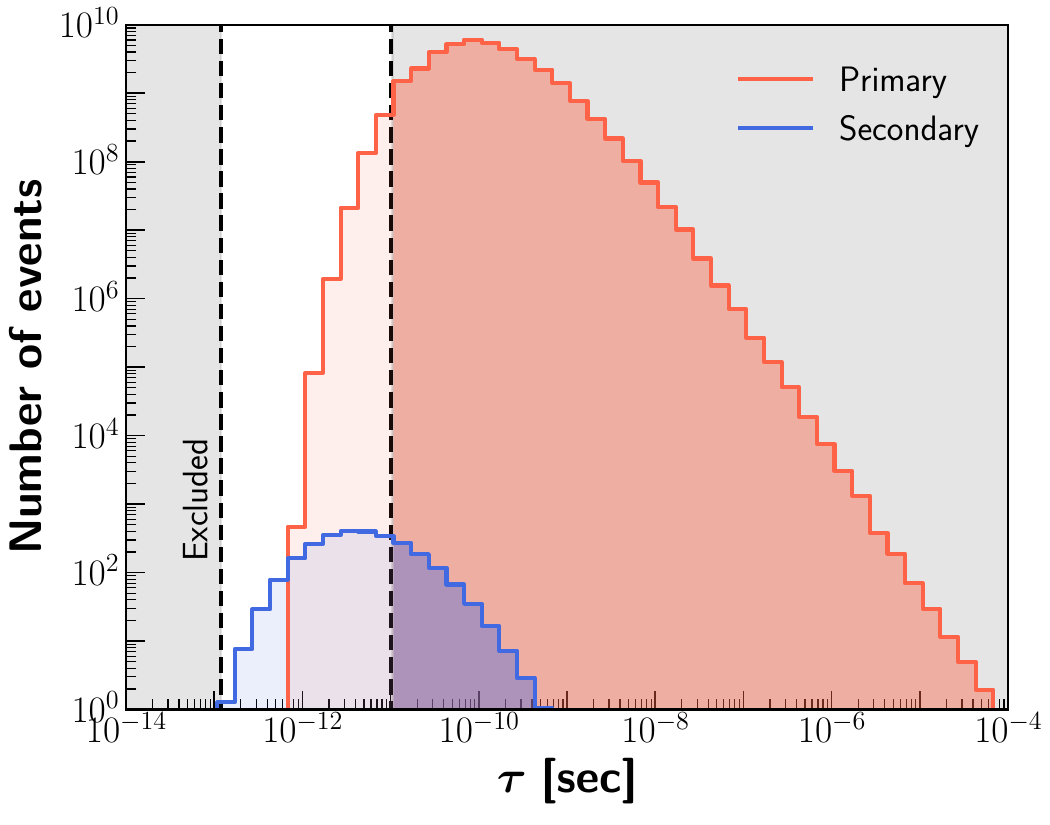}
\caption{SHiP\\
($m_{A^\prime}\simeq 20~\mev$, model \cref{sec:darkbrem})}\label{fig:histSHiP}
\end{subfigure}
\caption{Histograms with the expected number of events in the detectors: FASER 2 (left), MATHUSLA (center) and SHiP (right) coming from primary (red) and secondary (blue) production of a light long-lived particle. For the purpose of illustration, the plots were obtained for benchmark points indicated in the captions that correspond to the models described in \cref{sec:models}. The currently excluded ranges of lifetimes for these BPs are shown as grey-shaded regions.}
\label{fig:lifetimes}
\end{figure*}

As a result, the range of the LLP lifetimes, $\tau_{\textrm{LLP}}$, that can be probed is extended toward smaller values. In addition, for a range of intermediate values of $\tau_{\textrm{LLP}}$, both primary and secondary production give comparable contributions to the total number of events in the detector. We illustrate this in \cref{fig:lifetimes} for selected intensity frontier experiments and benchmark points (BPs). In the figure, we show the expected number of events associated with both the primary production at the interaction point (IP) and the secondary production mechanisms as a function of $\tau_{\textrm{LLP}}$. 

%------------------------
%------------------------

In this study, we analyze the prospects for such searches in some of the future  experiments that have been proposed to look for the LLPs and could naturally employ the aforementioned secondary production mechanism to increase their discovery potential. Among them, we study the FASER experiment~\cite{Feng:2017uoz,Ariga:2018zuc,Ariga:2018uku,Ariga:2018pin} that will be taking data during LHC Run 3 and can then be extended toward the High Luminosity LHC (HL-LHC) phase. We also discuss the prospects of two other experiments with a possible timeline similar, or identical, to HL-LHC, namely the proposed MATHUSLA~\cite{Chou:2016lxi,Alpigiani:2018fgd,Curtin:2018mvb} and SHiP~\cite{Bonivento:2013jag,Anelli:2015pba,Alekhin:2015byh} detectors. 

In the case of the proposed SHiP detector, we additionally discuss two other possible signatures of models with more than a single LLP that employ a dedicated neutrino subdetector. Interactions of LLPs can be discovered there based on their scatterings off electrons, as well as such scatterings followed by a prompt decay in the subdetector that leads to two simultaneous electromagnetic (EM) showers. We analyze how a combination of various signatures can deliver important information, on top of the one from the standard search for LLP decays in the decay vessel, that could help to resolve the nature of the LLP. 

This paper is organized as follows. In \cref{sec:models}, we introduce simplified BSM models of interest to us. In \cref{sec:experiments}, we discuss basic aspects of the LHC experiments under study, while more details of our analysis are given in \cref{sec:modeling}. The results for the secondary LLP production in scatterings off nuclei are presented in \cref{sec:results}, while additional signatures for scatterings off electrons in the SHiP neutrino subdetector are discussed in \cref{sec:resultselectrons}. We conclude in \cref{sec:conclusions}. More technical aspects of the study are described in the appendices. In \cref{sec:resultsaftercutschange} we discuss how the sensitivity reach from the secondary production can be affected by varying cuts imposed in the analysis. In \cref{sec:decays,sec:scatteringsigma}, respectively, we present the expressions relevant for the primary production and decay of LLPs, as well as the scattering cross sections corresponding to the secondary production. \Cref{sec:benchmarkgeometries,sec:physicscuts} are devoted to a more detailed discussion of the detector designs used in modeling, as well as relevant experimental cuts.

%============================================================
\section{Models\label{sec:models}}
%============================================================

In order to illustrate the impact of the secondary production mechanism on the sensitivity reach in intensity frontier searches, we consider some simplified BSM models that can lead to production of LLPs in scatterings in front of the detector and their subsequent decays within its fiducial volume.

To this end, we focus on popular scenarios with a light dark vector particle appearing in minimal extensions of the SM with a new $U(1)_d$ symmetry group. We assume that the resulting gauge boson, dubbed dark photon or $A^\prime$, couples to the SM solely via a kinetic mixing term $(\epsilon/2)\,F_{\mu\nu}F^{\prime\mu\nu}$, where $F^{\mu\nu}$ and $F^{\prime\mu\nu}$ are the field strength tensors of the SM photon and dark photon, respectively. The parameter $\epsilon$ can be naturally small when induced at a loop level due to the presence of new heavy charged particles~\cite{Holdom:1985ag}. After a field redefinition to remove the non-diagonal term in the field strength tensors, SM fermions acquire milli-charges under the $U(1)_D$ group with the corresponding interaction mediated by the dark photon (for recent reviews see~\cite{Raggi:2015yfk,Ilten:2018crw,Bauer:2018onh}). The relevant Lagrangian terms for the dark photon with mass $m_{A'}$ then read
\begin{equation}
\label{eq:LAprime}
\mathcal{L}\supset \frac{1}{2}\,m_{A'}^2\,A^{\prime\,2} - \epsilon\,e\,\sum_{f}{q_f\,\bar{f}\,\slashed{A}^\prime\,f},
\end{equation}
where the sum in the second term spans over SM fermions $f$ with electromagnetic charges $q_f$.

In the following, we will focus on scenarios with $m_{A^\prime}\sim \mev - \gev$. In this mass range, dark photons are promising targets for intensity frontier searches and one of just a few renormalizable BSM portals to study at a simplified level. In particular, the values of the kinetic mixing parameter in the allowed region of the parameter space of the dark photon model spanned by just two parameters, $m_{A'}$ and $\epsilon$, can be as large as $\epsilon\sim 10^{-3}$ and lie within reach of many current and future experiments~\cite{Beacham:2019nyx,Alimena:2019zri}. 

Once produced, dark photons can decay back into SM particles with the relevant decay length given in \cref{eq:AprimeGamma}. If dark photons are produced in secondary production processes right in front of the detector, this can allow one to probe boosted $A^\prime$s with the decay length $d_{A^\prime}$ of order meters
\be
\!\!
(c \tau \beta \gamma)_{A^\prime} \sim 1~\m \times \bigg[\frac{10^{-4}}{\epsilon}\bigg]^2 \!
\bigg[\frac{E_{A^\prime}}{100~\gev}\bigg] \!
\bigg[\frac{30~\mev}{m_{A^\prime}}\bigg]^2 \! .
\ee

Additional LLPs with mass of a similar order often arise in such models, e.g. as dark sector particles comprising DM, or in connection to the dark Higgs mechanism that can generate non-zero mass of $A'$. In such scenarios, dark photons can either decay visibly or decay predominantly into dark sector particles. Below, we briefly discuss several simple scenarios with $A^\prime$ accompanied by additional LLPs.

%+++++++++++++++++++++++++++++++++++++++
\subsection{Dark bremsstrahlung\label{sec:darkbrem}}
%+++++++++++++++++++++++++++++++++++++++

One of the most important motivations to search for new light subweakly coupled particles is the role they can play in cosmology and astrophysics, acting as mediators between the SM and DM (for a recent review see e.g.~\cite{Battaglieri:2017aum}). In particular, a light dark photon serves as an important example of such a portal that can yield correct relic density of thermally produced DM due to a generalized weakly interacting massive particle (WIMP) miracle~\cite{Boehm:2003hm,Feng:2008ya} or in a secluded WIMP DM scenario~\cite{Pospelov:2007mp}, as well as due to non-zero temperature effects in the forbidden DM regime~\cite{DAgnolo:2015ujb} (see also Ref.~\cite{Darme:2019wpd} for a recent study in this direction). In addition, depending on the hierarchy between DM and dark photon masses, this scenario can belong to a more general framework of self-interacting DM and can help to resolve problems in understanding DM distribution toward centers of galaxies and galaxy clusters~\cite{Kaplinghat:2015aga} (for review see~\cite{Tulin:2017ara}).

We focus on the model with fermionic DM $\chi$ coupling to SM via a dark photon portal described by \cref{eq:LAprime} and the following additional terms in the Lagrangian
\begin{equation}
\mathcal{L} \supset \bar{\chi}\left(i\,\slashed{D}-m_\chi\right)\chi,
\end{equation}
where $D_\mu = \partial_\mu - i\,g_D\,A^\prime_\mu$, $m_\chi$ is the DM mass and $g_D$ is the dark coupling constant that governs DM interactions with $A'$.\footnote{Here, as well as in the models described in \cref{sec:iDM,sec:darkHiggs}, an additional, close to unity and $\epsilon$-dependent, rescaling factor appears in the effective coupling $g_D$ due to $A^\prime$ field redefinition that leads to \cref{eq:LAprime}. It is, however, negligible for small values of $\epsilon$ considered in our study.} As a result, the parameter space of this model is spanned by four parameters, $m_{A^\prime}$, $m_\chi$, $\epsilon$ and $\alpha_D = g_D^2/(4\pi)$. In the case of a massive dark photon, that we focus on, dark fermion $\chi$ remains electrically neutral after gauge field transformations are applied~\cite{Babu:1997st}.

In the following, we set $\alpha_D=0.1$, a value that lies within perturbativity limits but at the same time is large enough so that the $\alpha_D$-dependent secondary production of LLPs can become sizable. For illustrative purposes, we also assume a fixed mass ratio $m_{\chi}:m_{A^\prime} = 0.6:1$. This corresponds to a particularly interesting mass regime in which $m_\chi < m_{A^\prime} < 2\,m_\chi$. Here, dark photons decay visibly into SM particles with possible striking experimental signatures in distant detectors. At the same time, the DM relic density is set by a freeze-out of direct $\chi\bar{\chi}$ annihilations into SM particles via intermediate $A^\prime$~\cite{Izaguirre:2015yja}, with only a small impact of very efficient annihilation into two dark photons relevant for the forbidden DM regime. Instead, in the case of a larger mass ratio, $m_\chi>m_{A^\prime}$, characteristic for secluded DM, one would typically obtain tiny DM relic density for a chosen value of $\alpha_D$. Further comments about DM relic density in this model are given in \cref{sec:results}. Once the aforementioned assumptions about the mass ratio and the value of $\alpha_D$ are taken into account, there remain only two free parameters of the model: $m_{A^\prime}$ and $\epsilon$. 

\begin{figure*}[tb]
\centering
\includegraphics[width=0.80\textwidth]{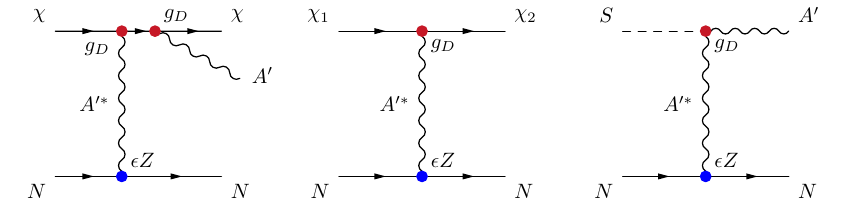}
\caption{Feynman diagrams relevant for secondary production processes discussed in the text. \textbf{Left:} dark bremsstrahlung of $A^\prime$ in DM scattering, \textbf{Center:}  upscattering of a lighter $\chi_1$ fermion into a heavier one $\chi_2$ in an inelastic DM model, \textbf{Right:}  scattering of a secluded dark Higgs boson with $A^\prime$ production.}
\label{fig:feynman}
\end{figure*}

While DM particles $\chi$ are stable, spectacular signatures of this model can come from dark photon decays inside the detector. Importantly, on top of $A^\prime$s produced at the primary IP, further dark photons can come from dark bremsstrahlung processes, $\chi\,T\rightarrow\chi\,T\,A^\prime$. Here, $\chi$ scatters off electron or proton/nucleus target $T$ in the material in front of the detector and radiates off the dark photon (see e.g. Refs.~\cite{Kahn:2014sra,deGouvea:2018cfv,DeRomeri:2019kic}). The relevant Feynman diagram is shown in the left panel of \cref{fig:feynman}. Alternatively, one can study scattering signatures of DM particles off electrons, $\chi\,e^-\rightarrow \chi\,e^-$, that can lead to excess in the number of high-energy EM showers in the detector with no significant nuclear recoil over expected neutrino-induced BG~\cite{Batell:2009di,Batell:2014mga,deNiverville:2016rqh,Battaglieri:2016ggd,Akesson:2018vlm,Buonocore:2018xjk}. 

%------------------------
%------------------------

%+++++++++++++++++++++++++++++++++++++++
\subsection{Inelastic dark matter\label{sec:iDM}}
%+++++++++++++++++++++++++++++++++++++++

Generalizing the aforementioned case of Dirac fermion DM, we also consider the scenario with a pair of Weyl (dark) fermions $\chi_L$ and $\chi_R$ with opposite chirality that share the same charge under the dark gauge group $U(1)_D$; see e.g. Refs.~\cite{Giudice:2017zke,Darme:2018jmx} for recent discussion. When $U(1)_D$ symmetry is spontaneously broken, Majorana mass terms can be generated  on top of the Dirac mass. The relevant Lagrangian can be written as $\mathcal{L}\supset \left(\Psi^T\,C\,M\,\Psi + \textrm{h.c.}\right)$, with $\Psi^T = (\chi_L,\,\chi^c_R)$ and the (real) mass matrix
\begin{equation}
M = \left(\begin{array}{cc}
m_L & M_\chi\\
M_\chi & -m_R
\end{array}\right),
\end{equation}
where the Majorana masses are denoted by $m_L$ and $m_R$ and the Dirac mass by $M_\chi$. By rotating to the mass basis, two dark fermion states appear with masses
\begin{equation}
m_{\chi_{1,2}} = \left|\frac{1}{2}\left(m_L+m_R \mp \sqrt{(m_L-m_R)^2 + 4\,M_\chi^2}\right)\right|.
\end{equation}
The effective Lagrangian then reads
\begin{equation}
\mathcal{L}\supset \sum_{i=1,2}{g_{ii}\,\bar{\chi}_i\gamma^\mu\chi_i\,A^
\prime_\mu} + \left(g_{12}\,\bar{\chi}_2\gamma^\mu\chi_1\,A^
\prime_\mu+\textrm{h.c.}\right),
\end{equation}
where we focus on the case with $g_{ii}\ll g_{12}$ and the dark photon coupling to the SM as shown in \cref{eq:LAprime}. In fact, the diagonal and non-diagonal couplings between dark fermions depend on their mixing angle $\theta$. They are proportional to $\cos{2\theta}$ and $\sin{2\theta}$, respectively, with the mixing angle defined as $\tan{2\theta} = 2\,M_\chi/(m_L+m_R)$. 

As can be seen, in the pseudo-Dirac limit with $M_\chi\gg m_L,m_R$, diagonal couplings are suppressed and two dark fermion states are characterized by a small mass splitting. This corresponds to a well-known scenario first discussed in the context of the DAMA anomaly in DM direct detection searches~\cite{TuckerSmith:2001hy}, which has recently received renewed attention due to possible interesting signatures in intensity frontier searches and other experiments from displaced decays of a heavier fermion, $\chi_2\rightarrow\chi_1\,e^+e^-$, see e.g. Refs.~\cite{Izaguirre:2015zva,Kim:2016zjx,Izaguirre:2017bqb,Darme:2017glc,Berlin:2018pwi,Berlin:2018bsc,Jordan:2018gcd,Berlin:2018jbm,Heurtier:2019rkz}. 

On the other hand, a similar suppression of diagonal couplings can be achieved even for larger Majorana masses by requiring $m_L\approx -m_R$. This then allows one to consider a larger mass splitting between $\chi_1$ and $\chi_2$,
\begin{equation}
\Delta_\chi = \frac{m_{\chi_2}-m_{\chi_1}}{m_{\chi_1}} \simeq \frac{2\,\min{\left\{M_\chi,m_L\right\}}}{|m_L - M_\chi|},
\end{equation}
while the mass eigenstates read $m_{\chi_{1,2}} \simeq \left|m_L\mp M_\chi\right|$. In particular, for $m_L\simeq -m_R\simeq M_\chi$, one obtains a very low mass of $\chi_1$, i.e. $m_{\chi_1}\approx 0$, while the $\chi_2$ mass can remain significantly larger, $m_{\chi_2}\simeq 2\,M_\chi$. On the other hand, for $\chi_1$ to be a cold DM candidate, we require $m_{\chi_1}$ to be not suppressed too much. This can be achieved by e.g. assuming $m_L\simeq -m_R\simeq 2\,M_\chi$ leading to $m_{\chi_2}\simeq 3\,M_\chi \simeq 3\,m_{\chi_1}$. In this case, for increasing the mass of dark fermions, other decay channels might open up on top of the leading one to an electron-positron pair. These include e.g. $\chi_2\rightarrow\chi_1\,\mu^+\mu^-$, as well as decays with hadronic particles in the final state.

While in the pseudo-Dirac limit $\chi_2$ can become long-lived due to a suppressed mass spectrum in its $3$-body decays, the lifetime of a heavier dark fermion becomes smaller for increasing $\Delta_\chi$. As a result, $\chi_2$s often struggle to reach a distant detector before decaying. This opens up a region in the parameter space of the model in which correct DM relic density can be obtained, while current bounds from beam-dump experiments are weakened~\cite{Izaguirre:2017bqb,Darme:2018jmx}.
  
Due to suppression of diagonal couplings, lighter dark fermions, when scattering off the electron or proton target, preferably produce a heavier state, $\chi_1\,T\rightarrow\chi_2\,T$, if kinematically allowed. This is illustrated in the central panel of \cref{fig:feynman}. If such upscattering occurs in front of the detector, subsequent decays of $\chi_2$ might lead to a spectacular signature inside the fiducial volume of the detector. For large mass splitting between both dark fermions, an approximate decay length of boosted $\chi_2$ reads
\begin{widetext}
\be
(c \tau \beta \gamma )_{\chi_2} 
\sim 1~\m \times 
\bigg[\frac{0.1}{\alpha_D}\bigg]
\bigg[\frac{5\times 10^{-4}}{\epsilon}\bigg]^2
\bigg[\frac{E_{\chi_2}}{100~\gev}\bigg]
\bigg[\frac{100~\mev}{m_{\chi_1}}\bigg]^5
\bigg[\frac{300~\mev}{m_{\chi_2}}\bigg]
\bigg[\frac{m_{A^\prime}}{400~\mev}\bigg]^4
\bigg[\frac{2}{\Delta_\chi}\bigg]^5,
\label{eq:chi2lifetime}
\ee
\end{widetext}
while a full expression for the relevant decay width is given in \cref{eq:chi2lifetimeapp}.

When presenting the results in \cref{sec:results}, we follow a simple mass scaling mentioned above with $m_{\chi_2}\sim 3\,m_{\chi_1}$ and take both masses in the $\mev - \gev$ range, which has been chosen for illustrative purposes. As the upscattering cross section decreases with growing dark photon mass, we additionally focus on the case when $m_{A^\prime}$ saturates the minimal value required for on-shell $A^\prime$ to decay invisibly into $\chi_1\chi_2$ pairs; i.e., we assume $m_{\chi_1}:m_{\chi_2}:m_{A^\prime} \sim 1:3:4$. In addition, similar to \cref{sec:darkbrem}, we assume $\alpha_D = 0.1$, where we define $\alpha_D = g_{12}^2/(4\pi)$. Hence, there remain only two free parameters of the inelastic DM (iDM) model that we vary freely when obtaining sensitivity reach plots: $m_{\chi_1}$ and $\epsilon$.

%+++++++++++++++++++++++++++++++++++++++
\subsection{Dark photon with secluded dark Higgs boson\label{sec:darkHiggs}}
%+++++++++++++++++++++++++++++++++++++++

A natural way to introduce a non-zero dark photon mass is to employ a dark Higgs mechanism in which $m_{A^\prime}$ is driven by a vacuum expectation value (vev), $v_S$, of a new SM-singlet complex scalar field $S$ added to the model. In comparison to the SM Higgs boson vev $v_h$, the vev of the new field is assumed to be small, $v_S\ll v_h$, as expected for a low-mass dark gauge boson. This new dark scalar, dubbed dark Higgs boson, can have non-negligible couplings to the SM fermions that either arise thanks to the mixing between $S$ and the SM Higgs boson $H$ or appear at a loop level with the exchange of intermediate $A^\prime$s. The relevant Lagrangian terms read~\cite{Batell:2009yf,Batell:2009di}
\begin{equation}
\mathcal{L}\supset |D_\mu\,S|^2 + \mu_S\,|S|^2 - \frac{\lambda_S}{2}\,|S|^4 - \frac{\lambda_{SH}}{2}\,|S|^2|H|^2,
\label{eq:LDHDP}
\end{equation}
where $D_\mu = \partial_\mu - i\,g_D\,A^\prime_\mu$, while dark photon coupling to the SM is given by~\cref{eq:LAprime}. The phenomenology of new BSM light scalars in connection to intensity frontier searches has been extensively studied in the literature; see, e.g. Refs.~\cite{Clarke:2013aya,Alekhin:2015byh,Feng:2017vli,Evans:2017lvd,Beacham:2019nyx} and references therein.

Assuming small mixing, $\lambda_{SH}\ll \lambda_S$, and by solving relevant tadpole equations, one can rewrite the dark scalar mass in terms of $v_S$ which reads $m_S^2 = 2\,\mu_S^2 - \lambda_{SH}\,v_h^2 = 2\,\lambda_S\,v_S^2$. At the same time, the dark photon mass induced by the vev of $S$ is given by $m_{A^\prime}^2 \simeq g_D^2\,v_S^2$. As a result, $m_S^2\sim m_{A'}^2\times\lambda_S/(2\pi\alpha_D)$, i.e., both dark scalar and dark vector masses are naturally of similar order. In particular, keeping a small value of $\lambda_{SH}$ allows one to suppress a contribution to the dark Higgs boson mass from the vev of the SM Higgs, $\lambda_{SH}\,v_h^2$.

In the following, we require the mixing term to be very small, $\lambda_{SH}\sim (m_S^2/v_h^2)\lesssim 10^{-6}$. This results in highly suppressed values of a mixing angle between the dark and SM Higgs bosons that typically lies below the reach of current and future searches. It then plays a negligible role in our phenomenological analysis, while we will comment on it when discussing current bounds on this scenario. 

In fact, the dominant couplings of such a secluded $S$ to the SM fermions arise via the dark photon portal and an unsuppressed coupling between $S$ and $A'$ that appears after $U(1)_D$ symmetry breaking, $\mathcal{L}\supset g_D\,m_{A^\prime}\,S\,A^{\prime\mu}A^\prime_\mu$. This can lead to efficient co-production of light scalars in any process leading to $A^\prime$ production, where $S$ can be bremmed off the vector leg.  As a result, a flux of dark scalars going toward the detector can be obtained along with dark photons produced at the primary IP.

Importantly, unlike with dark photons which can decay promptly depending on the value of the kinetic mixing parameter $\epsilon$, dark Higgs bosons in such a scenario are typically very long-lived if $m_S<m_{A^\prime}$. This is because their dominant decays into a pair of SM fermions, e.g. $S\rightarrow e^+e^-$, proceeds at a loop level through a triangle diagram with intermediate vector states~\cite{Batell:2009yf,Batell:2009di,Darme:2017glc}. A typical lifetime of $S$ then reads
\be
\!\!\tau_{S}
\sim 0.1~\second \times
\bigg[\frac{0.1}{\alpha_D}\bigg]
\bigg[\frac{10^{-3}}{\epsilon}\bigg]^4
\bigg[\frac{20~\mev}{m_S}\bigg]
\bigg[\frac{m_{A^\prime}}{30~\mev}\bigg]^2 \! .
\label{eq:Slifetime}
\ee
For larger values of $m_S$, the di-muon final state becomes possible, as well as 3- or even 2-body decays with one or two on-shell dark photons in the final state that reduce the lifetime of $S$. These, however, turn out to be irrelevant for the mass range of our interest and for the assumed mass ratio in our benchmark scenario $m_S = (3/4)\,m_{A^\prime}$, as discussed in \cref{sec:results}. We also fix the dark coupling constant to be $\alpha_D=0.1$, which leads to only two free parameters of the model varied to obtain the sensitivity reach plots: $m_{A^\prime}$ and $\epsilon$.

As evident from \cref{eq:Slifetime}, dark scalars in the considered scenario are effectively stable at collider scales. Thus, once a flux of dark Higgs bosons is produced, they will only rarely decay before reaching the detector. Instead, when traveling, they can scatter off nuclei and electrons producing secondary dark photons, $S\,T\rightarrow A^\prime\,T$, as illustrated in the right panel of \cref{fig:feynman}. The dark photons can then decay inside the detector. If such a secondary $A^\prime$ production can take place in the vicinity of the decay vessel, it allows one to probe small dark photon lifetimes.

%============================================================
\section{Experiments\label{sec:experiments}}
%============================================================

In order to illustrate the impact of secondary production of LLPs on the reach, we study the expected sensitivity for three representative proposed experiments. Since this effect becomes more evident for highly boosted particles, we focus first on the LHC experiments, namely FASER and MATHUSLA. We also study the possible impact of secondary production of LLPs on the reach plots relevant for the SHiP experiment.\footnote{This selection of the experiments is motivated by the scope of different experimental approaches they cover. We note, however, that the same ideas apply to other existing or proposed experiments such as CODEX-b~\cite{Gligorov:2017nwh,Aielli:2019ivi}, NA62~\cite{Dobrich:2018ezn} or SeaQuest~\cite{Berlin:2018pwi}.}

As discussed in more detail in \cref{sec:modeling} and \cref{sec:scatteringsigma}, the dominant contribution to the secondary production rate comes from $Z^2$-enhanced coherent scatterings off nuclei. They are characterized by a low-momentum transfer to the nuclear target. This is different from the neutrino-induced neutral hadron BG, which, in order for the recoiled hadron to be energetic, requires larger momentum transfer. 

As a result, for the signal of interest to us, no significant recoil energy is expected and veto layers in front of the decay vessel often remain inactive, unless the scattering process takes place right in front of them. We then always require in our analysis that $p_{\textrm{recoil}}<1~\gev$, while the typical recoil momentum is even smaller, $p_{\textrm{recoil}}\sim \mathcal{O}(100~\mev)$. In addition, we exclude all scattering processes happening in the last three hadronic interaction lengths, $3\,\lambda_{\textrm{had,int}}$, of the material lying in the most immediate neighborhood of the veto layers. While the last condition follows our generally conservative approach, we also discuss in \cref{sec:resultsaftercutschange} how the sensitivity reach can be changed by ameliorating, or strengthening, this cut.

\begin{figure*}[tb]
\centering
\begin{subfigure}{0.89\textwidth}
\centering
    \caption{FASER}\label{fig:expFASER}
    \includegraphics[width=1.0\textwidth]{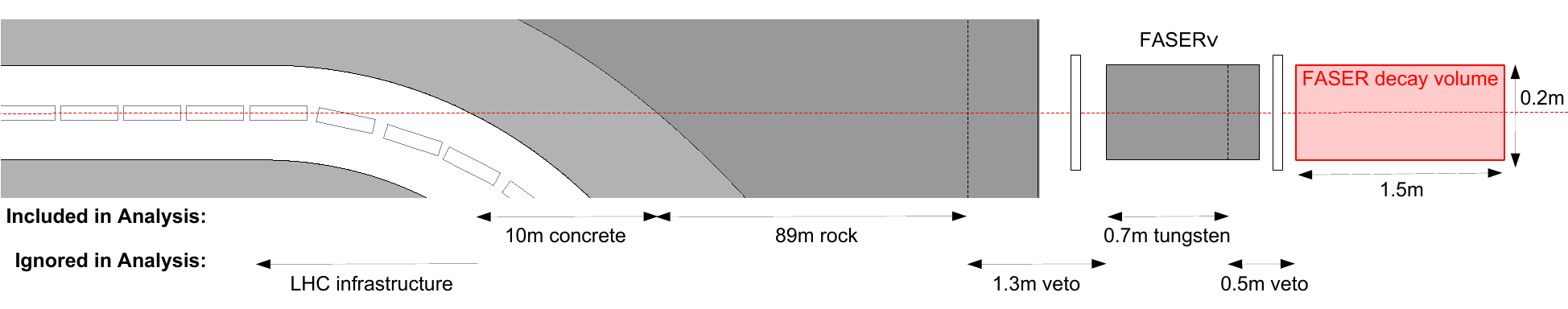}
\end{subfigure}
\begin{subfigure}{0.89\textwidth}
\centering
    \caption{FASER 2}\label{fig:expFASER2}
    \includegraphics[width=1.0\textwidth]{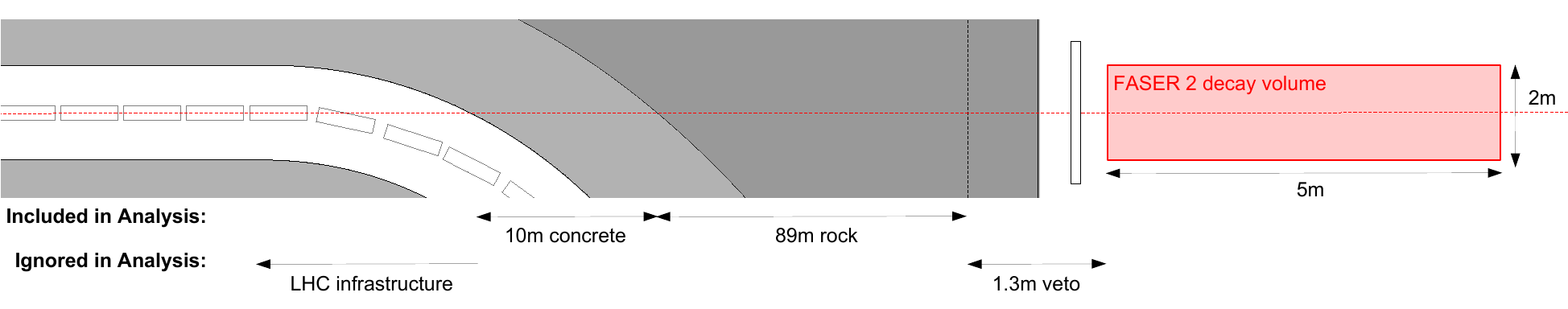}    \end{subfigure}
\begin{subfigure}{0.89\textwidth}
\centering
    \caption{MATHUSLA}\label{fig:expMATHUSLA}
    \includegraphics[width=1.0\textwidth]{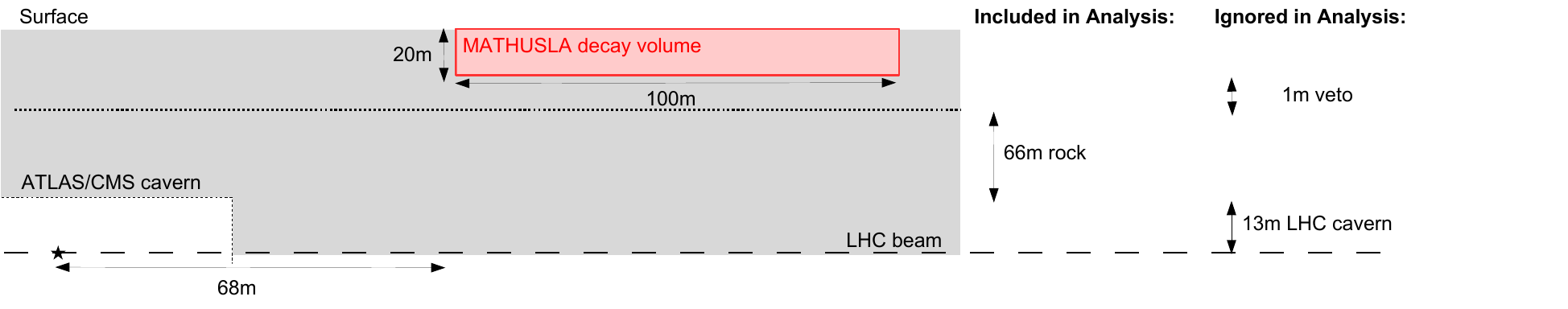}
\end{subfigure}
\begin{subfigure}{0.89\textwidth}
\centering
    \caption{SHiP}\label{fig:expSHiP}
    \includegraphics[width=1.0\textwidth]{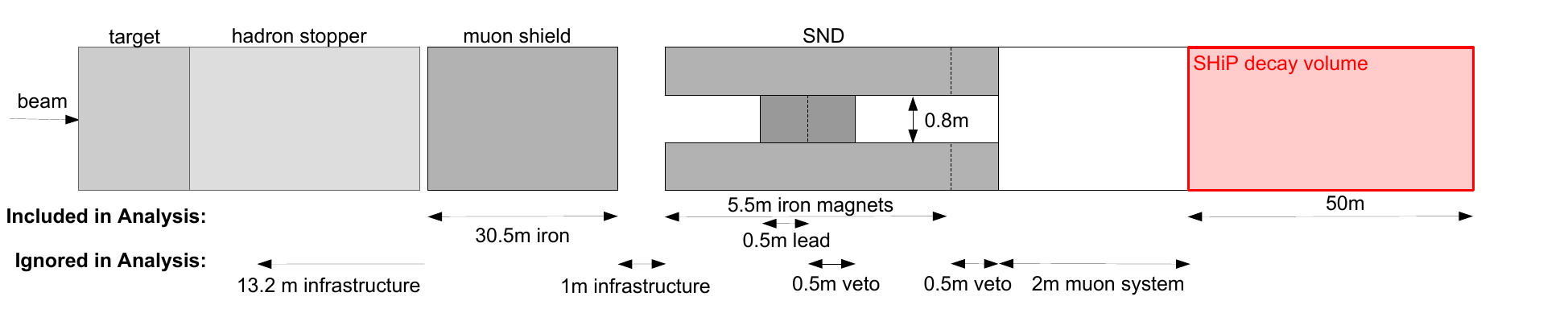}
\end{subfigure}
\caption{Simplified detector designs used in the study of primary and secondary production of LLPs with subsequent decay in the decay vessel. Elements of the detectors and parts of material in front were taken into account or excluded from the analysis, as indicated in the figures. From the top to bottom the figures correspond to the FASER, FASER~2, MATHUSLA and SHiP experiments.}
\label{fig:experiments}
\end{figure*}

The geometry for each experiment is illustrated in \cref{fig:experiments}. Below, we briefly describe basic details of the experiments that are relevant for our discussion.  A more detailed description of simplified detector designs employed in our study is given in \cref{sec:benchmarkgeometries}, and physics cuts applied to signal events in each experiment are discussed in \cref{sec:physicscuts}.

%------------------------
%------------------------

%+++++++++++++++++++++++++++++++++++++++
\subsection{FASER}
%+++++++++++++++++++++++++++++++++++++++

\paragraph{Basic detector design} The FASER experiment has been proposed~\cite{Feng:2017uoz} to search for LLPs that can be abundantly produced in the forward direction of the LHC and subsequently decay in a distant detector~\cite{Feng:2017vli,Feng:2017uoz,Kling:2018wct,Feng:2018noy,Ariga:2018uku}. Following a preparation of the detailed detector design~\cite{Ariga:2018zuc,Ariga:2018pin}, the experiment was approved to take data during LHC Run 3. In addition, a larger version of the detector has been considered~\cite{Ariga:2018uku}, dubbed FASER 2, that could collect data during the HL-LHC era. In the following, we will present the results for both these experiments, as well as for a small version of the FASER detector left for the entire HL-LHC era, that we further denote by FASER (HL-LHC).

The FASER detector consists of a cylindrically-shaped decay vessel with length $\Delta$ and radius $R$, followed by a spectrometer and calorimeter. At the front of the detector, there is a veto layer with a primary role to reject muon-induced BG. The detector is placed in the side LHC tunnel TI12 about $L=480~\textrm{m}$ away from the ATLAS IP along the beam collision axis. The actual detector position is shielded from the ATLAS IP by $10~\textrm{m}$ of concrete and about $90~\textrm{m}$ of rocks, on top of various elements of the LHC infrastructure. The detector geometries and integrated luminosities corresponding to all three aforementioned versions of the FASER experiment read as follows:
\begin{itemize}
\item \textbf{FASER:}
$\Delta=1.5~\m$, $\text{R}=10~\cm$, $\mathcal{L}=150~\ifb$,
\vspace{-0.1cm}
\item \textbf{FASER (HL):}
$\Delta=1.5~\m$, $\text{R}=10~\cm$, $\mathcal{L}=3~\iab$,
\vspace{-0.1cm}
\item \textbf{FASER 2:} 
$\Delta=5~\m$, $\text{R}=1~\m$, $\mathcal{L}=3~\iab$.
\end{itemize}

On top of the main FASER detector described above, recently an additional detector component has been proposed~\cite{Abreu:2019yak}, dubbed FASER$\nu$, to be placed in front of the main detector with a primary role to study interactions of high-energy neutrinos. This is an emulsion detector covering the transverse size of FASER with tungsten layers interleaved with emulsion films. The total length of tungsten in FASER$\nu$ is $1~\textrm{m}$. Both in front of FASER$\nu$ and after the emulsion detector, there are scintillator layers acting as a veto for muon-induced BG.

As far as the secondary production of LLPs is concerned, FASER can employ all the material in front of the detector that could serve as a target. In particular, these include the aforementioned concrete shielding and rock, but also FASER$\nu$ subdetector. In the following, we will take into account secondary production processes happening in all these places. To this end, we will assume that both FASER (Run 3) and FASER (HL-LHC) are equipped with the FASER$\nu$ subdetector, while FASER 2 does not contain it due to its larger size that needs to be fitted in the TI12 tunnel. Further details about detector design are shown in \cref{fig:expFASER,fig:expFASER2} for FASER and FASER 2, respectively.

\paragraph{LLP decay signature in the decay vessel} The LLPs produced at the ATLAS IP or in secondary processes in the material in front of FASER, can decay inside the fiducial volume of the detector. The expected BG in searches for such decays can be suppressed to negligible levels by employing basic signal characteristics. First, a high-energy cut on visible energy in the detector, $E_{\textrm{visible}}>100~\textrm{GeV}$, can be applied to reject all soft BG particles with only a small impact on the expected number of signal events. In addition, pointing and timing information can be used to further associate two charged tracks detected from LLP decay with $pp$ collisions happening at the ATLAS IP. As a result, FASER can operate in an essentially BG free environment~\cite{Ariga:2018zuc}. In the following, we also assume $100\%$ detection efficiency for easier comparison with other experiments.

%+++++++++++++++++++++++++++++++++++++++
\subsection{MATHUSLA}
%+++++++++++++++++++++++++++++++++++++++

\paragraph{Basic detector design} Similar to FASER, the MATHUSLA experiment has been proposed~\cite{Chou:2016lxi} to take advantage of possible abundant production of long-lived BSM particles in $pp$ collisions at the LHC that could have escaped detection in existing searches~\cite{Curtin:2018mvb}. It employs a large-scale detector to be placed above the CMS IP to take data during the HL-LHC era. Here, we follow the proposed MATHUSLA100 design~\cite{Alpigiani:2018fgd} with additional excavation performed to position the detector partially underground, closer to the CMS IP, in order to maximize the physics reach~\cite{Lubatti:2019vkf}. 

The main part of MATHUSLA is an air-filled decay volume of the size of $20~\textrm{m}\times 100~\textrm{m}\times 100~\textrm{m}$ which is followed by a tracking system that spans over the entire area of the detector and is placed on top of the decay volume. In addition, scintillator layers are placed at the bottom and on the sides of the decay volume to veto charged particles entering the detector. The precise geometry of the decay volume that we use reads
\be
80~\textrm{m} < x& < 100~\textrm{m},\\
-50~\textrm{m} < y& < 50~\textrm{m},\\
68~\textrm{m} < z& < 168~\textrm{m},
\label{eq:MATHUSLAgeometry}
\ee
where $x$ corresponds to an upward direction, while $z$ is the direction along the LHC beam pipe and the origin of the coordinate system is placed in the CMS IP. When modeling secondary production of LLPs, we take into account the rock separating MATHUSLA from the CMS cavern. We illustrate this design in \cref{fig:expMATHUSLA}.

\paragraph{LLP decay signature in the decay vessel} In our analysis, we take into account all the LLP decay events producing two charged SM tracks that happen inside the decay volume of MATHUSLA. We assume $100\%$ efficiency of detection and the daughter track separation for energies $p_{\textrm{daughter}}>1~\textrm{GeV}$. Position and timing information about the tracks is used to identify the vertex in the decay volume. This, along with the direction reconstruction of the coming LLP, allows one to greatly reduce various sources of BG including cosmic rays, muons from $pp$ collisions at the CMS IP, and atmospheric-neutrino-induced BG. Since the direction of the LLP is only mildly changed by the recoil in the secondary production processes, as we discuss in \cref{sec:scatteringsigma}, we will assume in the following that the search of our interest can be performed with zero BG.

%+++++++++++++++++++++++++++++++++++++++
\subsection{SHiP} 
%+++++++++++++++++++++++++++++++++++++++

\paragraph{Basic detector design} The proposed SHiP detector~\cite{Bonivento:2013jag} is a fixed-target experiment that will use the SPS beam of $400~\textrm{GeV}$ protons incident on a target material made out of titanium-zirconium doped molybdenum alloy and tungsten. A nominal value of $N_{\textrm{POT}}=2\times 10^{20}$ protons on target (POT) allows one to potentially produce a large number of LLPs that could subsequently decay in a distant decay vessel. 

In between the target and decay vessel, there is a place for a hadronic stopper and an active muon shield with an essential role to reduce muon-induced BG to negligible levels. The active muon shield is followed by the Scattering and Neutrino Detector (SND) which has been designed to study the interactions of SM neutrinos and light DM particles. After the SND, a $50$ m long decay vessel begins. Decays of LLPs into charged SM particles are detected by observing the resulting tracks in the Decay Spectrometer (DS).

Since the initial release of the technical proposal~\cite{Anelli:2015pba}, the SHiP detector design underwent revision primarily in order to reduce the cost and weight of the active muon shield while maintaining its assumed high-quality performance. In our analysis, we follow the recent technical update~\cite{Ahdida:2654870}, while we note that further possible modifications to the design might require updating the results in the future. We simplify a complicated design of the planned SHiP detector. However, when doing so, we keep its key components that could play an essential role for the secondary production of LLPs. A schematic drawing of the SHiP detector design is shown in \cref{fig:expSHiP}.

The most important part of the detector with regards to secondary production of LLPs is the SND and its surrounding magnet. Scattering processes occurring there can lead to LLPs produced only several meters in front of the decay vessel. The SND consists of an emulsion detector which is followed by the SND muon system. Since the latter can partially act as a front veto for the decay vessel, we exclude from our analysis all scattering events happening in the material lying in its close neighborhood, within $3\,\lambda_{\textrm{had,int}}$. Notably, on top of absorption of soft particles with $p<1~\gev$ in the emulsion detector in front of the SND muon system, charged soft particles produced in the emulsion detector can also be swept away by the SND magnet and then never reach the SND muon system. This then prevents the events from being rejected as combinatorial BG. 

The emulsion detector in the SND is also equipped with electronic tracking layers that can time stamp the events. These could detect even soft activity, corresponding to energy $\mathcal{O}(100~\textrm{MeV})$, in the case of secondary production taking place inside the emulsion detector, even though the SND muon system will not be activated. In the following, we assume that events will not be rejected based solely on this soft activity in the emulsion detector. On the other hand, even if such a rejection is present, additional secondary production processes can take place in the surrounding magnets and in the muon shield that will not be vetoed. These events will then always contribute to the secondary production rate.

\paragraph{LLP decay signature in the decay vessel} A detailed reconstruction of signal events in SHiP with two charged and energetic tracks from LLP decays within the decay volume, as well as BG discrimination, employs a number of observables including e.g. the momentum of detected tracks and their impact factor with respect to the target at the IP. In the following, we apply a simplified acceptance procedure that is primarily based on the momentum of the visible tracks coming from the LLP decay. In particular, we require each visible track to have $p\gtrsim 1~\textrm{GeV}$. 

In fact, similar to MATHUSLA, we find that the precise value of this low momentum threshold does not play an important role, at least as far as secondary production of LLPs is concerned. In this regime, even setting the lower limit for the visible energy at the level of $E_{\textrm{vis}}>10~\textrm{GeV}$ leads to very similar results, as discussed in \cref{sec:resultsaftercutschange}. 

When obtaining the sensitivity reach plots, we assume that SHiP can operate in zero BG environment and can detect signal events with $100\%$ efficiency. See e.g. Ref.~\cite{Ahdida:2654870} for further discussion about BG in SHiP and Ref.~\cite{Anelli:2015pba} for more realistic studies of the efficiency.

%============================================================
\section{Details of modeling\label{sec:modeling}}
%============================================================

As discussed above and illustrated in Fig.~1, models with more than a single LLP can effectively lead to the production of BSM species in both initial $pp$ and $pN$ interactions at the LHC or in the target material, as well as in scattering processes taking place more closely to the decay vessel. We refer to the former as primary production, while to the latter as secondary.

Once produced, LLPs can travel some distance and, eventually, decay in the detector leading to a visible signal of two oppositely-charged tracks. The probability of this happening in the decay vessel depends on the decay length of a boosted LLP, as well as on the geometrical acceptance of the detector. 

Below, we briefly summarize our analysis, while further details, including the expressions for relevant branching ratios, decay widths and scattering cross sections are given in \cref{sec:decays,sec:scatteringsigma}. In \cref{tab:processes} we list the processes that we consider in the models listed above. For each we identify both $\textrm{LLP}_1$ and $\textrm{LLP}_2$. In particular, the $\textrm{LLP}_2$ particles produced either at the IP or in secondary production can eventually decay in the detector, leading to observable signatures. On the other hand the intermediate particles denoted by $\textrm{LLP}_1$ are either stable or semi-stable and do not generate any decay signatures. The primary production of both types of LLPs at the IP is typically associated with rare decays of mesons originating from $pp$ and $pN$ collisions, although other processes are also possible, as discussed below.

\begin{table*}
\centering
%\begin{ruledtabular}
\renewcommand{\arraystretch}{1.3}
\begin{tabular}{|c|c|c|c|c|c|c|c|}
\hline
\hline
 \multicolumn{3}{|c|}{Model description} &  \multicolumn{3}{c|}{Primary production (IP)} & Secondary production & Decays
 (detector)\\
  Model &  $\textrm{LLP}_1$ & $\textrm{LLP}_2$ & $pp\!\rightarrow\!\textrm{mesons}$& $\ldots\!\rightarrow\!\underline{\textrm{LLP}_1}$ & $\ldots\!\rightarrow\!\underline{\textrm{LLP}_2}$ & $\textrm{LLP}_1\!\rightarrow\! \underline{\textrm{LLP}_2}$ & $\textrm{LLP}_2 \!\rightarrow\! \textrm{SM}$\\
\hline
\hline
Dark bremsstrahlung  & $\chi$ & $A^\prime$ & $\pi^0,\eta$& $\textrm{meson}\!\rightarrow\! \gamma\,\underline{\chi}\underline{\chi}$ & $\textrm{meson}\!\rightarrow\! \gamma\,\underline{A^\prime}$ & $\chi\,T\!\rightarrow\! \chi\,\underline{A^{\prime}}\,T$ & $A^\prime\!\rightarrow\! e^+e^-$\\
(\cref{sec:darkbrem}) & &  &  &  &  & & \\
\hline
Inelastic dark matter & $\chi_1$ & $\chi_2$ & $\pi^0,\eta,\eta',$ &  \multicolumn{2}{c|}{$\textrm{meson, brem, Drell-Yan}\!\rightarrow\!A^\prime$} & $\chi_1\,T\!\rightarrow\!\underline{\chi_2}\,T$ & $\chi_2\!\rightarrow\!\chi_1\,\ell\ell$\\
(\cref{sec:iDM}) &  &   & $\rho,\omega$& $A^\prime\!\rightarrow\! \underline{\chi_1}\,\chi_2$ & $A^\prime\!\rightarrow\! \chi_1\,\underline{\chi_2}$ & & \\
& & & & $\chi_2\!\rightarrow\!\underline{\chi_1}\,\ell\ell$ & & & \\
\hline
Secluded dark Higgs & $S$ & $A^\prime$ & $\pi^0,\eta$ & $\textrm{meson}\!\rightarrow\! (\gamma)A^\prime\underline{S}$ & $\textrm{meson}\!\rightarrow\! (\gamma/\pi)\underline{A^\prime}$ & $S\,T\!\rightarrow\!\underline{A^\prime}\,T$ & $A^\prime\!\rightarrow\! e^+e^-$\\
(\cref{sec:darkHiggs}) &  & & $\rho$  &  &  & & \\
\hline
\hline
\end{tabular}
%\end{ruledtabular}
\caption{A list of light long-lived particles: $\textrm{LLP}_1$ (intermediate ones)  and $\textrm{LLP}_2$ (decaying in the detector) for each of the models discussed in \cref{sec:darkbrem,sec:iDM,sec:darkHiggs}. In each case we show
the relevant production and decay processes that we consider in the analysis. Primary production of both $\textrm{LLP}_1$ and $\textrm{LLP}_2$ takes place at the IP, as illustrated in \cref{fig:idea}. Secondary production away from the IP corresponds to scatterings of $\textrm{LLP}_1$ off electron or nuclei target $T=e^-,N$ that lead to $\textrm{LLP}_2$ particles produced in front of the detector. A pair of SM particles in the final state of a heavier dark fermion decay in the model with inelastic dark matter, $\chi_2\rightarrow \chi_1\,\ell\ell$, typically corresponds to $e^+e^-$ or $\mu^+\mu^-$, but can also denote hadronic final states, if kinematically available.
\label{tab:processes}}
\end{table*}

%+++++++++++++++++++++++++++++++++++++++
\subsection{Primary production of the LLPs\label{sec:primaryprod}}
%+++++++++++++++++++++++++++++++++++++++

Light new physics particles can be produced in high-energy hadronic interactions in a number of processes. The dominant production channels depend on the BSM model of interest and on the mass of the LLP. We begin our discussion by considering the primary production of dark photons. As shown in \cref{tab:processes}, in models involving dark bremsstrahlung of $A'$ or containing a secluded dark Higgs boson, these processes determine the primary flux of $\textrm{LLP}_2$ identified with a dark photon itself. In addition, the intermediate production of on-shell $A^\prime$s is a crucial step leading to primary fluxes of $\chi_1$s and $\chi_2$s in the scenario with inelastic DM.
\newline

\paragraph{Dark photons} Given our focus on $\textrm{MeV}-\textrm{GeV}$ mass range dark photons, the dominant production processes relevant for our simulations are the following:

\begin{description}
\item[Meson decays] Light dark photons can most efficiently be produced in rare decays of mesons, if kinematically available. We simulate meson distributions produced in $pp$ collisions at the LHC and $pN$ collisions with the molybdenum target at SHiP with the Monte-Carlo (MC) event generator \texttt{EPOS-LHC}~\cite{Pierog:2013ria}, as implemented in the \texttt{CRMC} simulation package~\cite{CRMC}. In our simulations, we take into account possible rare BSM decays of pions, $\eta$ and $\eta'$ mesons, as well as vector mesons $\rho$ and $\omega$. We focus on the dominant decay channels $\pi^0,\eta,\eta'\rightarrow\gamma A'$ and $\rho,\omega\rightarrow\pi A'$.

\item[Proton bremsstrahlung of $A^\prime$] Dark photons can also be efficiently produced due to bremsstrahlung in coherent proton scatterings. This is especially relevant for $A^\prime$ heavier than the threshold for production in rare pion and $\eta$ meson decays.

We model the bremsstrahlung of $A^\prime$ following the Fermi-Weizsacker-Williams (FWW) approximation and taking into account an additional proton form factor to allow for off-shell mixing with vector mesons $\rho$ and $\omega$, see e.g. Refs.~\cite{Faessler:2009tn, Blumlein:2013cua, deNiverville:2016rqh, Feng:2017uoz} for a recent discussion. The mixing with vector mesons leads to an increased production rate of dark photons in proton bremsstrahlung for $m_{A'}\sim 775~\textrm{MeV}$. 

In the case of MATHUSLA, focus on the high-$p_T$ regime of the LHC makes it more challenging to directly apply the FWW formalism. Instead, here we approximate the relevant contribution to the dark photon spectrum by rescaling the spectrum of vector meson $\rho$ by the appropriate mixing angle $\theta_V$ as discussed in Appendix D of Ref.~\cite{Berlin:2018jbm}. We note, however, that this contribution plays a subdominant role with respect to the one from meson decays in the region of parameter space of the inelastic DM model probed by MATHUSLA. 

\item[Hard processes] For even heavier dark photons, with masses $m_{A'}\gtrsim 1.5~\textrm{GeV}$, a hard scattering contribution from Drell-Yan $A'$ production can start to play a dominant role. We take this into account, although it only concerns a small part of the parameter space of the inelastic DM model under study that is relevant for SHiP. 

\item[Other (subdominant) processes] Dark photons could also be produced in various other processes, e.g. in subsequent showers, but these have been found to be subdominant and, therefore, we neglect them in the following. As discussed below, secondary production becomes the most prominent in regions of the parameter space characterized by a relatively short lifetime where a large boost factor is required for the LLP to reach the detector before decaying. It is then sufficient for us to focus on the high-energy part of the LLP spectrum that is dominantly associated with the initial proton interactions in the target. 
\end{description}

The primary production of other LLPs listed in \cref{tab:processes} employs intermediate spectra of mesons or $A^\prime$s that are generated as discussed above. Below, we describe the relevant processes leading to a flux of these LLPs traveling from the IP toward the detector, for each of the models under study, cf. \cref{sec:darkbrem,sec:iDM,sec:darkHiggs}.

\paragraph{Dark bremmstrahlung} For the benchmark scenario described in \cref{sec:darkbrem}, a flux of dark matter $\chi$ particles going toward the detector comes primarily from $3$-body decays of light pseudoscalar mesons via off-shell dark photon, i.e. $\pi^0,\eta\rightarrow \gamma\,A^{\prime\,\ast}\rightarrow \gamma\,\chi\chi$~\cite{DeRomeri:2019kic}. The contribution from other processes such as proton bremsstrahlung could become important for heavier $\chi$, but it is subdominant in the mass range of our interest. 
\newline

\paragraph{Inelastic dark matter} The benchmark model with inelastic DM discussed in \cref{sec:iDM} is characterized by the dark photon mass exceeding the masses of two dark fermions, $m_{A^\prime}>m_{\chi_1}+m_{\chi_2}$. In this case, the dominant $2$-body decays of on-shell dark photons into the $\chi_1\chi_2$ pair become possible. The parent dark photon flux is governed by one of the production processes discussed above, depending on the $A^\prime$ mass.

In addition, since $\chi_2$s are not stable, they decay into $\chi_1$ and, typically, an electron-positron pair. These decays allow one to detect $\chi_2$s, when they happen inside the decay vessel. On the other hand, when $\chi_2$s decay before reaching the decay vessel, the resulting lighter fermions additionally contribute to the flux of $\chi_1$ relevant for secondary production discussed below. We take such displaced $\chi_2$ decays into account in our simulations.
\newline

\paragraph{Secluded dark Higgs boson} As discussed in \cref{sec:darkHiggs}, the secluded dark Higgs boson that couples to the SM predominantly via the dark photon, can be efficiently co-produced in any of the $A^\prime$ production mechanisms described above. In particular, for the mass range of our interest, the most important such production comes from $3$-body meson decays $\pi^0,\eta\rightarrow \gamma\,A^{\prime\,\ast}\rightarrow\gamma A' S$ with intermediate dark photon, as well as from $2$-body decays of vector mesons such as $\rho\to S\,A^\prime$.

%+++++++++++++++++++++++++++++++++++++++
\subsection{Secondary production of the LLPs\label{sec:secondaryprod}}
%+++++++++++++++++++++++++++++++++++++++

On top of the LLPs produced in the vicinity of the primary proton IP, in models of our interest it is also possible for them to appear in secondary production processes in material closer to the decay vessel. The relevant Feynman diagrams are shown in \cref{fig:feynman} in which an incoming particle $LLP_1$, that corresponds to $\chi$, $\chi_1$ or $S$ in the models described in \cref{sec:darkbrem,sec:iDM,sec:darkHiggs}, respectively, produces an outgoing species $LLP_2$ with possibly much smaller lifetime, where $LLP_2 = A^\prime$ or $\chi_2$ in the considered scenarios.

These scatterings can occur on both electrons and nuclei with the latter giving the dominant contribution, especially in the regime of coherent scatterings with $Z^2$ enhancement. In addition, the coherent scatterings are associated with a small nuclear target recoil, $p_{\textrm{recoil}}\sim \mathcal{O}(100~\textrm{MeV})$, that typically does not cause any veto activation, as well as only mildly affects the momentum of LLPs. While we perform MC simulations to account for the latter effect, we note that a very good approximation of the sensitivity reach can be obtained when working in the simplified, collinear regime with $p_{\textrm{LLP}_1}\approx p_{\textrm{LLP}_2}$. A more detailed discussion of the relevant scattering cross sections is given in \cref{sec:scatteringsigma}. 

%+++++++++++++++++++++++++++++++++++++++
\subsection{Event rate}
%+++++++++++++++++++++++++++++++++++++++

When LLPs produced in one of either primary or secondary production processes decay inside the decay vessel, this can lead to a visible signal in the detector. The number of expected signal events depends on the relevant production rates, as well as on the decay in volume probability that takes into account the acceptance factor $\mathcal{A}$. The latter depends on the geometry of the detector, as well as on the efficiency to generate and detect visible charged tracks satisfying experimental cuts.

In this study, we perform full numerical MC simulations, which takes into account the interaction kinematics and experimental geometry, to obtain the sensitivity reach plots presented in \cref{sec:results}. Although in a strict sense not identical, it is illustrative to consider the probabilities $\mathcal{P}_{\textrm{prim.}}$ for a LLP$_2$ produced at the IP and $\mathcal{P}_{\textrm{sec.}}$ for an LLP$_1$ with subsequent interaction producing a LLP$_2$ to lead to the signal in the detector. 

In the case of primary production, the decay in volume probability of the LLP with momentum $\vec{p}$ can be written as
\begin{equation}
\mathcal{P}_{\textrm{prim.}}(\vec{p}) = e^{-L/d}\,\left(1 - e^{-\Delta/d}\right)\,\mathcal{A}(\theta,\phi),
\label{eq:probprimary}
\end{equation}
where $L$ is the distance to the beginning of the decay vessel and $\Delta$ is the length of the vessel. The decay length of the LLP in the ultrarelativistic regime reads $d =c \tau \beta \gamma  \simeq c\tau E/m$ where $\tau$ is the LLP lifetime, $m$ corresponds to its mass, while the LLP energy is given by $E$. The acceptance factors $\mathcal{A}$ relevant for each of the experiments are included in numerical simulations. 

For the LLPs coming from secondary production, the decay in volume probability needs to be convoluted with the scattering rate, which can differ for the various materials in which secondary production occurs. For a single incoming $LLP_1$, the relevant probability to produce a $LLP_2$ which then decays inside the detector reads
\begin{equation}
\mathcal{P}_{\textrm{sec.}}(\vec{p}) = \int_{x_\textrm{min}}^{x_\textrm{max}}{\frac{dx}{\ell_{\textrm{int.}}}\,e^{-x/d}\,\left(1-e^{-\Delta/d}\right)}\,\mathcal{A}(\theta,\phi).
\label{eq:probsecondary}
\end{equation}
Here the integration limits $x_{\textrm{min}}$ and $x_{\textrm{max}}$ correspond to the distance to the decay vessel. They are dictated by the geometry of the detector and surrounding material, as well as by veto-related requirements discussed above. The interaction length is given by $\ell_{\textrm{int.}}^{-1} = \sigma(E) \times (\rho/m_T)$, where $\sigma$ is the scattering cross section per nucleus relevant for the secondary production mechanism, $\rho$ is the density of material, $m_T$ is the mass of the target nucleus and in \cref{eq:probsecondary} we assumed that $\ell_{\textrm{int.}}\gg (x_{\textrm{max}} - x_{\textrm{min}})$ which is always the case for scenarios of our interest. Contributions associated with various detector components are then summed up to obtain the total event rate.

As manifest in \cref{eq:probsecondary}, secondary production processes are typically subdominant with respect to primary ones due to additional suppression for the small scattering cross section. However, since secondary production can take place much closer to the detector, $x_{\textrm{min}}\ll L$, it allows one to probe the short lifetime regime where the contribution from primary production is already highly suppressed by the exponential factor in \cref{eq:probprimary}, $\exp{\left(-L/d\right)}\ll 1$. 

%============================================================
\section{Results for Scattering off Nuclei\label{sec:results}}
%============================================================

\begin{figure*}[tb]
\centering
\includegraphics[width=0.48\textwidth]{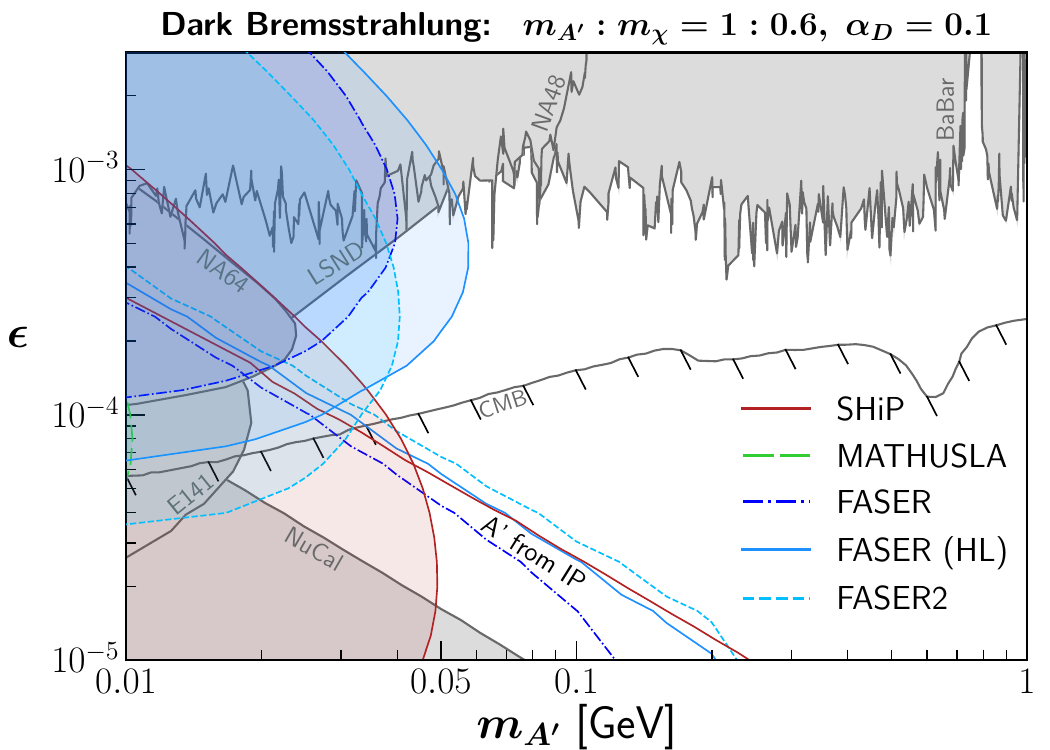}
\includegraphics[width=0.48\textwidth]{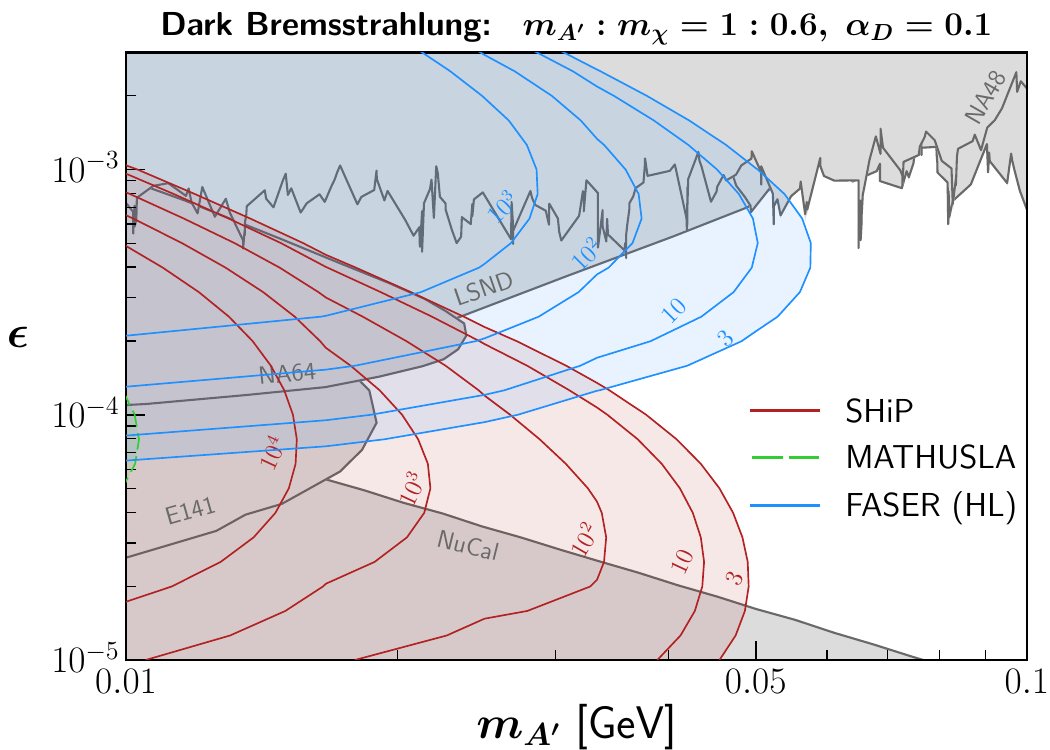}
\caption{Results for the model with a single dark matter particle and a dark photon produced due to dark bremsstrahlung, as discussed in \cref{sec:darkbrem}. In the left panel, the sensitivity reach of different experiments is shown: FASER during LHC Run 3 (dash-dotted blue line), FASER during HL-LHC era (solid blue), FASER 2 (dashed blue), MATHUSLA (long dashed green) and SHiP (solid red). The reach corresponding to the secondary production of LLPs is shown with colorful shaded regions. Additional contributions from primary production are indicated by the label ``$A^\prime$ from IP''. Current exclusion bounds are shown as grey-shaded regions. The CMB constraint on $\epsilon$ corresponding to the scenario with thermally produced $\chi$ dark matter is shown as the grey-line-bounded region. In the right panel, lines with different numbers of events relevant for the secondary production and for FASER (HL-LHC), MATHUSLA, and SHiP experiments are shown. The color coding is the same as in the left panel.
}
\label{fig:result_darkbrem}
\end{figure*}

\begin{figure*}[tb]
\centering
\includegraphics[width=0.48\textwidth]{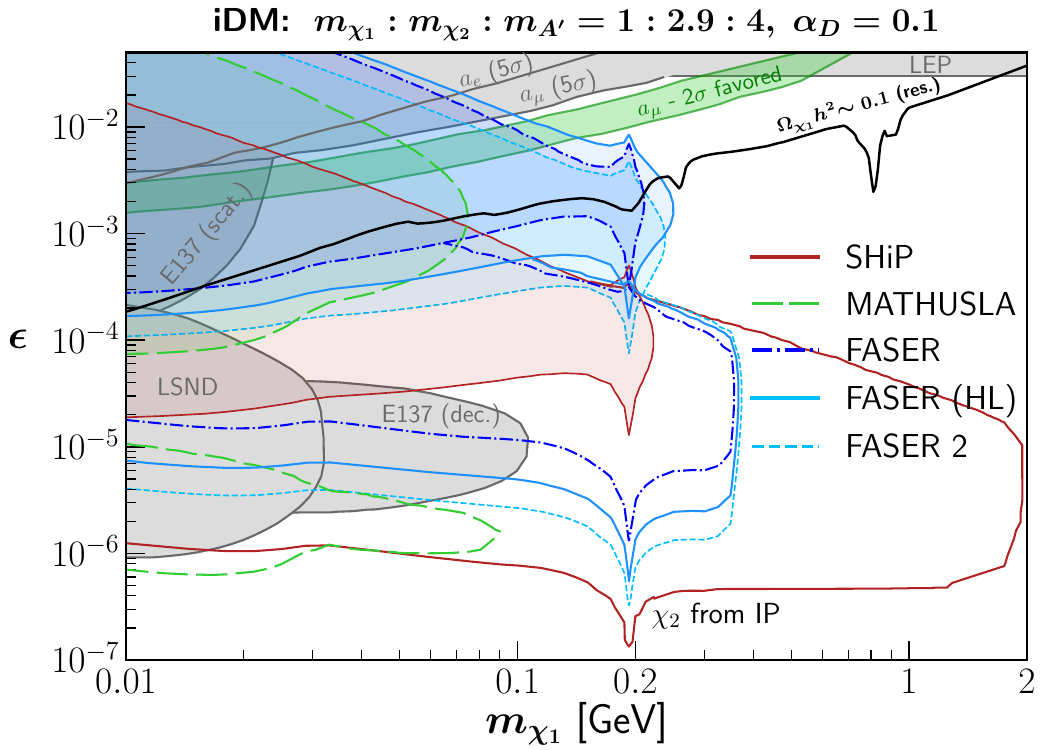}
\includegraphics[width=0.48\textwidth]{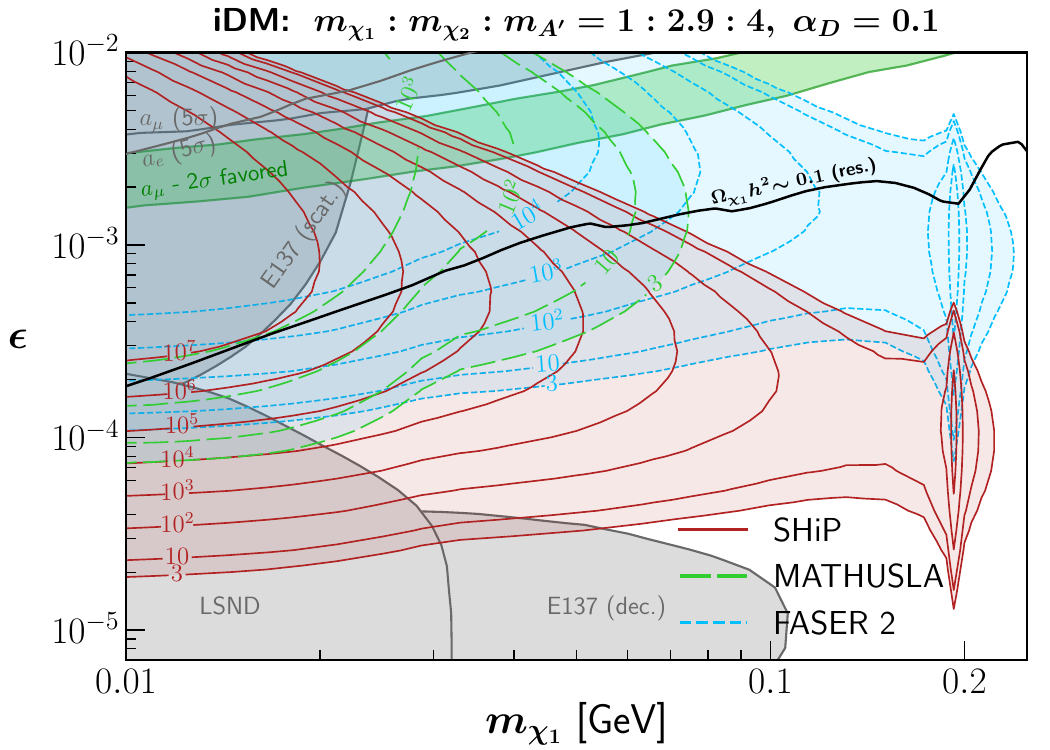}
\caption{Same as \cref{fig:result_darkbrem}, but for a model with inelastic DM discussed in \cref{sec:iDM}. In the right panel, blue lines with different numbers of events correspond to the FASER 2 experiment.}
\label{fig:result_iDM}
\end{figure*}

\begin{figure*}[tb]
\centering
\includegraphics[width=0.48\textwidth]{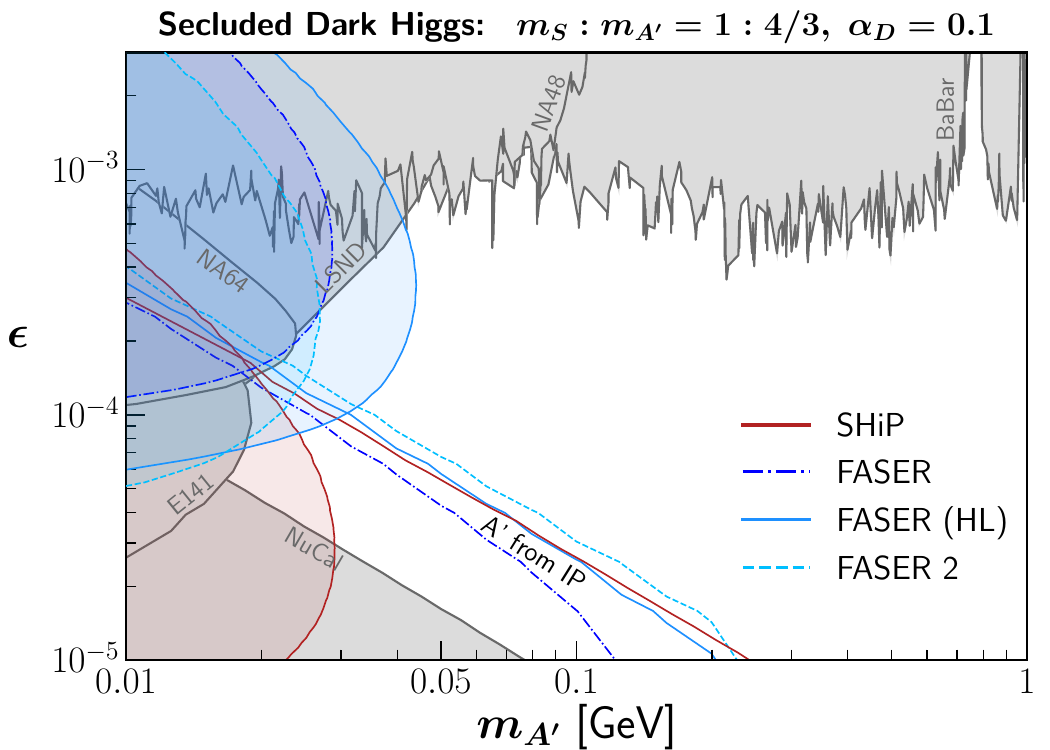}
\includegraphics[width=0.48\textwidth]{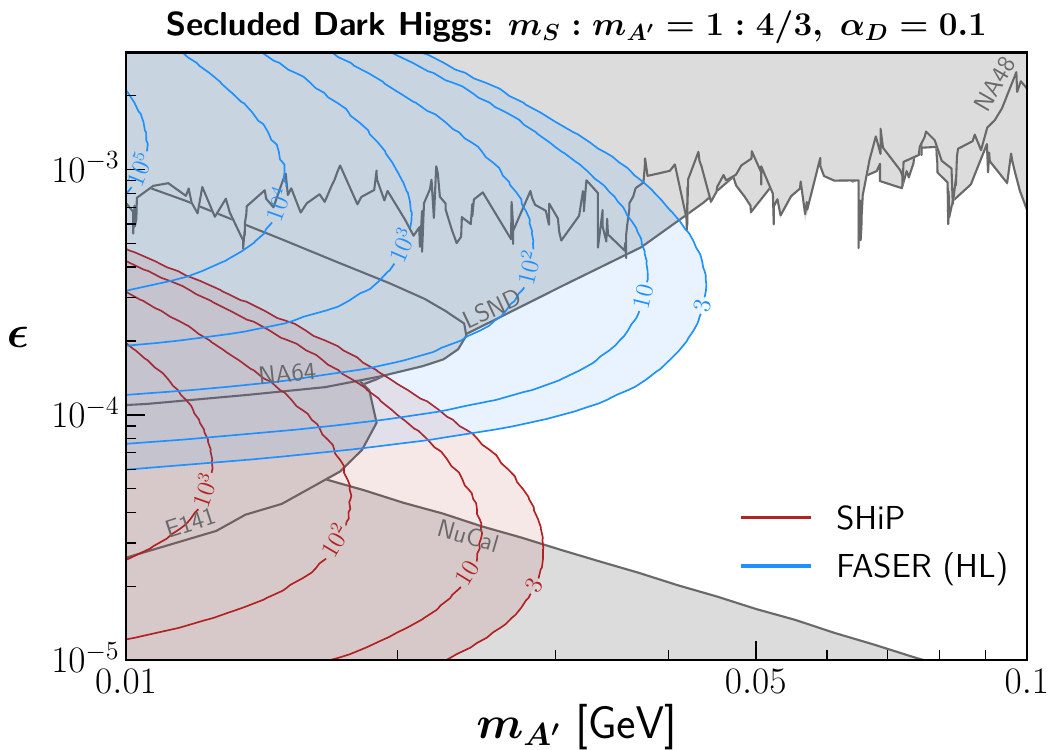}
\caption{Same as \cref{fig:result_darkbrem}, but for a model with a secluded dark Higgs boson discussed in \cref{sec:darkHiggs}.}
\label{fig:result_dhdp}
\end{figure*}

In order to illustrate the interplay between the primary and secondary production mechanisms, we have studied the sensitivity reach for selected LLP models in the FASER, MATHUSLA, and SHiP experiments. The respective results are shown in \cref{fig:result_darkbrem} for the model with $A^\prime$ produced from dark bremsstrahlung, in \cref{fig:result_iDM} for the model with inelastic DM, and in \cref{fig:result_dhdp} for the model with the secluded dark Higgs boson, which are described in \cref{sec:darkbrem,sec:iDM,sec:darkHiggs}, respectively.

%------------------------
%------------------------

%------------------------
%------------------------

%------------------------
%------------------------

%+++++++++++++++++++++++++++++++++++++++
\subsection{Current bounds from past and existing experiments}
%+++++++++++++++++++++++++++++++++++++++

Before analyzing the reach of future experiments, we first discuss current bounds on the benchmark scenarios under study. These are associated with both searches for dark photons and for other LLPs present in the models. Below, we briefly summarize these constraints, beginning from bounds on visibly decaying vanilla $A^\prime$s, and then discussing constraints relevant for each of the three considered models, respectively. The most stringent bounds are shown in \cref{fig:result_darkbrem,fig:result_iDM,fig:result_dhdp} as grey-shaded regions. A more comprehensive discussion can be found e.g. in Ref.~\cite{Beacham:2019nyx} and references therein.

\paragraph{Dark photon with visible decays} Current constraints on visibly decaying dark photons are relevant for models with a dark bremsstrahlung of $A^\prime$ and with a secluded dark Higgs boson. The upper limit on the kinetic mixing parameter, $\epsilon\lesssim 10^{-3}$, shown in \cref{fig:result_darkbrem,fig:result_dhdp}, come from null searches for narrow resonances in $e^+e^-\to \gamma\,A^{\prime}\to \gamma\,\ell^+\ell^-$ events and rare pion decays, $\pi^0\to \gamma\,e^+e^-$, reported by the BaBar~\cite{Lees:2014xha} and NA48/2~\cite{Batley:2015lha} collaborations, respectively. On the other hand, bounds on lower values of $\epsilon$ have been derived from reinterpretation of old results of a number of fixed-target experiments, including E141~\cite{Riordan:1987aw} and NuCal~\cite{Blumlein:2013cua}, as well as the limits obtained by current searches in e.g. the NA64~\cite{Banerjee:2018vgk} detector. Depending on the value of $m_{A'}$, these bounds can effectively exclude part of the dark photon parameter space with $10^{-7} \lesssim \epsilon\lesssim-10^{-4}$.

We now discuss current constraints for each of the three models described in \cref{sec:darkbrem,sec:iDM,sec:darkHiggs}.

\paragraph{Dark bremsstrahlung} The bounds on the model discussed in \cref{sec:darkbrem} follow the above discussion relevant for visibly decaying dark photon. An additional constraint shown in \cref{fig:result_darkbrem} corresponds to the search for light DM particles scattering off electrons in LSND that follow Refs~\cite{Batell:2009di,deNiverville:2011it,deGouvea:2018cfv}.

We note that in the benchmark scenario considered in our study, the region of the parameter space relevant to secondary production of the dark photon in the dark bremsstrahlung process corresponds to the too low thermal DM relic density of $\chi$. In this scenario, $\chi$ would be one of the DM components that constitutes only a fraction of the total DM abundance. This inefficiency of purely thermal production of $\chi$ could, in principle, be compensated in non-standard cosmological scenarios or by adding an additional non-thermal component to its relic density in further extensions of the model. However, it is important to note that the model with Dirac fermion DM efficiently annihilating via vector portal into SM particles is tightly constrained by the Planck CMB data~\cite{Slatyer:2009yq,Slatyer:2015jla}. We indicate this in \cref{fig:result_darkbrem} by showing the lower bound on $\epsilon$ as a function of $m_{A^\prime}$ that is relevant for the scenario with the relic abundance of $\chi$ coming only from the standard freeze-out mechanism.

A notable exception to this obstacle could be a scenario in which total $\chi$ relic density is generated due to an initial asymmetry between $\chi$ and $\bar{\chi}$~\cite{Nussinov:1985xr,Kaplan:2009ag}. In this case, $\chi$ could even correspond to the entire DM relic density provided that the symmetric component is sufficiently suppressed~\cite{Lin:2011gj}. On the other hand, it would then be a subject to additional constraints from DM direct detection (DD) searches, as discussed e.g. in Ref.~\cite{deGouvea:2018cfv}. These will partially exclude the region of the parameter space relevant for larger values of $\epsilon$ and small $m_{A^\prime}$, which is shown as allowed in \cref{fig:result_darkbrem}. In addition, some larger part of this region would then be covered by future DM DD searches and would provide a complementary method for discovery, focusing on the DM particle $\chi$, with respect to the intensity frontier searches of $A^\prime$ discussed in this study. We do not show corresponding bounds and future sensitivities in \cref{fig:result_darkbrem} as their presence depends on additional assumptions about the cosmological scenario.

\paragraph{Inelastic DM} In the case of the inelastic DM model, and for sufficiently long-lived $\chi_2$, the most relevant bounds come from null searches for invisibly decaying $A^\prime$ reported by the BaBar~\cite{Lees:2017lec} and NA64~\cite{Banerjee:2017hhz} collaborations. These correspond to the dark photon decaying promptly into a $\chi_1\chi_2$ pair with $\chi_2$ decaying outside the detector. However, for a large mass splitting between the dark fermions, $\Delta_{\chi}\sim 2$, the lifetime of $\chi_2$ is typically too small for the heavier fermion to decay outside the detector, as can be deduced from \cref{eq:chi2lifetime}. In this case, the invisible bounds could only apply to a small region of the parameter space relevant for $m_{{\chi_1}}\sim \mathcal{O}(10~\mev)$ and, generally, remain less important than other constraints discussed below. In addition, constraints on visibly decaying dark photons also do not apply to this scenario. This is because three-body decays of $\chi_2$ into $\chi_1$ and $e^+e^-$ pair would not lead to a narrow resonance peak in the electron-positron spectrum seen in the detector, opposite to e.g. $A^\prime$ decays into $e^+e^-$ pairs with no missing energy.   Interestingly, it has recently been pointed out that a displaced vertex trigger could be employed to increase the sensitivity of the Belle-II experiment to such a scenario~\cite{Duerr:2019dmv}. While such a signature could provide an independent way of probing the relevant region of the parameter space, it is left for future studies to assess its relevance for very short-lived $\chi_2$s that often decay in the closest neighborhood of the nominal interaction point of the experiment.

The dominant current bounds are then associated with measurements of the anomalous magnetic moment of electron~\cite{Parker:2018vye,Davoudiasl:2018fbb} and muon~\cite{Bennett:2006fi,Pospelov:2008zw} that constrain large deviations of these quantities from the SM predictions, cf. Refs~\cite{Blum:2018mom,Keshavarzi:2018mgv,Davier:2019can} for recent discussion, as well as from electroweak precision observables at LEP and LHC~\cite{Hook:2010tw,Curtin:2014cca}. In addition, we also recast bounds on $\chi_1$ particles scattering off electrons in the E137 experiment~\cite{Batell:2014mga,Bjorken:1988as}, see also e.g. Ref.~\cite{Mohlabeng:2019vrz} for recent relevant discussion, assuming $E_{e^-,\mathrm{rec.}}\gtrsim 1~\gev$ threshold for the recoil energy. Notably, the bound from lighter fermion scattering in the LSND, which is discussed above for the model with a single DM particle, is suppressed in the case of inelastic DM due to a large mass splitting between $\chi_1$ and $\chi_2$, as well as a very small self-coupling of the lighter fermion.

On the other hand, heavier fermions produced at the IP could travel to the detector and leave a decay signature. This requires the $\chi_2$ decay width to be  suppressed enough, so that it does not decay too early. The corresponding bounds from decays in the E137~\cite{Bjorken:1988as} and LSND~\cite{Auerbach:2001wg} detectors are shown in \cref{fig:result_iDM}, following Ref.~\cite{Darme:2018jmx}. Importantly, they constrain the region of the parameter space relevant for values of the kinetic mixing parameter in the range $10^{-6}\lesssim \epsilon\lesssim 10^{-4}$, while some regions of the parameter space with larger $\epsilon$, but lying below the upper bound $\epsilon\lesssim 10^{-3}$, are left unconstrained.

Last, but not least, we note that in the model with inelastic DM the aforementioned stringent bounds from CMB can be avoided as annihilations of Majorana fermions, $\chi_1\chi_1\to \textrm{SM}$, are $p$-wave suppressed. Similarly, there are no such bounds from coannihilation processes between $\chi_1$ and $\chi_2$, as heavier $\chi_2$ particles decay prior to a time of recombination.

When presenting the reach plots, we have assumed fixed mass ratios among all three dark species, $m_{\chi_1}:m_{\chi_2}:m_{A^\prime} = 1:2.9:4$ that were chosen for illustrative purposes. We note, however, that small changes in these benchmark ratios would have a mild impact on the reach plots, as long as the dark photon decays predominantly into a $\chi_1\chi_2$ pair. On the other hand, the relic density of $\chi_1$ DM could change significantly, even by a few orders of magnitude, depending on how close are the assumed ratios to satisfy the resonance condition, $m_{\chi_1}+m_{\chi_2}\simeq m_{A^\prime}$ (see e.g. Ref.~\cite{Feng:2017drg} for recent related discussion). The line with the correct value of $\Omega_{\chi_1}h^2$ could then correspond to a wide range of $\epsilon$, from values even lower than the region corresponding to the $(g-2)_\mu$ anomaly, up to large ones that are already excluded. For illustrative purposes, we present in \cref{fig:result_iDM} the line corresponding to the scenario with the resonance, following Ref.~\cite{Darme:2018jmx}. We take into account hadronic annihilation final states relevant for sufficiently large $m_{\chi_1}$; cf Ref.~\cite{Izaguirre:2015zva}.

\paragraph{Dark photon and secluded dark Higgs boson} The most important bounds on the model with the secluded dark Higgs boson come from the aforementioned searches for visibly decaying dark photons. Additional searches for $S$ produced in rare decays of kaons, heavier mesons, or the SM Higgs boson that are based on the mixing between $S$ and $H$ do not constrain our parameter space of interest. This is because this mixing is highly suppressed for a secluded dark Higgs boson.

The rate of light meson decays into $S$ could, however, be enhanced thanks to their coupling to dark photons. This allows one to study corresponding constraints on such a scenario from subsequent delayed decays of $S$ into $e^+e^-$ pairs inside the detector, or from $S$ scattering off electrons, $S\,e^-\to A^\prime\,e^-$ followed by a prompt $A^\prime$ decay. 

The constraints based on the scattering of $S$ can be derived similarly to the aforementioned bounds on fermionic DM from the LSND experiment. One important difference in this case is, though, the subsequent decay of $A^\prime$ into $e^+e^-$ that could additionally contribute to the EM signal in the detector. Given low energies relevant for LSND, typically $E_{\mathrm{LLP}}\lesssim \mathcal{O}(100~\mathrm{MeV})$, these decays are prompt and, therefore, could be reconstructed as a single-electron signal together with the recoiled electron. Assuming that this is the case and that the entire EM energy lies within the range relevant for LSND cuts~\cite{Aguilar:2001ty}, $18~\mathrm{MeV}\lesssim \left(E_{e^-}(\mathrm{recoil})+E_{e^-}+E_{e^+}\right) \simeq E_{S}\lesssim 50~\mathrm{MeV}$ or similarly in between $60$ and $200~\mathrm{MeV}$~\cite{Athanassopoulos:1997er}, we have recasted the relevant bounds from Refs~\cite{deGouvea:2018cfv,Darme:2017glc} and show them in \cref{fig:result_dhdp} as the region excluded by the LSND. 

In fact, the bounds that we present for LSND should be considered as an approximate indication of the already excluded region of the parameter space. A more detailed analysis would take into account that the opening angle between electrons and positrons in the final state could exceed $12^\circ$, resulting in distinguishable lepton tracks which no longer mimic neutrino-electron elastic scatterings~\cite{Athanassopoulos:1996ds}. Such events with two or even three lepton tracks would, however, also be subject to constraints from the LSND, although deriving these bounds would require access to the experimental data and would then go beyond the scope of this study; see also Ref.~\cite{Izaguirre:2017bqb} for relevant discussion.

On the other hand, the constraints based on delayed $S$ decays into $e^+e^-$ inside the detector are less severe. This is due to a very long lifetime of $S$, as indicated in \cref{eq:Slifetime}. In addition, these bounds are also sensitive to a small mixing between $S$ and $H$ that could affect the exact value of $\tau_S$ without having an impact on the main results presented below. For this reason, we only qualitatively discuss several such possible constraints, while we do not show them in \cref{fig:result_dhdp}. 

In particular, as discussed in Refs.~\cite{Morrissey:2014yma, Darme:2017glc}, relevant constraints from electron beam-dump experiments are typically subdominant with respect to the ones that employ protons impinged on the target material. As an example, decays of $S$ inside the LSND or MiniBoone detectors, that are interpreted as single electron events satisfying relevant cuts, can constrain an additional slice of the allowed parameter space, on top of standard dark photon searches. For negligible $S$-$H$ mixing, this bound is, however, generally subdominant with respect to the aforementioned constraint from $S$ scattering in the LSND. The same is true for possible bounds from $S$ decays in the old high-energy proton beam-dump experiments, e.g.  BEBC~\cite{CooperSarkar:1985nh} or CHARM~\cite{Bergsma:1985qz}.

Importantly, due to the long lifetime of $S$, additional constraints can arise from Big Bang Nucleosynthesis (BBN) (see~\cite{Iocco:2008va,Cyburt:2015mya} for recent reviews). They are associated with possible co-production of $S$ in scattering processes producing light dark photons in the early universe. The dominant contributions come from annihilation processes $q\bar{q},l^+l^-\rightarrow A'^\ast\rightarrow A'S$ and could lead to a large number of long-lived dark Higgs bosons produced, depending on the value of $\epsilon$ and $\alpha_D$. On the other hand, a small but not negligible contribution from $S$-$H$ mixing could allow one to always keep $\tau_S$ below the value relevant for BBN, i.e. $\tau_S\lesssim 0.1-1~\sec$, while maintaining a secluded regime in terms of collider searches.

%+++++++++++++++++++++++++++++++++++++++
\subsection{Sensitivity reach of future experiments}
%+++++++++++++++++++++++++++++++++++++++

The sensitivity reach of all the experiments and models under study is shown in the left panels of \cref{fig:result_darkbrem,fig:result_iDM,fig:result_dhdp}. The reach corresponding to the secondary production of LLPs is additionally marked with colorful shaded regions to distinguish it from the one relevant for the primary production. As can be seen, both types of production mechanisms can simultaneously cover some parts of the parameter space of the models, in which case the respective numbers of events add up to the total visible signal in the detector.

On top of this, some regions of the parameter space can only be covered by one type of the production process. This illustrates the complementarity between both mechanisms that allows one to probe additional scenarios that would otherwise seem to lie beyond the reach of the given experiment. It is important to note that a distinction that we focus on corresponds to only the production mechanisms, while the signature in the detector is the same in both cases. In practice, this means that neglecting the impact of the secondary production could affect the interpretation of the results of the experiments. 

Notably, the secondary production is most relevant for regions of parameter space with large couplings between the LLPs and the SM, which brightens the prospects for potential simultaneous discovery of such LLPs in many experiments in the near future. In order to properly interpret such a co-discovery, it would be essential to model secondary production of LLPs.

Focusing on the secondary production, we show in the right panels of \cref{fig:result_darkbrem,fig:result_iDM,fig:result_dhdp} the number of event contours for the experiments under study. As can be seen, depending on the model, we can typically expect up to $\mathcal{O}(10^3)$ events of this type, but this number can grow to even $> 10^6$ in certain scenarios, especially for the SHiP experiment. Below, we comment on the relevance of the secondary production mechanism for each of the experiments under study.

%...........................
\subsubsection{FASER (Run 3, HL-LHC) and FASER 2}
%...........................

The secondary production mechanism can improve the sensitivity of the FASER experiment both during LHC Run 3, as well as for the HL-LHC era. For all three considered benchmark models, the secondary production extends the FASER reach toward smaller LLP lifetimes, or larger values of $\epsilon$. In the case of the inelastic DM benchmark model, even the FASER detector operating during LHC run 3 can probe a large region of currently unconstrained parameter space. 

Note that the reach of FASER extends further to larger coupling in comparison to the reach of FASER~2, even though the latter detector has a larger size. This is due to the additional dense material in the tungsten-based neutrino detector, FASER$\nu$, placed in front of the FASER decay volume. If such a block of tungsten, or a similarly dense material, would be placed in front of FASER~2, it would also have a positive impact on the physics reach provided that muon-induced BG can be successfully rejected by the front veto. This might be a useful observation for the detector design, especially in case some hints of new physics appear in the near future that correspond to the models with smaller LLP lifetimes than the FASER experiment can typically cover. 

Interestingly, the regions of parameter space that are accessible at FASER through secondary production mechanisms can be related to outstanding problems in particle physics and anomalies: 
i) In the case of the inelastic DM model, FASER probes the currently unexcluded region of the parameter space that yields the correct value of the DM relic density of $\chi_1$, as discussed in \cite{Darme:2018jmx}. 
ii) In addition, the secondary production opens up the possibility to probe an unconstrained region of the parameter space of the inelastic DM model corresponding to the observed discrepancy between the measurements and SM predictions of the muon anomalous magnetic moment~\cite{Bennett:2006fi,Pospelov:2008zw}. 
iii) Furthermore, the probed region of the dark photon parameter space can correspond to the existing anomaly observed in decays of the excited state of beryllium-8 with a mass of $m_{A^\prime}=17~\mev$ ~\cite{Krasznahorkay:2015iga}, as discussed e.g. in Ref.~\cite{Beacham:2019nyx}. However, realistic BSM models that could accommodate for this anomaly~\cite{Feng:2016jff, Feng:2016ysn} require going beyond the simplest dark photon case and modifying its couplings to nucleons. This would also affect the reach of both FASER detectors. 

Notably, these particularly motivated regions of parameter space can already be probed by the FASER detector operating during the LHC Run 3, with an additional positive role of FASER$\nu$, while this would not be possible if only the primary production at the IP were taken into account.

%...........................
\subsubsection{MATHUSLA}
%...........................

The primary target of the MATHUSLA detector, which is to be placed off the LHC beam collision axis, is new physics particles produced in rare decays of heavier SM species, e.g. $B$ mesons or the SM Higgs boson, or in hard $pp$ scatterings that are the most relevant for larger LLP masses. It is then not a surprise that MATHUSLA has no reach in the currently allowed region of the parameter space of the models discussed above with below-GeV-mass $A^\prime$ decaying in the detector. 

Interestingly, however, in the model with inelastic DM illustrated in \cref{fig:result_iDM}, MATHUSLA can have a non-negligible reach for both the primary and the secondary production of LLPs. This is due to a larger angular spread in a DM flux produced with one additional decay of relatively heavier dark photon, $A^\prime\rightarrow\chi_1\chi_2$, when compared to the $A^\prime$ flux from e.g. pion decays. On top of this, the primary production benefits from a relatively large lifetime of $\chi_2$.

As a result, two distinct regions of the parameter space can be seen corresponding to both types of a production mechanism with the secondary one covering the region with the correct value of DM relic density and, partially, the one corresponding to the aforementioned $(g-2)_\mu$ anomaly. 

%...........................
\subsubsection{SHiP}
%...........................

The secondary production in case of the SHiP experiment has also a positive impact on the sensitivity to BSM searches. It can lead not only to an improved discovery prospects but also to a potentially a large number of such events. This is particularly relevant for the model with inelastic DM, as shown in the right panel of \cref{fig:result_iDM}. Taking into account the secondary production could then also become crucial for the model parameter reconstruction, in case of discovery.

The sensitivity reach of SHiP for both models focusing on the dark photon decaying in the detector can also be improved; cf \cref{fig:result_darkbrem,fig:result_dhdp}. Importantly, when obtaining the results, we have adapted a quite conservative approach to exclude all the secondary processes happening in the SHiP detector more closely than about $2.5~\m$ away from the decay vessel or so, as detailed in \cref{sec:experiments,sec:benchmarkgeometries,sec:physicscuts}. On the other hand, if these cuts can be weakened, the sensitivity of SHiP with respect to the secondary production will be further enhanced.

Interestingly, on top of the interplay between the secondary and primary production of LLPs, in the case of the SHiP experiment, additional signatures of the models with more than a single LLP can arise by analyzing LLP scatterings off electrons. We discuss them in more detail in the section below.

%============================================================
\section{Scattering off Electrons\\ in SHiP SND \label{sec:resultselectrons}}
%============================================================

We have so far been focusing on two possible production mechanisms of LLPs, but only on one standard LLP decay signature in the detector with two high-energy charged tracks emerging from the vertex in the decay vessel. On the other hand, models with more than a single LLP also offer additional types of experimental signatures that can be studied contingent on a specific detector design. 

In particular, the presence of electronic tracker layers in the emulsion detector in the SHiP's SND can be markedly advantageous for BSM searches. Depending on their time resolution, they could time stamp the scattering events in the SND by detecting the corresponding recoil products. Such events could then be analyzed either separately or in conjunction with a subsequent LLP decay. Interactions of LLPs with the light electron target typically generate large recoil energy of $e^-$, as discussed in \cref{sec:scatteringsigma}. This leads to a detectable EM cascade in the SND with no hadronic counterpart, for which the expected BG is greatly reduced. 

\begin{figure*}[ht]
\centering
\includegraphics[width=1.0\textwidth]{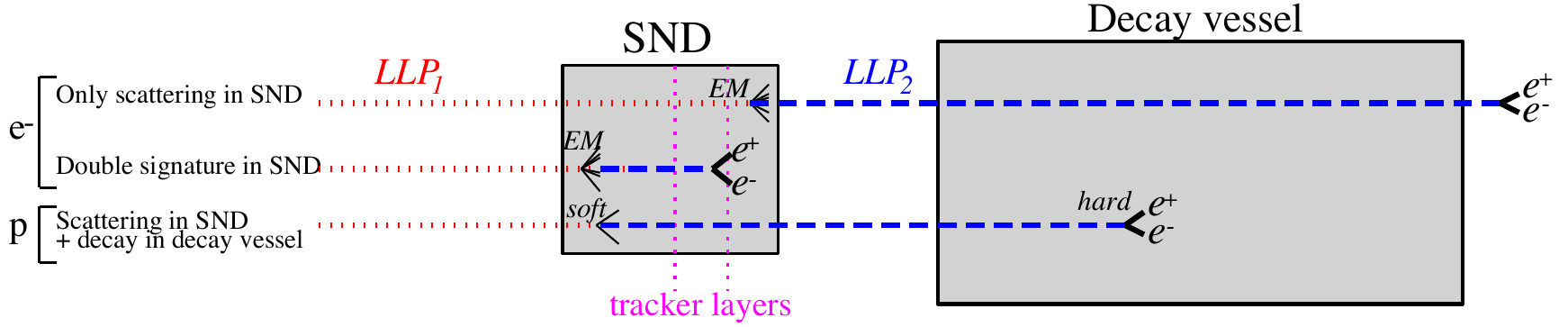}
\caption{A schematic drawing of the two signatures relevant for LLP scatterings off electrons in the SND that are discussed in the text. From top to bottom: only scattering signal with no decay in the decay vessel and the double signature of both scattering and decay that occur in the SND. At the bottom, we also show the standard signature based on scatterings of the LLP off nuclei with soft target recoil followed by a decay into two high-energy charged tracks in the decay vessel. The color coding for light long-lived particles is the same as in \cref{fig:idea}. Two of the tracker layers in the SND are schematically shown as vertical dotted purple lines.}
\label{fig:SNDidea}
\end{figure*}

While the actual capabilities of the SHiP experiment with respect to such signatures will depend on the final detector design, below we briefly discuss two such possible search strategies employing scatterings off electrons in the SND with possible secondary production of LLPs. They are also schematically shown in \cref{fig:SNDidea}.

%------------------------
%------------------------

\begin{description}
\item [(Only) Electron scattering signature]
The search for light dark sector particles scattering off electrons is one of the main aims of the SND~\cite{Anelli:2015pba}. The expected number of relevant BG events has been reported to be about $800$ for $2\times 10^{20}$ POT~\cite{Ahdida:2654870}. These events are mostly associated with quasi-elastic electron neutrino and anti-neutrino charged current (CC) interactions with nuclei. While it might be possible to further reduce this background, for example by using more sophisticated tracking algorithms to identify soft protons in $\nu_e\,n\rightarrow e^-\,p$ interactions, there remain truly irreducible backgrounds related to elastic scattering off electrons $\nu e \to \nu e$, or quasi-elastic scattering of anti-neutrinos, $\bar{\nu}_e p \to e^+ n $, with neutrons in the final state that easily avoid detection.

For our models of interest, a similar pure scattering signature is possible, especially if subsequent decay of the LLP from secondary production is delayed and happens outside the detector. In a simplified geometry employed in our study, this corresponds to decay of the LLP at a distance at least about $71~\textrm{m}$ away from the end of the SND, where we have assumed that the SHiP Decay Spectrometer has the total length of about $15~\textrm{m}$. When obtaining the result, we follow Refs.~\cite{Anelli:2015pba,Buonocore:2018xjk} and apply cuts on the recoil energy,  $1~\textrm{GeV} < E_e < 20~\textrm{GeV}$, and on the recoil electron angle, $10~\textrm{mrad} < \theta_e < 20~\textrm{mrad}$.

\item [``Double'' signature inside the SND] Yet another spectacular signature of models with more than a single LLP can be associated with a simultaneous generation of two resolvable collinear EM showers with no hadronic recoil counterpart, that both can be seen inside the emulsion detector. These can be associated with the secondary production process followed by a prompt LLP decay. Importantly, neutrino-induced combinatorical BG to such a process can be greatly suppressed.

While an optimized description of this search will require a separate study, here we discuss one such simple strategy in which both EM showers should satisfy the aforementioned cuts relevant for the pure scattering signature. In addition, we require both showers to be initiated at positions in the emulsion detector that are not too close to each other. As a benchmark gap between these positions, we choose a distance of $10~\textrm{cm}$. This typically allows for each of the showers to be detected by different tracking layers, as these are placed about every $8~\textrm{cm}$. Notably, a distance of $10~\textrm{cm}$ corresponds to more than $15$ radiation lengths in lead. This significantly reduces any possible overlap between the showers in the emulsion films. 

On the other hand, we note that requiring such a large gap between the two showers might be a too conservative approach. This is due to generally excellent spatial resolution of the emulsion detectors that could be combined with the signal detected by the tracker layers. We then also illustrate below how weakening of this cut might allow one to study an even broader class of BSM scenarios.
\end{description}

\begin{figure*}[ht]
\centering
\begin{subfigure}{0.48\textwidth}
\centering 
\includegraphics[width=1.0\textwidth]{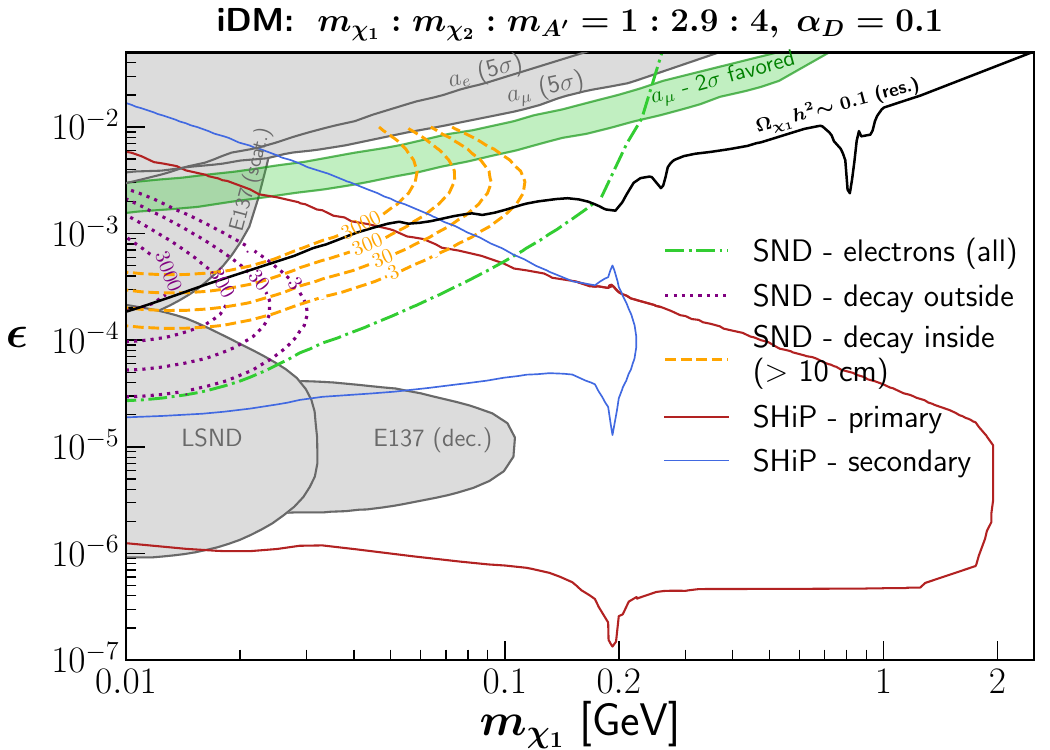}
\caption{inelastic DM}\label{fig:SNDiDM}
\end{subfigure}
\begin{subfigure}{0.48\textwidth}
\centering
\includegraphics[width=1.0\textwidth]{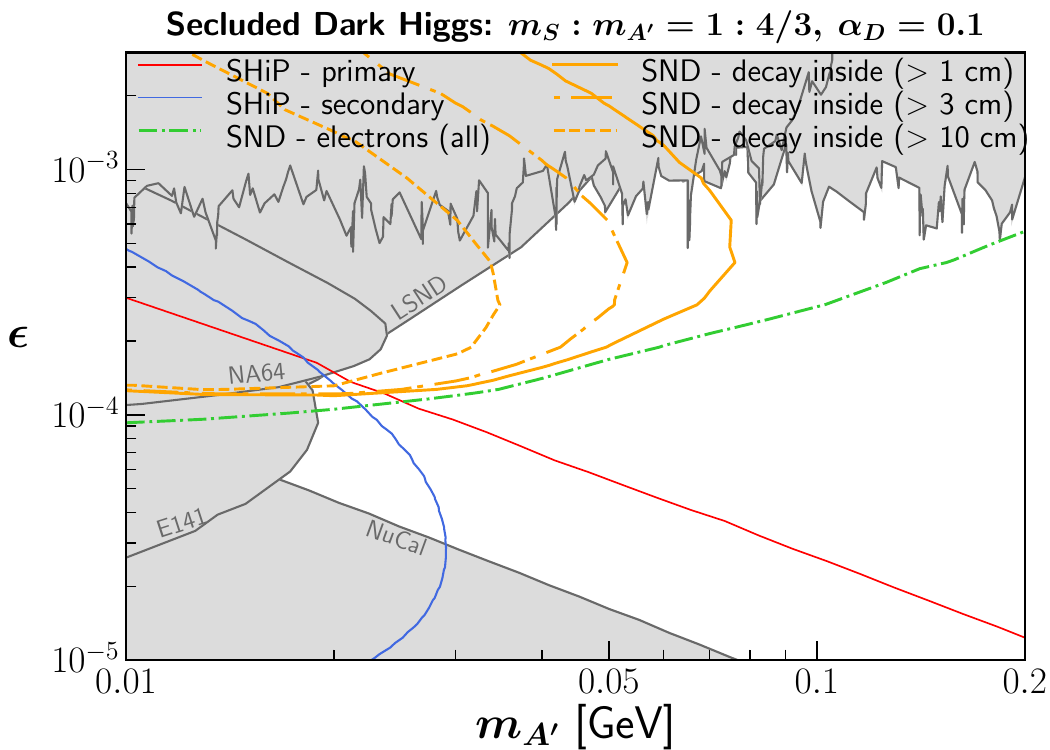}
\caption{dark photon with secluded dark Higgs boson}\label{fig:SNDDHDP}
\end{subfigure}
\caption{Results corresponding to additional signatures relevant for LLP scatterings off electrons in the SND obtained for the model with inelastic DM (left) and the one employing a dark photon with a secluded dark Higgs boson (right). Both plots have been obtained for benchmark scenarios defined in \cref{sec:models}. The green dash-dotted lines correspond to $N_{\textrm{ev}}=3$ electron scattering events in the SND, while solid red and blue lines to the reach from the standard decay in volume signature and LLPs coming from primary or secondary production, respectively. In the left panel, the purple dotted (dashed golden) lines were obtained for the pure scattering (double) signature discussed in the text and correspond to a different number of events as indicated in the plot. In the right panel, the dashed (dash-dotted, solid) golden lines represent the reach of the double signature with $N_{\textrm{ev}}=3$ events.}
\label{fig:SNDresults}
\end{figure*}

While the signature based solely on scatterings in the SND for models with a single dark matter particle has already been discussed in the literature (see e.g. Refs~\cite{Batell:2009di,Batell:2014mga,deNiverville:2016rqh,Battaglieri:2016ggd,Akesson:2018vlm,Buonocore:2018xjk}), here we present exemplary results of such searches for the model with inelastic DM, as well as for the one with a dark photon and secluded dark Higgs boson. The relevant results corresponding to benchmark scenarios introduced in \cref{sec:iDM,sec:darkHiggs} are shown in \cref{fig:SNDiDM,fig:SNDDHDP}, respectively. In these plots, a region above dot-dashed green lines corresponds to at least three electron scattering events happening in the SND. Hence, these lines indicate an absolute lower limit on $\epsilon$ as a function of the relevant mass that could be probed by any signature based on scatterings off electrons in the SND in the absence of BG. This can be compared with the reach from the standard decay in the volume signature discussed in \cref{sec:results}. The corresponding results for both primary and secondary production of LLPs are shown as solid red and solid blue lines in \cref{fig:SNDresults}, respectively.

%------------------------
%------------------------

The results corresponding to the aforementioned pure scattering search are shown for the inelastic DM model in \cref{fig:SNDiDM} as dotted purple lines with the fixed number of expected electron-scattering signal events, $N=3,30,300,3000$. Notably, the number of events in this case can reach up to $\mathcal{O}(1000)$ in the allowed region of the parameter space. This can be compared with order $100$ events required to exceed the expected level of BG by $5\sigma$ where the error has been roughly estimated as $\sqrt{N_{\textrm{BG}}}$ for the aforementioned number of BG events. These events are associated with $\chi_1$ upscattering to $\chi_2$ taking place in the SND, while $\chi_2$ survives for long enough so that it will decay outside the decay vessel and DS. For a sufficiently long-lived heavier fermion, an additional contribution to the total number of signal events would come from $\chi_2$ downscattering to $\chi_1$ in the SND.

A complementary search can be performed by employing the double signature of both upscattering and decay happening inside the SND with at least $10~\textrm{cm}$ gap in between. In this case, the maximum distance between the secondary production point and decay is the length of the SND, $\Delta_{\textrm{SND}}\sim 1~\textrm{m}$. This allows one to probe a regime of a smaller lifetime. As a result, the corresponding dashed golden lines with different number of events $N_{\textrm{ev}}=3,30,\ldots$, cover a distinct region in the parameter space of the model, with larger values of $\epsilon$, than in the pure scattering signature case discussed above.

As seen in \cref{fig:SNDiDM}, a combination of various search strategies employing the SND and the decay spectrometer, either separately or in a joint signature, could further shed light on the nature and lifetime of the LLP.

A somewhat different scenario is illustrated in \cref{fig:SNDDHDP} for the model with a secluded dark Higgs boson. Here, the LLP produced in secondary production processes in the SND is a very short-lived dark photon that promptly decays back to an electron-positron pair. As a result, no events with $A^\prime$ being able to travel outside the detector are expected. Actually, even the $10~\textrm{cm}$ gap in between the two showers in the SND might be too large as dark photons will typically decay very fast. We then show several lines corresponding to $3$ events with the double signature for a selection of minimal distances between the scattering and decay vertices ranging from $1~\textrm{cm}$ to $10~\textrm{cm}$. As expected, the smaller is the allowed distance between the beginning of the two showers, the better is the reach, at least assuming zero BG. On the other hand, for too small such a distance, both showers could no longer be effectively resolved. In this case, we would again enter the regime of an effectively pure scattering signature with a much larger expected BG. This would significantly affect the sensitivity reach. We leave for future studies a determination of the effective gap for which the showers can be distinguished based on an interplay between the signals measured by the tracker layers and in the emulsion films.

Last, but not least, we note that, if a sufficient number of events is observed with simultaneous EM showers that both can be resolved, more information useful for the model parameter reconstruction could be obtained from analyzing the distribution of the gaps between the vertices in such events. 

%============================================================
\section{Conclusions\label{sec:conclusions}}
%============================================================

One of the most promising ways to find new light long-lived particles is to search for their highly displaced decays in distant detectors separated from the primary interaction point. However, these potentially very clean searches are limited by the lifetime of LLPs that need to travel the entire distance to the detector. On the other hand, in models with more than a single LLP, often additional production mechanisms naturally appear due to efficient couplings between the BSM species. These can lead to the secondary production of LLPs in the close proximity of the decay vessel. As a result, smaller than usual LLP lifetimes can be probed, while the search still benefits from the shielding from the SM background, similar to the case of LLPs produced at distant IP.

In this study, we have illustrated such a possibility for several theoretically motivated models extending the SM by additional $U(1)_d$ group and the corresponding gauge boson, namely the dark photon $A^\prime$. Particular models of our interest include the following: dark photon being a mediator between the SM and DM, with either a single Dirac fermion DM particle or a pair of Majorana dark fermions, as well as $A^\prime$ gaining a non-zero mass via the dark Higgs mechanism and the vev of the secluded dark Higgs boson. 

While these models serve as an example, the secondary production can play an important role in many types of BSM scenarios with light new particles. In particular, in a more realistic setup, $A^\prime$ could both play a role of DM mediator and benefit from the presence of an additional scalar particle in the model. This would lead to combined signatures from different scenarios discussed in our study; cf Refs.~\cite{Darme:2017glc,Darme:2018jmx} for recent studies in this direction. In addition, more than a single LLP can arise e.g. in the context of the Twin Higgs scenario~\cite{Chacko:2005pe}, supersymmetry (see e.g. Ref.~\cite{Choi:2019pos}) or models with a linear dilaton action with possibly light Kaluza-Klein (KK) gravitons and scalars~\cite{Giudice:2017fmj}. Such models have also been proposed to accommodate for past and present experimental anomalies, cf. e.g. Ref.~\cite{Bertuzzo:2018itn} for a recent such study connected to the MiniBooNE excess and LSND anomaly. Remarkably, even in the presence of just a single LLP, the model can still be subject to effective secondary production e.g. if the BSM species couple to the SM neutrinos; cf. Ref.~\cite{Coloma:2017ppo} for such a study for heavy neutral leptons and the IceCube detector.

Importantly, the secondary production is the most important for the regions of the parameter space of the models that are often the most appealing due to prospects for a simultaneous discovery in many different experiments. By taking into account this production mechanism, intensity frontier searches can offer an independent way of verifying possible future hints of new physics that would otherwise seem to be beyond their reach. 

This can also be true for the existing discrepancies between the theoretical predictions and measurements. In particular, we show how the secondary production of LLPs in front of the FASER or MATHUSLA detectors can allow them to probe the $(g-2)_\mu$ anomaly in the unconstrained region of the parameter space of the inelastic DM model. Notably, FASER capabilities with respect to this scenario are enhanced by a dedicated neutrino subdetector, FASER$\nu$, placed in front of the decay vessel.

Similar such Scattering and Neutrino Detector (SND) can also substantially add to discovery prospects of SHiP. On top of already carefully studied signature based on LLP scatterings in the subdetector, we also propose to search for simultaneous production and decay of the LLP in the SND with two separate and collinear electromagnetic showers that are coincident in time. In case of discovery, employing a combined information about all the signal events detected in both the SND and the decay vessel could shed more light on the nature of the LLP.

Last but not least, while we analyze the impact of the secondary production for only selected intensity frontier experiments, namely FASER, MATHUSLA, and SHiP, the same production mechanism can be relevant for other similar experiments e.g. CODEX-b~\cite{Gligorov:2017nwh,Aielli:2019ivi}, NA62~\cite{Dobrich:2018ezn}, SeaQuest~\cite{Berlin:2018pwi}, and further discussed i.a. in Ref.~\cite{Beacham:2019nyx}. We encourage the experimental collaborations to take this possibility into account in detailed modeling, or even at the level of detector design, as it can crucially extend the sensitivity reach to new physics to promising scenarios with connection to current, or possible future, hints of new physics.

%============================================================
\acknowledgments
%============================================================

We thank Kevin Kelly for providing the MG model file used in \cite{deGouvea:2018cfv} and Luc Darm\'{e} for useful discussions. ST expresses special thanks to the Mainz Institute for Theoretical Physics (MITP) of the Cluster of Excellence PRISMA+ (Project ID 39083149) and the GGI Institute for Theoretical Physics for their hospitality and support. 
KJ and LR are supported in part by the National Science Centre (NCN) research grant No. 2015/18/A/ST2/00748. LR is also supported  by  the  project ``AstroCeNT:  Particle  Astrophysics Science and Technology Centre'' carried  out  within  the  International  Research  Agendas  programme  of the Foundation for Polish Science financed by the European Union under the European Regional Development Fund. FK~is supported by U.S.~Department of Energy Grant DE-AC02-76SF00515. F.K. performed part of this work at the Aspen Center for Physics, which is supported by National Science Foundation grant PHY-1607611. ST is supported by the Lancaster-Manchester-Sheffield Consortium for Fundamental Physics under STFC grant ST/P000800/1. ST is partially supported by the Polish Ministry of Science and Higher Education through its scholarship for young and outstanding scientists (decision no. 1190/E-78/STYP/14/2019).

%============================================================
%============================================================
\onecolumngrid
\appendix
\onecolumngrid
%============================================================
%============================================================

%============================================================
\section{Sensitivity reach for secondary LLP production and varying experimental cuts\label{sec:resultsaftercutschange}}
%============================================================

The results presented in \cref{sec:results} correspond to a particular set of cuts that we impose when estimating the number of signal events, as detailed in \cref{sec:experiments,sec:benchmarkgeometries,sec:physicscuts}. In particular, for the energy thresholds we follow standard values relevant for each experiment, while as far as secondary production of LLPs is concerned in front of veto layers, we require the recoil momentum to be low, $p_{\textrm{recoil}}<1~\textrm{GeV}$, and reject all the events happening within the last three hadronic interaction lengths ($3\,\lambda_{\textrm{had,int}}$) of the layers. This corresponds to about $1~\textrm{m}$ of standard rock, $0.5~\textrm{m}$ of lead, and $30~\textrm{cm}$ of tungsten~\cite{Tanabashi:2018oca}. Here, we discuss how the result can be affected when one sharpens these cuts. For illustrative purposes, we focus on the model with inelastic DM discussed in \cref{sec:iDM}.

\begin{figure}[tb]
\centering
\includegraphics[width=0.325\textwidth]{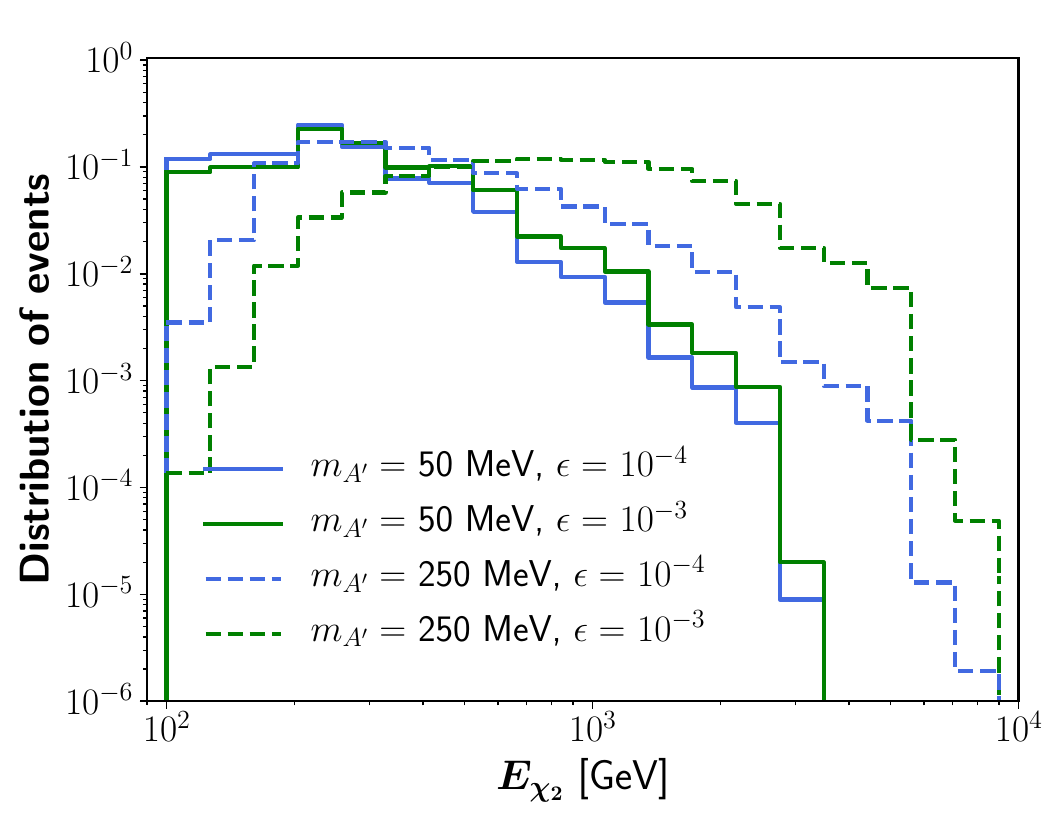}
\includegraphics[width=0.325\textwidth]{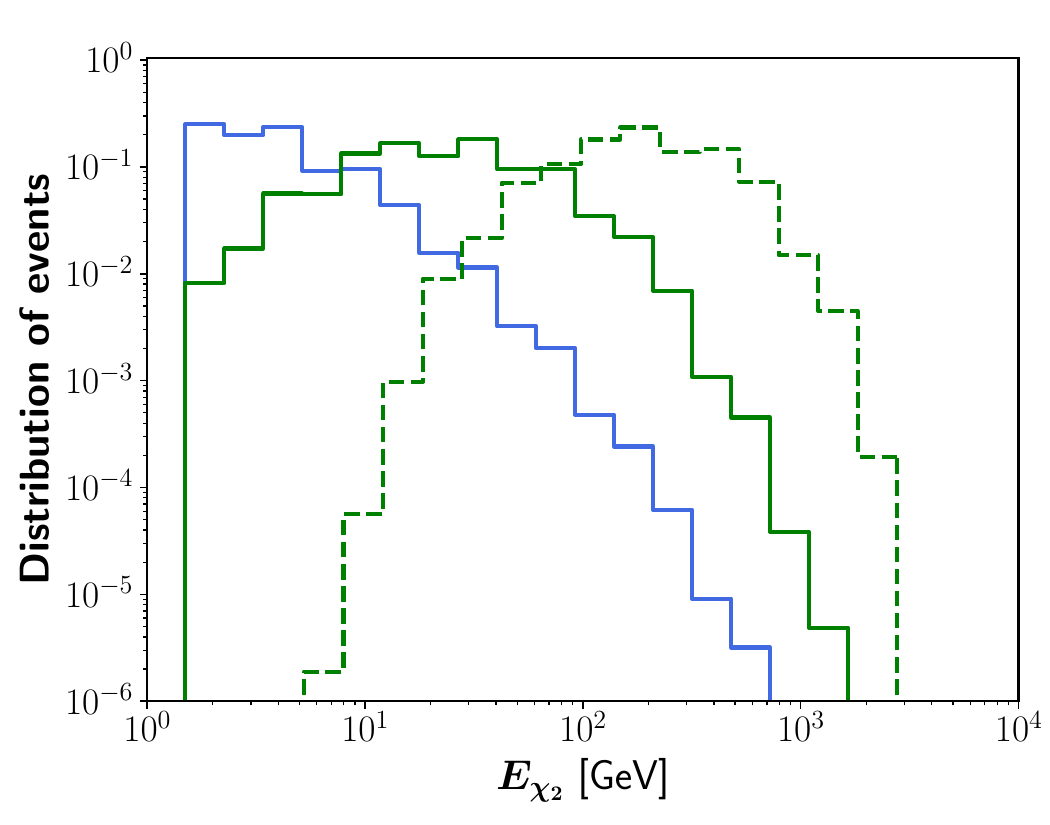}
\includegraphics[width=0.325\textwidth]{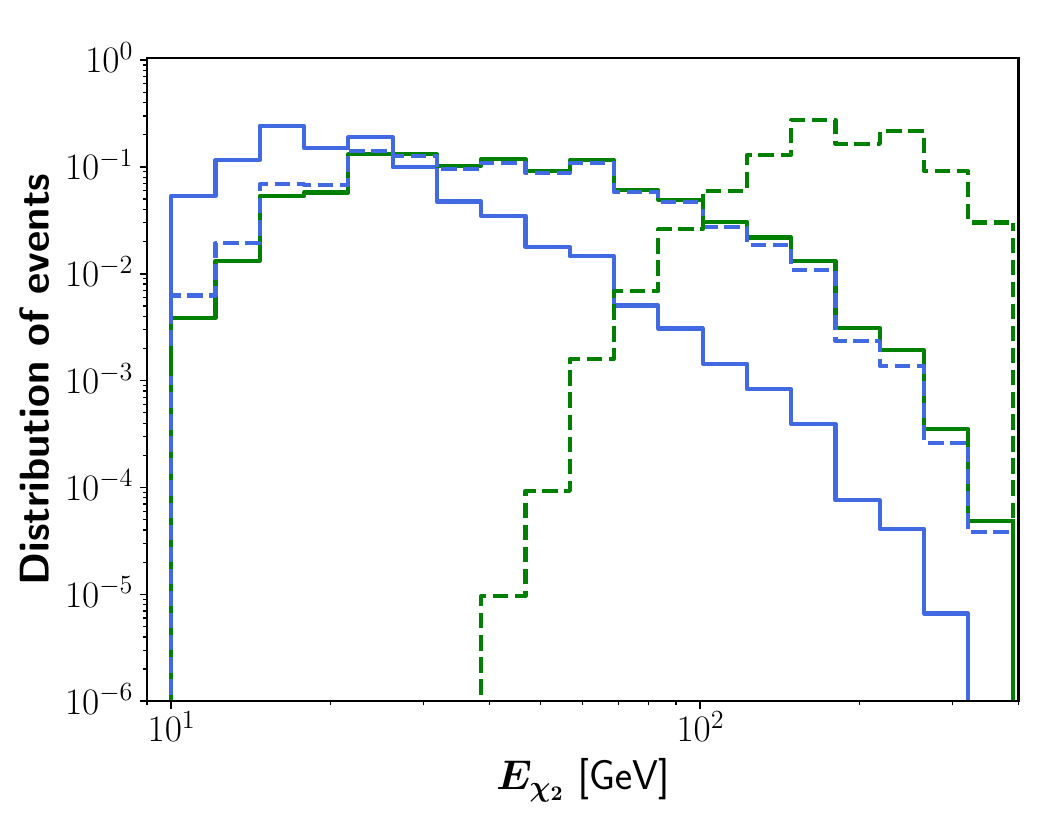}
\begin{subfigure}{0.325\textwidth}
\centering
\includegraphics[width=1.0\textwidth]{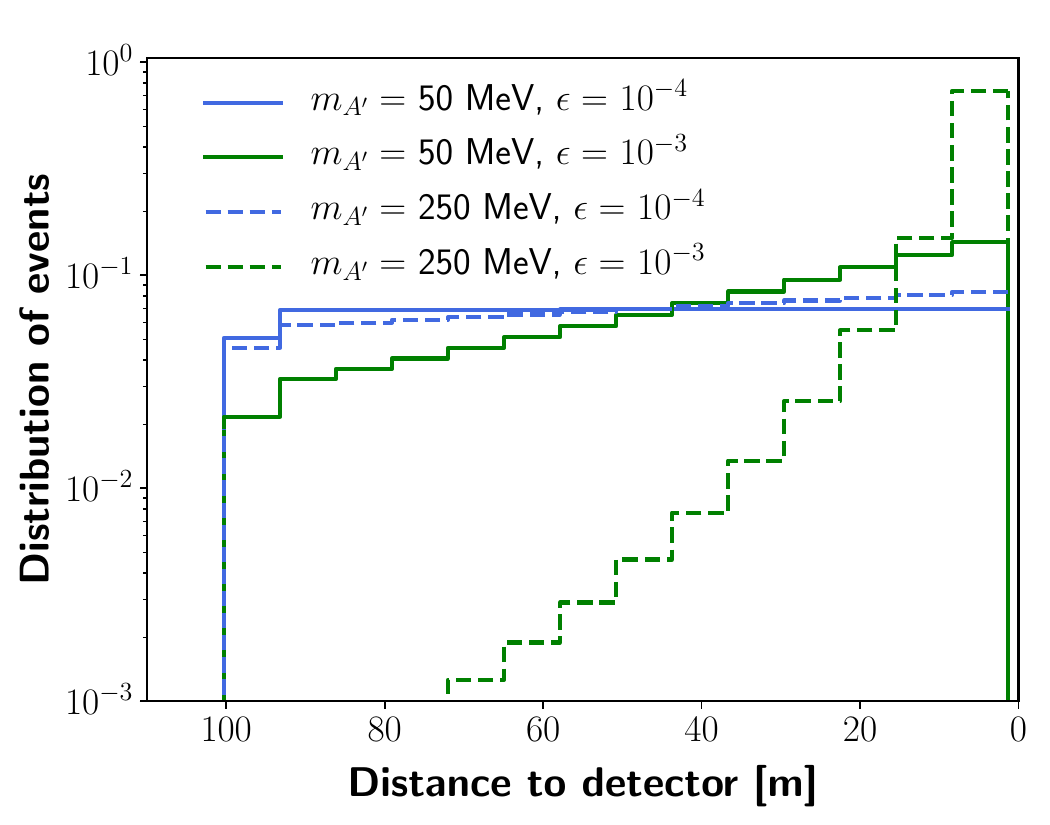}
\caption{FASER 2}\label{fig:FASER2histograms}
\end{subfigure}
\begin{subfigure}{0.325\textwidth}
\centering
\includegraphics[width=1.0\textwidth]{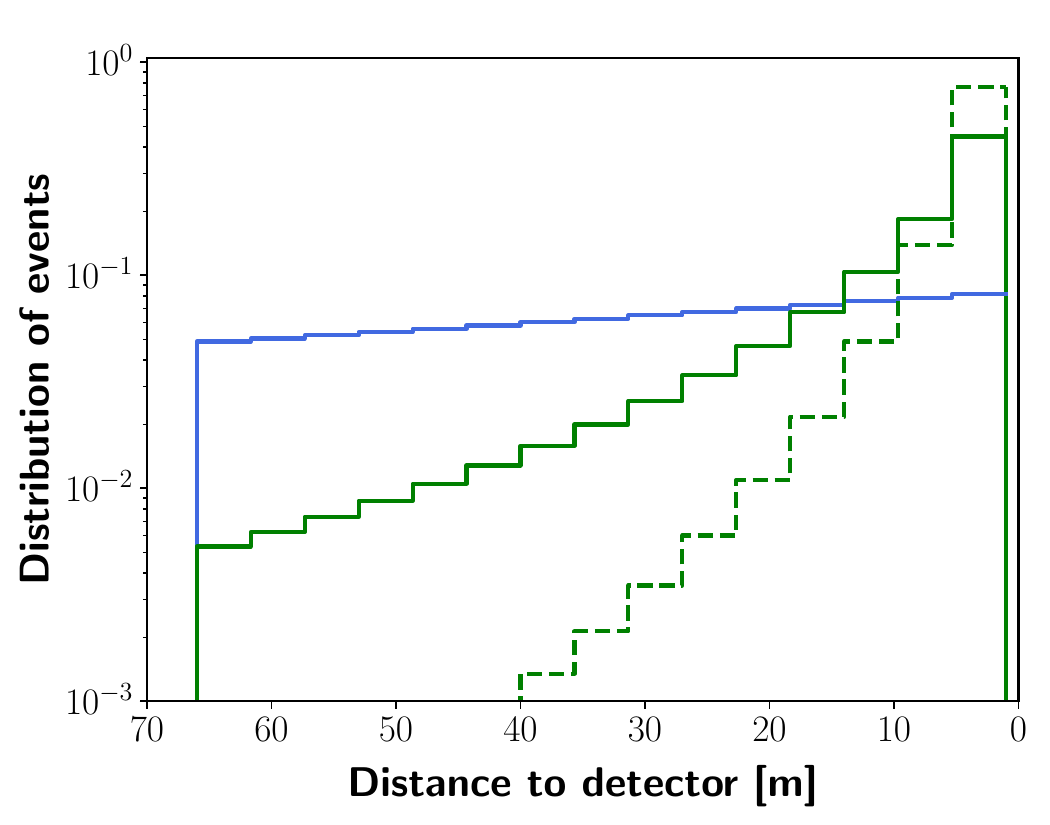}
\caption{MATHUSLA}\label{fig:MATHUSLAhistograms}
\end{subfigure}
\begin{subfigure}{0.325\textwidth}
\centering
\includegraphics[width=1.0\textwidth]{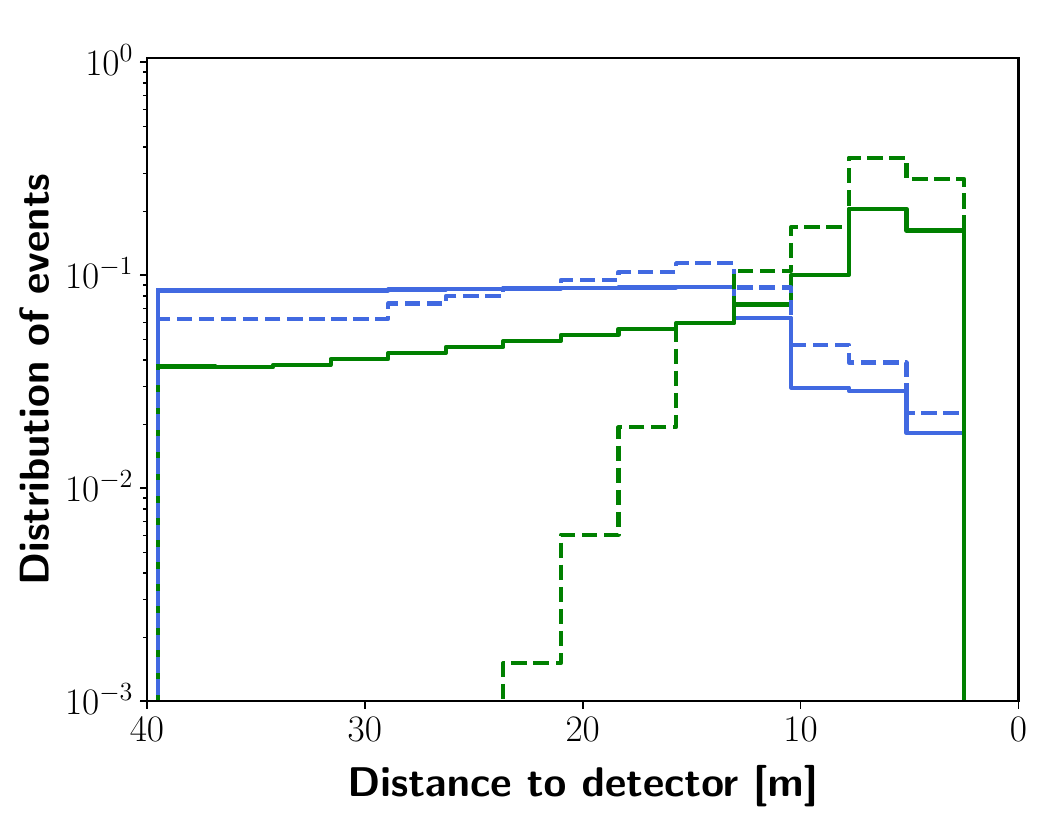}
\caption{SHiP}\label{fig:SHiPhistograms}
\end{subfigure}
\caption{Energy distribution of the events coming from the secondary production in front of the detector (top) and similar distribution for the position of the upscattering and its distance to the detector (bottom) for the FASER 2 (left), MATHUSLA (center), and SHiP (right) experiments. In the top panels, the energy of the decaying particle $\chi_2$ is shown on the horizontal axis. The histograms were obtained for the model with inelastic DM and the benchmark scenarios indicated in the plots with the mass ratios between all the LLPs that are given by $m_{\chi_1}:m_{\chi_2}:m_{A^\prime} = 1:2.9:4$. }
\label{fig:eventsdist}
\end{figure}

It is instructive to first analyze the example energy distribution of the events seen in the detectors under study when they are associated with the secondary production of $\chi_2$ in front of the decay vessel. This is shown in the top panels of \cref{fig:eventsdist} for the FASER 2, MATHUSLA and SHiP experiments, and for the benchmark scenarios with $m_{A^\prime} = 50$ or $250~\mev$ and $m_{\chi_1}:m_{\chi_2}:m_{A^\prime} = 1:2.9:4$, as well as for $\epsilon = 10^{-3}$ or $10^{-4}$. As expected, the energy of the events grows with a decreasing lifetime of $\chi_2$, i.e. with growing $\epsilon$ and LLP masses. This is due to larger boost factors required for $\chi_2$ to reach the decay vessel before decaying. For this reason, one expects that the region of the parameter space of the model that is the most sensitive to increasing the energy threshold will correspond to lower values of $\epsilon$ and lighter LLP particles. Notably, these are typically scenarios for which the primary production of LLPs dominate over the secondary one.

%------------------------
%------------------------

In the bottom panels of \cref{fig:eventsdist}, we show the distribution of the distance between the position of $\chi_1$ upscattering to $\chi_2$, and the decay vessel. As can be seen, for a smaller lifetime of $\chi_2$, the impact of the last part of the material right in front of the decay vessel becomes more important. This is illustrated as more steeply growing histograms for increasing $\epsilon$ and the LLP masses. Hence, one can expect that excluding from the analysis larger parts of the material in front of the detector would most straightforwardly affect this region of the parameter space of the model. 

In the opposite limit, for small $\epsilon$ and light LLPs, the distribution of the upscattering events becomes flat in the distance to the detector. In this regime, the lifetime of $\chi_2$ is large enough for it to travel long distances. As a result, in this regime, the primary production of $\chi_2$ dominates over the secondary one. A somewhat unexpected behavior of the solid and dashed blue lines in the figure relevant for the SHiP experiment that corresponds to such a long-lifetime regime and describes the diminishing number of events from the region in the most immediate vicinity of the decay vessel is connected to the geometry of the detector and the presence of the SND with smaller average density than surrounding magnets and the muon shielding.

\begin{figure}[t]
\centering
\begin{subfigure}{0.325\textwidth}
\centering
\includegraphics[width=1.0\textwidth]{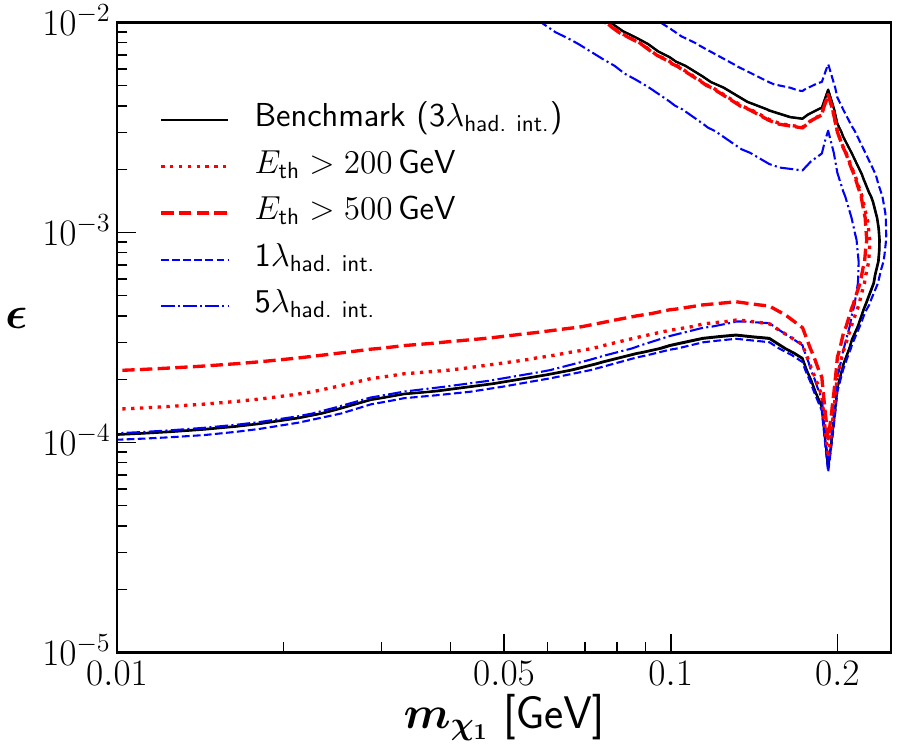}
\caption{FASER 2}\label{fig:FASER2cuts}
\end{subfigure}
\begin{subfigure}{0.325\textwidth}
\centering
\includegraphics[width=1.0\textwidth]{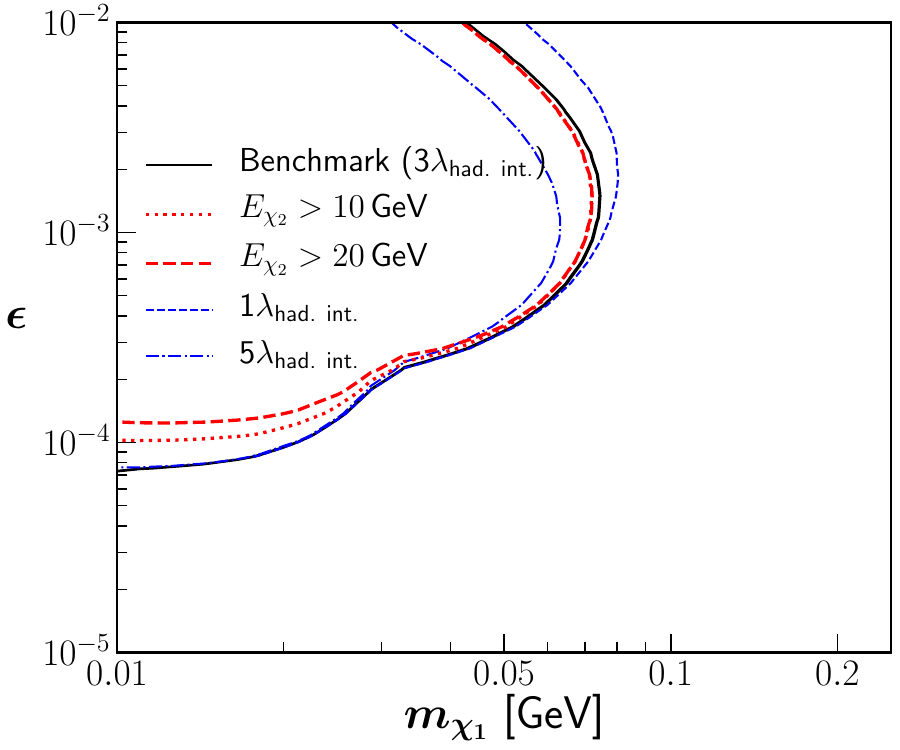}
\caption{MATHUSLA}\label{fig:MATHUSLAcuts}
\end{subfigure}
\begin{subfigure}{0.325\textwidth}
\centering
\includegraphics[width=1.0\textwidth]{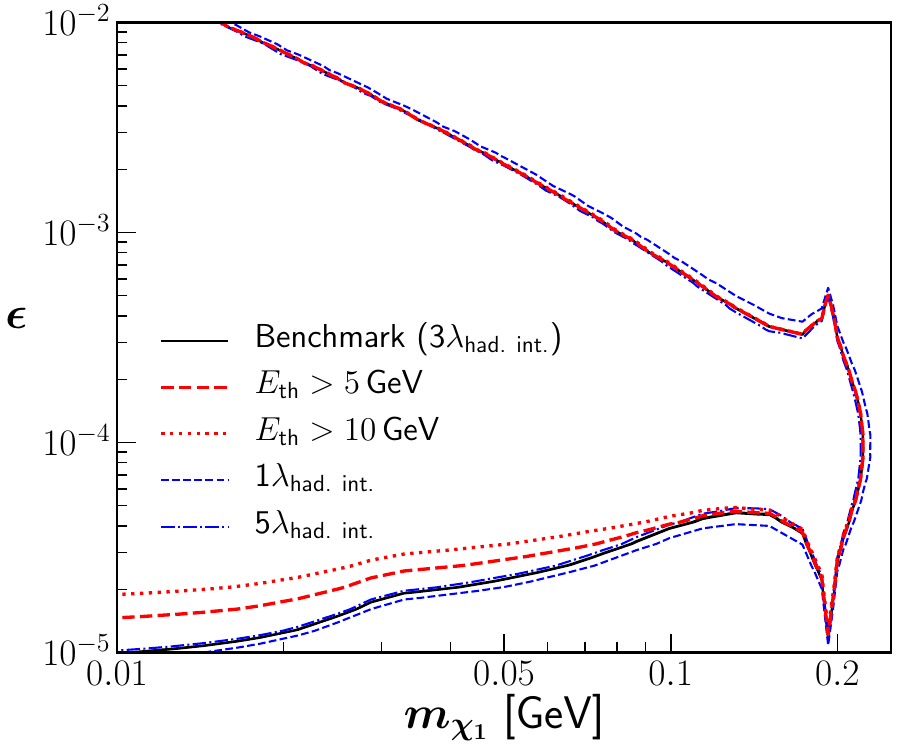}
\caption{SHiP}\label{fig:SHiPcuts}
\end{subfigure}
\caption{Impact of stronger experimental cuts on the sensitivity reach due to secondary production illustrated for inelastic DM model for FASER 2 (left), MATHUSLA (center), and SHiP (right) detectors. The default results presented in \cref{sec:results} are shown with solid black lines. The long-dashed and dotted red lines correspond to increased experimental energy thresholds as indicated in the figures. The dashed and dash-dotted blue lines illustrate the impact of changing the length of material in front of the veto layers, given as a multiple of the hadronic interaction length, $\lambda_{\mathrm{had,int}}$, that is excluded from the analysis.
}
\label{fig:Cuts}
\end{figure}

We illustrate all these effects in \cref{fig:Cuts} where we present the sensitivity reach plots for the considered experiments. In this figure, the solid black lines correspond to the default results discussed in \cref{sec:results}, while other lines were obtained by varying the length of material in front of the decay vessel acting as an effective veto (blue lines) and varying the energy thresholds (red lines) .

%------------------------
%------------------------

In our default analysis setup we require the secondary production to occur at least $3\, \lambda_{\mathrm{had,int}}$ before the front vetoes. However, this cut is rather conservative and could be relaxed. We require the target nucleus to have recoil momenta below $p_{\text{recoil}} < 1~\gev$, resulting in velocities $v/c< 0.1$. At these velocities the nucleus will experience an energy loss $dE/dx \gg 10~\mev~\cm^2/\g$. Therefore the nucleus will stop after traveling a distance $\Delta x = p_{\text{recoil}} / (dE/dx) \ll 100~\g/\cm^2$ which is smaller than $\lambda_{\mathrm{int,had}}$ for all considered targets. Changing the effective length of the front veto has a large effect on the sensitivity reach for LLPs with short lifetimes. This is illustrated by the blue lines in \cref{fig:Cuts}, where we exclude the last $\lambda_{\mathrm{had,int}}$ ($5\,\lambda_{\mathrm{had,int}}$) of material, resulting in an improved (reduced) reach at large couplings $\epsilon$. 

The impact of varying the energy thresholds affects mostly the reach in lower values of the kinetic mixing parameter. Here, even soft LLPs can reach the detector before decaying due to generally larger lifetime $\tau_{\chi_2}$. Therefore, increasing the energy threshold can have an impact on the number of signal events. However, for FASER 2 and SHiP, this mostly affects the region of the parameter space where sensitivity is, either way, driven by the contribution due to the primary production. This is not the case for the MATHUSLA detector where both primary and secondary production regimes are clearly separated. On the other hand, the MATHUSLA detector does not have the energy measurement, and the effective energy threshold is only associated with the detection capabilities of soft particles.

%============================================================
\section{Decay branching ratios and widths relevant for the analysis\label{sec:decays}}
%============================================================

%+++++++++++++++++++++++++++++++++++++++
\subsection{Primary production of LLPs\label{sec:decayBRsprod}}
%+++++++++++++++++++++++++++++++++++++++

\begin{description}
\item [Dark Photon Production]
    If the dark photon is light, it is mainly produced via the decay of a pseudoscalar meson $P=\pi^0,\eta,\eta'$ into a photon and a dark photon. The corresponding branching fraction is given by 
    \be
    \mathrm{B}_{P\to \gamma A^\prime} 
    = 2\,\epsilon^2\,\mathrm{B}_{P\to \gamma\gamma} \,
    \lambda^{\frac32}(m_P^2,m_{A^\prime}^2,0)/\lambda^{\frac32}(m_P^2,0,0)
    = 2\,\epsilon^2\,\mathrm{B}_{P\to \gamma\gamma}\,\left(1-m_{A^\prime}^2/m_P^2\right)^3.
    \ee
    Here we have used the K{\"a}ll{\'e}n function $\lambda(a,b,c)=(a-b-c)^2 - 4bc$, where $m_P$ is the pseudoscalar meson mass and $\mathrm{B}_{P\to \gamma\gamma}$ is the branching fraction of the di-photon decay channel of $P$. Additionally, the dark photon can be produced in decays of a vector meson $V=\rho,\omega$ into a pseudoscalar meson $P$ and a dark photon (such as $\omega \to \pi^0 A^\prime$), or via the decay of a pseudoscalar meson $P$ into a vector meson $V$ and a dark photon (such as $\eta' \to \rho^0 A'$). The corresponding branching fractions are given by \cite{Berlin:2018pwi}
    \be
    \mathrm{B}_{V\to P A^\prime} 
    &= \epsilon^2\,\mathrm{B}_{V\to P\gamma} \,
    \lambda^{\frac32}(m_V^2,m_P^2,m_{A^\prime}^2)/\lambda^{\frac32}(m_V^2,m_P^2,0) 
    \quad \text{and}\\
    \mathrm{B}_{P\to V A^\prime} 
    &= \epsilon^2\,\mathrm{B}_{P\to V\gamma} \,
    \lambda^{\frac32}(m_P^2,m_V^2,m_{A^\prime}^2)/\lambda^{\frac32}(m_P^2,m_V^2,0)\, .
    \ee
\item [Fermionic Dark Matter] 
    For heavy dark fermions, $m_1+m_2> m_{A^\prime}$, the differential branching fraction of a pseudoscalar meson $P$ decaying into $\gamma\chi_1\chi_2$ via intermediate off-shell dark photon $A^{\prime\,\ast}$ is given by~\cite{Kahn:2014sra,DeRomeri:2019kic}
    \begin{align}
    \label{eq:piontoDM}
    \frac{d^{2} \mathrm{B}_{P \rightarrow \gamma \chi_1 \chi_2}}{d s \, d \theta}=\mathcal{S}\epsilon ^2\alpha_D\,
    \mathrm{B}_{P \rightarrow \gamma \gamma}   \!\times \!
    \frac{\sin\!\theta\,\lambda^{\frac 12}(s,m_1^2,m_2^2)}{4\pi s^2}\!\times\!
    \frac{-\lambda(s,m_1^2,m_2^2)\sin^2\theta
    +2s[s\!-\!(m_1\!-\!m_2)^2]}{(s-m_{A'}^2)^2 + m_{A'}^2 \Gamma_{A'}^2}
    \!\times\!\left[\!1\!-\!\frac{s}{m_P^{2}}\!\right]^{3} \! .
    \end{align}
    Here, $s$ is the squared four-momentum of $A^{\prime\ast}$, while the angle $\theta$ corresponds to the momentum of $\chi_1$ in the rest frame of $A^{\prime\,\ast}$ that is measured with respect to the boost direction of the off-shell dark photon. The case with a single dark fermion corresponds to $\mathcal{S} = 1/2$ and $m_1=m_2\equiv m_\chi$, while for distinct dark fermions $\mathcal{S} = 1$. 
    
    For light dark fermions with $m_1+m_2< m_{A^\prime}$, the dominant production mode of $\chi_1\chi_2$ pairs is due to prompt decays of on-shell dark photons, $A^\prime\rightarrow \chi_1\chi_2$, with the relevant branching ratio of $\textrm{B}(A'\rightarrow \chi_1\chi_2)\simeq 1$ dictated by the hierarchy between the coupling constants, $\alpha_D\gg \alpha\,\epsilon^2$. 
\item [Secluded dark Higgs boson]
    The branching fraction of a pseudoscalar meson decay $P\to \gamma A^{\prime\,\ast}\to \gamma\,S\,A^{\prime}$ reads~\cite{Darme:2017glc}
    \begin{equation}
    \frac{d^{2} \mathrm{B}_{P\to \gamma S A'}}{d s \, d \theta}
    =\epsilon^{2} \alpha_{D}\, \mathrm{B}_{P \to \gamma \gamma} \!\times\! 
    \frac{\sin\!\theta\,\lambda^{\frac 12}(s,m_{A^\prime}^2,m_S^2)}{8\pi s^2}\,
    \!\times\!
    \frac{8 m_{A^\prime}^2 s+\lambda(s,m_{A^\prime}^2,m_S^2) \sin^2\theta }
    {(s-m_{A^\prime}^2)^{2}+m_{A^\prime}^2 \Gamma_{A^\prime}^2} 
    \!\times\!
    \left[1\!-\!\frac{s}{m_P^{2}}\right]^{3} ,
    \end{equation}
    where $s$ and $\theta$ are defined similarly to \cref{eq:piontoDM} but with $\chi$ replaced by $S$. Additionally, the branching fraction relevant for the vector meson, e.g. $\rho$, into a dark Higgs boson and a dark photon, $\rho\rightarrow SA^\prime$, is equal to~\cite{Batell:2009di,Darme:2017glc}
    \begin{equation}
    \mathrm{B}_{\rho \rightarrow S A'}=
    \epsilon^{2} q_{S}^{2} \alpha_{D} \,
    \mathrm{B}_{\rho \rightarrow e^{+} e^{-}} 
    \!\times\! 
    \frac{\lambda^{\frac 12}(m_\rho^2,m_{A^\prime}^2,m_S^2)}
    {\alpha_{e m}m_{\rho}^2} 
    \!\times\!
    \frac{12 M_{A'}^{2}m_{\rho}^{2}+\lambda(m_\rho^2,m_{A^\prime}^2,m_S^2)}
    {(m_{\rho}^{2}-m_{A^\prime}^{2})^{2}+m_{A^\prime}^{2} \Gamma_{A^\prime}^{2}}\, .
    \end{equation}

\end{description}

%+++++++++++++++++++++++++++++++++++++++
\subsection{Decays of LLPs\label{sec:decayBRsdec}}
%+++++++++++++++++++++++++++++++++++++++

\begin{figure}[tb]
\centering
\includegraphics[width=0.49\textwidth]{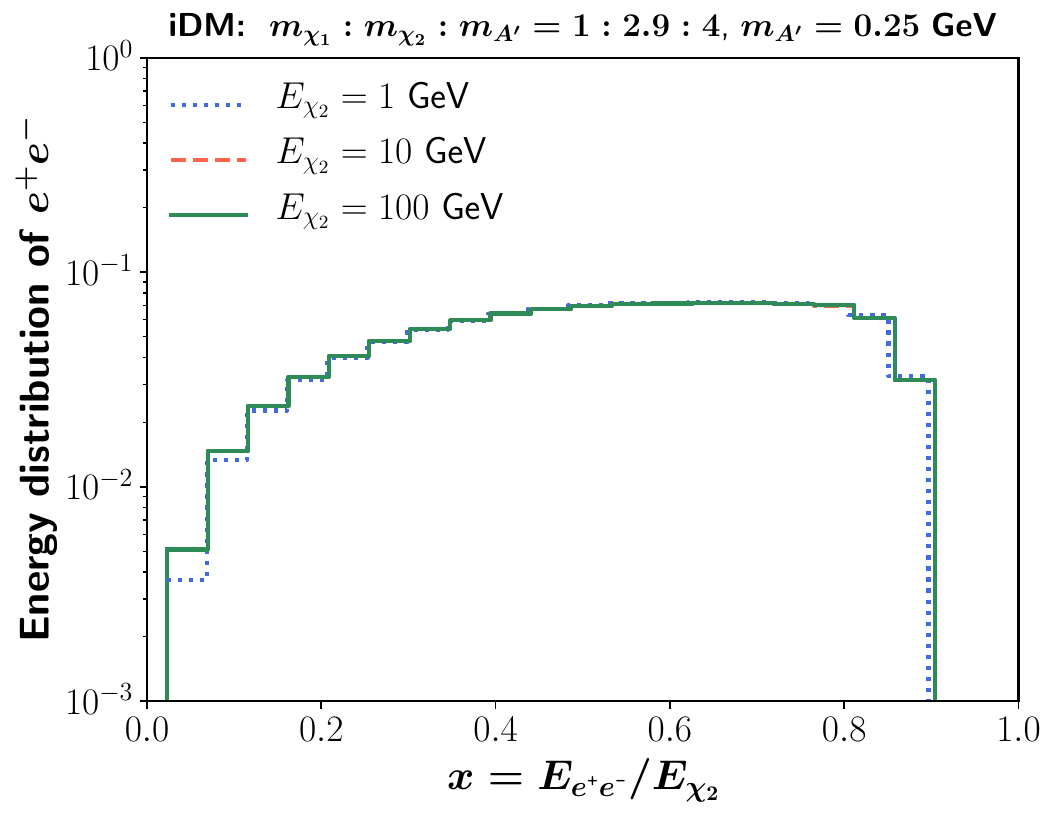}
\includegraphics[width=0.49\textwidth]{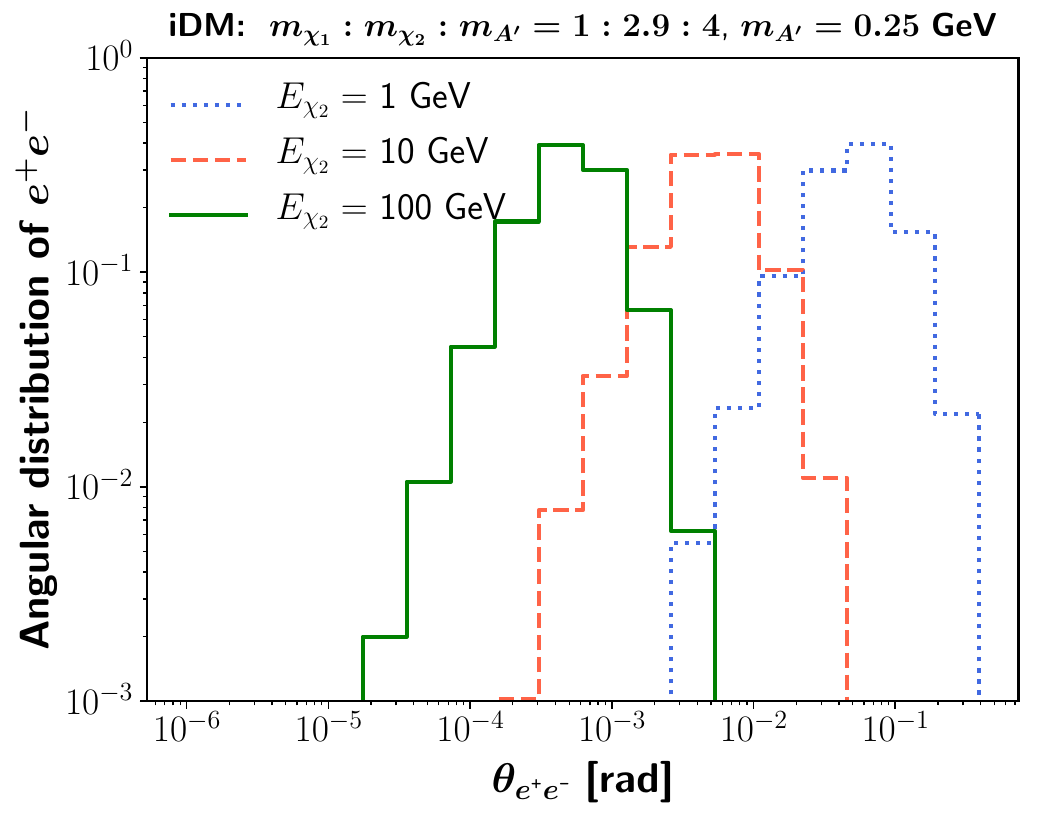}
\caption{The energy (left) and angular (right) distribution of an $e^+e^-$ pair from $\chi_2$ decay with respect to the energy and direction of the parent particle. The histograms are shown for three different energies of $\chi_2$: $E_{\chi_2} = 1~\gev$ (dotted blue), $10~\gev$ (dashed red), and $100~\gev$ (solid green), and for the benchmark point parameters indicated in the figures.
}
\label{fig:chi2decaydist}
\end{figure}

\begin{description}
\item [Dark Photon Decay]
    The decay width of an on-shell dark photon into SM particles is
    \begin{equation}
    \Gamma_{A^\prime} = \frac{\Gamma_{A^\prime\rightarrow e^+e^-}}{\textrm{B}_e(m_{A^\prime})},
    \quad \textrm{where}\quad
    \Gamma_{A^\prime\rightarrow e^+e^-} = \frac{\epsilon^2\,e^2\,m_{A^\prime}}{12\,\pi}
    \! \times \!
    \bigg[1-\frac{4 m_e^2}{m_{A^\prime}^2}\bigg]^{\frac 12}\!\!\!
    \! \times \!
    \bigg[1+\frac{2\,m_e^2}{m^2_{A^\prime}}\bigg],
    \label{eq:AprimeGamma}
    \end{equation}
    where $m_e$ is the electron mass and $\textrm{B}_e = \textrm{B}(A^\prime\to e^+e^-)$ is the branching fraction of a decay into an electron-positron pair. For $m_{A^\prime}$ below the di-muon threshold, $B_e=1$ which is typically the case in our analysis. For heavier dark photons, decays into $\mu^+\mu^-$, as well as hadronic final states start to play an important role~\cite{Buschmann:2015awa}.
\item [Decay of $\chi_2$ in inelastic Dark Matter Models]
    The relevant differential branching fraction for decay into an electron-positron pair, $\chi_2\rightarrow\chi_1\,e^+e^-$, can be found e.g. in Ref.~\cite{Giudice:2017zke}. In the case of heavier dark fermions and a larger mass splitting between them, decays into other SM particles become kinematically allowed. We take this into account by including the branching fraction of an off-shell dark photon decaying into an electron-positron pair, $\textrm{B}_{e}(m_{A^{\prime\ast}}=m_{ee})$,  evaluated at the invariant mass of the electron pair $m_{ee}$. The decay width is then given by
    \begin{equation}
    \Gamma_{2}=\frac{g_{12}^{2} \epsilon^{2} \alpha}{64 \pi^{2} m_{\chi_2}^{3}}
    \!\times\!
    \int_{s_{2}^{-}}^{s_{2}^{+}} \!\! ds_{2}
    \int_{s_{1}^{-}}^{s_{1}^{+}} \!\! ds_{1} \, \frac{4|A|^{2}}{(m_1^{2}\!+\!m_2^{2}\!+\!2 m_{e}^{2}\!-\!s_{1}\!-\!s_{2}\!-\!m_{A'}^{2})^{2}+m_{A'}^{2} \Gamma_{A^{\prime}}^{2}} 
    \!\times\! \frac 1{\textrm{B}_{e}(m_{A^{\prime\ast}}\!=\!m_{ee})},
    \label{eq:chi2lifetimeapp}
    \end{equation}
    where $\Gamma_{A^\prime}$ is given in \cref{eq:AprimeGamma} and
    \begin{align} 
    |A|^{2} = (s_1+s_2-2m_1m_2-2m_e^2)[(m_1+m_2)^2+4 m_{e}^2]
    + 2 (m_e^2+m_1 m_2)^2 -s_{1}^{2}-s_{2}^{2}
    \end{align}
    and integration limits are
    \be 
    s_{1}^\pm =
    m_1^{2}\!+\!m_{e}^{2}+\frac{1}{2 s_{2}} 
    \Big[(m_2^{2}\!-\!m_{e}^{2}\!-\!s_{2})
    (m_1^{2}\!- &m_{e}^{2}\!+\!s_{2}) \pm
    \lambda^{\frac 12}(s_{2}, m_2^{2}, m_{e}^{2}) 
    \lambda^{\frac 12}(s_{2}, m_1^{2}, m_{e}^{2}) \Big] \ , \\  
    s_{2}^+=(m_2 - m_{e})^{2} \quad &\text{and} \quad s_{2}^-=(m_1 + m_{e})^{2} \ . 
    \ee

    In \cref{fig:chi2decaydist} we show sample distributions of the electron pair energy as well as the angle of electron pair momentum with respect to the direction of $\chi_2$. As can be seen in the left panel, in the boosted regime, where $E_{\chi_2}\gg m_{\chi_2}$, the distribution of the energy fraction of $\chi_2$ that goes into the visible final state, $x=E_{e^+e^-}/E_{\chi_2}$, only mildly depends on $E_{\chi_2}$. In particular, for a larger mass splitting between the dark fermions characteristic for scenarios of our interest, the visible energy often corresponds to more than a half of the energy of $\chi_2$, with a kinematical upper limit $E_{e^+e^-}\lesssim \left[ 1-(m_{\chi_1} / m_{\chi_2})^2 \right] \, E_{\chi_2}$. The relevant value for the assumed mass ratio between both dark fermions is given by $E_{e^+e^-}\lesssim 0.9\,E_{\chi_2}$. The upper limit on $E_{e^+e^-}$ becomes more stringent in the case of lower energies, $E_{\chi_2}\sim m_{\chi_2}$, due to a non-zero mass of $\chi_1$ in the final state.
    
    In the right panel, we show the angle between electron pair momenta relative to the direction of decaying $\chi_2$. Notably, the visible charged tracks are typically collimated along the direction of $\chi_2$. This is especially true for energies $E_{\chi_2}\gtrsim 100~\gev$, i.e. above the energy threshold of FASER. On the other hand, for energies $E_{\chi_2}\sim 10~\gev$, which are more relevant for the secondary production in the SHiP and MATHUSLA experiments, the deflection angle of order few tens of $\mrad$s leads to the impact parameter with respect to the IP of order meters, given the distance between the IP and the decay vessel exceeding $50~\m$. This might partially limit the actual reach of both experiments in this model once a more detailed experimental analysis is performed that considers a pointing requirement. 
    This effect is even more important for some scenarios with lower energy $\mathcal{O}(\mathrm{GeV})$, corresponding to the regions of the parameter space dominated by the primary production of LLPs.
\end{description}

%------------------------
%------------------------

%============================================================
\section{Scattering cross sections for secondary production of LLPs\label{sec:scatteringsigma}}
%============================================================

In this section, we will give the expressions for the scattering cross sections relevant for secondary production of new light particles, $\mathrm{LLP}_1+T\to \mathrm{LLP}_2+T$, where the target can be electron or nuclei, $T=e,N$. For the models discussed in \cref{sec:darkbrem,sec:iDM,sec:darkHiggs}, $\mathrm{LLP}_1 = \chi,\chi_1$ or $S$, while $\mathrm{LLP}_2 = A^\prime,\chi_2$ or $A^\prime$, respectively. Notably, all the considered models are characterized by somewhat different kinematics when both the secondary production and subsequent decay  of $\mathrm{LLP}_2$ in the detector are taken into account:
\begin{itemize}
    \item $2\rightarrow 3$ scattering followed by a $2$-body decay (relevant for a scenario with dark bremsstrahlung; cf \cref{sec:darkbrem}),
    \item $2\rightarrow 2$ scattering followed by a $3$-body decay (inelastic DM; cf \cref{sec:iDM}),
    \item $2\rightarrow 2$ scattering followed by a $2$-body decay ($A^\prime$ with secluded dark Higgs boson; cf \cref{sec:darkHiggs}).
\end{itemize}
In the following, we present the expressions for the scattering cross sections relevant to our analysis.

%+++++++++++++++++++++++++++++++++++++++
\subsection{$2\to 2$ scatterings}
%+++++++++++++++++++++++++++++++++++++++

In case of models with inelastic DM and secluded dark Higgs boson, secondary production proceeds via $2\to 2$ scattering processes illustrated in the center and right panels of \cref{fig:feynman}, respectively. The relevant differential cross section in the lab frame reads
\begin{equation}
\frac{d \sigma}{d E_T}
= \frac{m_{T}}{8 \pi \lambda\left(s, m_{T}^{2}, m_{1}^{2}\right)} \overline{|\mathcal{M}|}^{2},
\end{equation}
where $E_{T}$ and $m_{T}$ is the  energy and mass of the recoiling target, while $m_{1}\equiv m_{\mathrm{LLP}_1}$ is the mass of the initial state scattered LLP. The integration limits are:
\begin{equation}
E_{T}^{ \pm}
=\frac{s+m_{T}^{2}-m_2^{2}}{2 s}\times (E_{1}+m_{T}) 
\pm \frac{\lambda^{\frac 12}(s,m_T^2,m_2^2)}{2 s}\times p_{1},
\end{equation}
where $E_1$ and $E_2$ are energies of the initial and final state LLPs, respectively, and $m_2 \equiv m_{\mathrm{LLP}_2}$.

%...........................
\subsubsection{Scatterings off electrons}
%...........................

Let us first consider the case of scattering off electrons. 
\begin{description}
\item [Inelastic DM]
    Following Ref.~\cite{Kim:2016zjx}, the squared matrix element for upscattering $\chi_1 e \to \chi_2 e$ is given by
    \begin{equation}
     \overline{ |\mathcal{M}|^{2}} = \frac{8\left(\epsilon\,e\,g_{12}\right)^2 m_e}
     {(2\,m_e\left(E_{\chi_2}-E_{\chi_1}\right)-m_{A^{\prime}}^2)^{2}} \times \mathcal{M}_0,
    \end{equation}
    where the amplitude is defined as
    \begin{equation} 
    \mathcal{M}_{0}
    =m_e\left(E_{\chi_{1}}^{2}+E_{\chi_{2}}^{2}\right)
    -\left(\delta m_{x}\right)^{2}\left(E_{\chi_{2}}
    -E_{\chi_{1}}+m_e\right) / 2 +m_e^{2}\left(E_{\chi_{2}}-E_{\chi_{1}}\right)
    +m_{\chi_{1}}^{2} E_{\chi_{2}}-m_{\chi_{2}}^{2} E_{\chi_{1}} ,
    \label{eq:M0}
    \end{equation}
    with $\delta m_\chi = m_{\chi_2}-m_{\chi_1}$.
\item [Secluded dark Higgs boson]
    Following Ref.~\cite{Morrissey:2014yma}, the squared matrix element for the process $S e \to A^\prime e$ is
    \begin{equation}
     \overline{ |\mathcal{M}|^{2}} = \frac{4\left(\epsilon\,e\,g_{12}\right)^{2} m_e}{\left\{2m_e\left(E_2-E_1\right)-m_{A'}^{2}\right\}^{2}}  \left[ E_1\,(2\,E_2\,m_e + m_{A'}^2)+E_2\,(m_{S}^2-2\,m_{A'}^2)-2\,m_{A'}^2\,m_e)\right],
    \end{equation}
    where $E_1 = E_S$ and $E_2 = E_{A^\prime}$.
\end{description}
%

%...........................
\subsubsection{Scatterings off nuclei}
%...........................

The coupling between the dark photon and protons is dependent on nuclear form factors. This introduces a nontrivial dependence on the momentum transfer squared, $q^2=-Q^2<0$, as dictated by the nucleon electromagnetic current, $\mathcal{J}_\mu = \bar{u}(p_4)\,\left[F_1\,\gamma_\mu - \left(\sigma_{\mu\nu}\,q^\nu/2\,m_p\right)\right]\,u(p_2)$, where $F_1(q^2)$ and $F_2(q^2)$ are the Dirac and Pauli form factors, respectively. These are usually expressed through the Sachs form factors, $G_E = F_1 - \tau\,F_2$ and $G_M = F_1+F_2$, where we defined $\tau = Q^2/(4\,m_p^2) > 0$. It is customary to express $G_E$ employing the dipole approximation, which is particularly useful in the regime of low momentum exchange relevant for our analysis, which is characterized by $Q^2\lesssim 1~\gev$,
\begin{equation}
G_E=\left(1+Q^2 / 0.71 \gev^2\right)^{-2}=G_M/\mu_p,
\end{equation}
where $\mu_p=2.79$ is the proton magnetic moment and in the last step we have followed the usual approximation that a simple scaling $G_M\simeq \mu_p\,G_E$ holds for small, but even non-zero, values of $Q^2$. For our purposes, the most convenient parametrization consists of form factors which take the form
\be
G_{1} := \tau\,G_{M}^{2} = \tau\,\mu_{p}^{2}\,G_{E}^{2} 
\qquad \text{and} \qquad
G_{2} := \frac{G_{E}^{2}+\tau G_{M}^{2}}{1+\tau} \stackrel{\tau \lll 1}
{\simeq} G_{E}^{2}\left[1+\tau\left(\mu_{p}^{2}-1\right)\right].
\label{eq:G1G2}
\ee
As follows from \cref{eq:G1G2}, for a small momentum transfer, $\tau\ll 1$, one obtains $G_2\gg G_1$ and the term in the cross section proportional to $G_1$ can typically be neglected. The term proportional to $G_2$, instead, plays the dominant role. 

In addition, for sufficiently small momentum transfer, the internal structure of the nuclei is not probed and the dominant contribution to the cross section comes from coherent scatterings off entire nuclei. In the following, we implement approximate nuclear form factors that lead to a relatively smooth transition between the incoherent and coherent regimes~\cite{Tsai:1973py,Kim:1973he,Bjorken:2009mm}
\begin{equation}
G_{2, \mathrm{tot}}(t)=G_{2, \mathrm{el}}(t)+G_{2, \mathrm{inel}}(t),
\label{eq:G2tot}
\end{equation}
where $G_{2,\mathrm{el}}$ and $G_{2,\mathrm{inel}}$ are the form factors corresponding to the coherent and incoherent regimes, respectively. They are given by
\be
\!\!\!
G_{2, \mathrm{el}}(t) = 
Z^2 \bigg[\frac{a^2 t}{1\!+\!a^2 t}\bigg]^{\!2} \!
\bigg[\frac{1}{1\!+\!t d}\bigg]^{\!2} \text{ and }\   
G_{2, \mathrm{inel}}(t)=
Z \bigg[\frac{a^{\prime 2} t}{1\!+\!a^{\prime 2} t}\bigg]^{\!2}\! \bigg[\frac{G_{E}^2\!+\!\tau G_{M}^2}{1\!+\!\tau}\bigg] 
\!\simeq\! 
Z \bigg[\frac{a^{\prime 2} t}{1\!+\!a^2 t}\bigg]^{\!2} \!
\bigg[\frac{1\!+\!(t/4 m_p^2)(\mu_p^2\!-\!1)}
{(1\!+\!t /0.71\gev^2)^4}\bigg],
\ee
where the relevant atomic form factors have parameters $a=111\,Z^{-1 / 3} / m_{e}$ and $a^\prime = 773\,Z^{-2/3}/m_e$ for coherent and incoherent form factors, respectively, while the coherent nucleus form factor is characterized by $d=0.164\,\gev^{2} A^{-2 / 3}$ with $Z$ ($A$) being the atomic (mass) number of the nucleus. 

We note that our $G_2$ form factor differs slightly from the relevant expression in Ref.~\cite{Bjorken:2009mm} due to a spurious square factor in the nuclear part of the form factor appearing in that reference. We also note that, given our focus on $Q^2\lesssim 1~\gev$, we do not have to consider the regime of deep inelastic scattering (DIS). 

\begin{description}
\item [Inelastic DM]
    For the upscattering process $\chi_1 N \to \chi_2 N$, the squared matrix element is given by
    \begin{equation}
    \overline{ |\mathcal{M}|^{2}} 
    = \frac{8(\epsilon e g_{12})^2 m_p}
    {(2m_p (E_{\chi_2}-E_{\chi_1})-m_{A^{\prime}}^2)^2}
    \times \left[\frac{1}{2} \mathcal{M}_1
    G_{2}+\frac{G_{1}}{\tau}(\mathcal{M}_0-\frac{1}{2}
    \mathcal{M}_{1})\right].
    \end{equation}
    where $\mathcal{M}_0$ is given by \cref{eq:M0} and $\mathcal{M}_1$ reads
    \begin{equation}
    \mathcal{M}_{1} =m_{T}\Big(
    \big[E_{\chi_1}+E_{\chi_2}-(m_{\chi_2}^2-m_{\chi_1}^2)/(2 m_T)\big]^2 +(E_{\chi_1}-E_{\chi_2}+2 m_T)
    \big[(E_{\chi_2}-E_{\chi_1})-\delta_{m}^2 /(2 m_{T}) \big]
    \Big) .
    \end{equation}
\item [Secluded dark Higgs Boson]
    Finally, for the process $S N \to A^\prime N$ the squared matrix element is given by
    \be
    \overline{|\mathcal{M}|}^{2}=&  \frac{2\left(\epsilon\,e\,g_{12}\right)^2}
    {\left\{2\,m_{p}\left(E_{\chi_{2}}-E_{\chi_{1}}\right)-m_{A^{\prime}}^{2}\right\}^{2}}\bigg[-\left(m_S^{2} Q^{2}+\frac{1}{4}\left(Q^{2}-m_S^{2}+m_{A'}^{2}\right)^{2}+3\,m_{A'}^{2}\,Q^{2}\right)\frac{G_1}{\tau}\\ 
    &+\left\{\left(2\,E_{1}\,m_{T}-\frac{1}{2}\left(Q^{2}-m_S^{2}+m_{A'}^{2}\right)\right)^{2}-m_{A'}^{2}\left(Q^{2}+4\,m_{T}^{2}\right)\right\}
    \left[G_2 - G_1\frac{2\,(\mu_p-1)}{\mu_p^2} \right] \bigg].
    \ee
\end{description}
As discussed above, the scattering cross section is dominated by the contributions proportional to $G_2$, effectively replaced by $G_{2,\mathrm{tot}}$ given in \cref{eq:G2tot}, while the terms proportional to $G_1$ can be neglected in the case of low momentum exchange. 

%...........................
\subsubsection{Energy and angular distribution in the secondary production} 
%...........................

\begin{figure}[tb]
\centering
\includegraphics[width=0.49\textwidth]{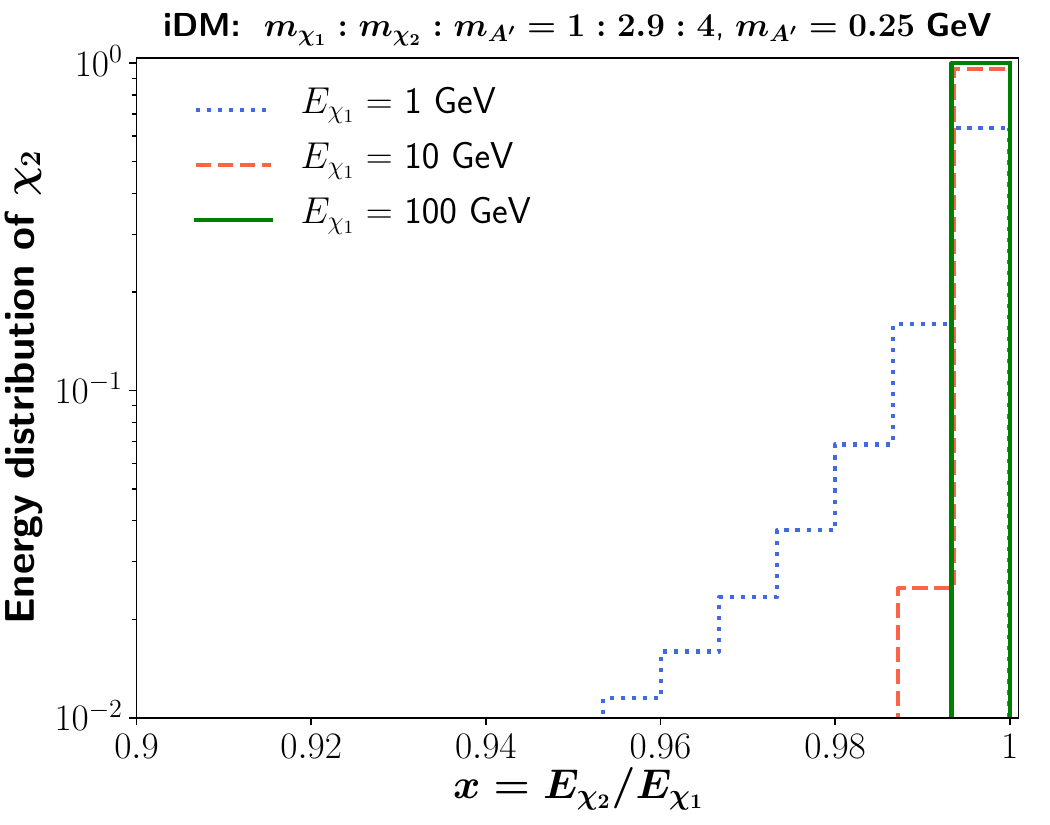}
\includegraphics[width=0.49\textwidth]{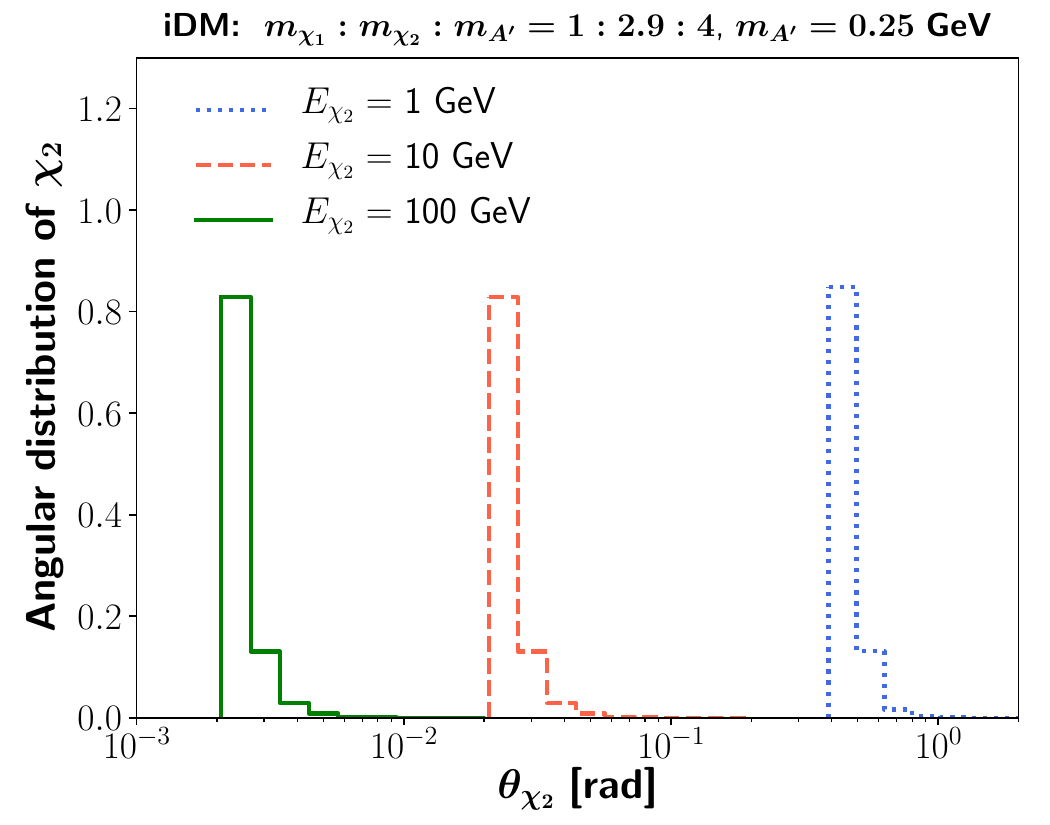}
\caption{The energy and angular distribution of $\chi_2$ produced in the upscattering of $\chi_1$ in the inelastic dark matter model with respect to the energy and direction of the initial state particle $\chi_1$. The color coding and the choice of the benchmark parameters are as in \cref{fig:chi2decaydist}.
}
\label{fig:scatdist}
\end{figure}

Importantly, as illustrated in \cref{fig:scatdist} for the case of inelastic DM, the energies and direction of the LLP is not much affected by the recoil, i.e. $p_{\chi_2}\approx p_{\chi_1}$. In particular, the outgoing LLP tends to inherit most of the energy of the incoming high-energy particle, as shown in the left panel.

On the other hand, even a small deflection angle $\theta_{\chi_2}$ can play a role in the physics analysis. In particular, for energies $E_{\chi_2}\sim 100~\gev$ relevant for FASER, a typical value of $\theta_{\chi_2}\sim 3~\mrad$ deduced from the right panel of \cref{fig:scatdist} leads to a displacement that exceeds the $10~\cm$ radius of the detector, if the upscattering happens more than about $30~\m$ away from the decay vessel. In this case, it is particularly important to properly model the deflection in the secondary production during the simulation. Instead, for larger energies or LLPs with a smaller lifetime, which are produced in the last few meters in front of the decay vessel, as well as for the large-size FASER 2 detector, small values of the deflection angle play a less important role in modeling.

Similarly for SHiP, assuming that the interaction takes place close to the decay vessel at a distance about $50~\m$ away from the IP, and that $E_{\chi_2}\sim 10~\gev$, a typical deflection angle, $\theta_{\chi_2}\simeq 30~\mrad$, leads to an impact parameter with respect to the target of order $150~\cm$. This value is smaller than $250~\cm$ relevant for the loose selection cuts, as mentioned in \cref{sec:physicscuts}, and it would be further suppressed for larger energies of $\chi_1$.  On the other hand, a small but non-zero value of the deflection angle might render it impossible to employ the tight selection cuts typical for e.g. dark photon searches. This would be further complicated by the presence of an additional missing energy in the final state, as discussed in \cref{sec:decayBRsdec}. 

In addition, if the secondary production takes place away from the decay vessel, than the displacement $\mathcal{O}(\m)$ could result in LLP missing the vessel. This, however, mainly concerns scenarios with relatively long-lived LLPs for which the dominant production mechanism is, either way, the primary production.

%------------------------
%------------------------

%+++++++++++++++++++++++++++++++++++++++
\subsection{$2\to 3$ scattering}
%+++++++++++++++++++++++++++++++++++++++

In case of the dark photon mediated single dark matter model, the secondary production of the dark photon proceeds through the dark bremsstrahlung processes illustrated in the Feynman diagram in the left panel of \cref{fig:feynman}. In this work, the elastic $2 \to 3$ scattering of the dark matter off the nucleus is modeled numerically using \texttt{MadGraph~5}~\cite{Alwall:2014hca}. We have created a simplified model containing the DM particle $\chi$, a nucleus $N$, and the dark photon $A^\prime$, where the effective coupling of the dark photon to the nucleus takes into account the elastic form factor $G_2(t)$ defined in \cref{eq:G2tot}. We simulate the interaction separately for all considered materials and for various energies of the incoming dark matter particle. The simulated events are then convoluted with the incoming DM flux, where we sample over the interaction location and weight by the decay in volume probability. 

%============================================================
\section{Benchmark detector designs\label{sec:benchmarkgeometries}}
%============================================================

Here we summarize main features of the simplified benchmark detector designs for the experiments considered in our study. These are also illustrated in \cref{fig:experiments}. \textbf{Bold case} is used for these pieces of material, for which we model secondary production of LLPs in scatterings of other light particles. Any possible secondary production processes happening in other parts of the detectors are not taken into account.

More detailed experimental setups have been discussed for FASER/FASER 2 detectors in Refs.~\cite{Ariga:2018uku,Ariga:2018pin,Abreu:2019yak}, for MATHUSLA in Refs.~\cite{Alpigiani:2018fgd,Curtin:2018mvb}, while in the case of SHiP in Refs.~\cite{Anelli:2015pba,Ahdida:2654870}. In the following, we employ simplified geometries that encompass the most important aspects of detector designs under study and, therefore, remain sufficient for the purpose of our phenomenological analysis.

These benchmark geometries might be further updated in the future, especially for experiments to be performed at the time of the HL-LHC era, in which case sensitivity plots presented in this study would have to be updated. We note, however, that the results presented in \cref{sec:results} are typically not sensitive to possible (mild) changes in detector designs. The most important are potential modifications of detector design in the most immediate neighborhood of the decay vessel. Depending on the nature of these changes, they could either improve or worsen reach in the regions of parameter space of BSM models corresponding to LLPs with the shortest lifetime. Even such modifications, however, would often not have significant impact on the reach, as we briefly discuss in \cref{sec:resultsaftercutschange}.

\begin{description}
\item [FASER (both Run3 and HL-LHC)] (beginning with most distant places from the detector)
\begin{itemize}
    \item We do not model in details all the LHC infrastructure that can be found in between ATLAS IP and concrete shielding close to the place when the arc section of the LHC tunnel begins. While such detailed modeling would go beyond this study, we also note that the most relevant sections of the LHC tunnel for LLP production considered here are either ATLAS IP (primary production) or the rock close to FASER (secondary production).
    \item The first piece of material that we include in our simulations is \textbf{a region with $\mathbf{10}~\m$ of concrete shielding} that starts about $100~\m$ in front of the tunnel TI12, where the FASER detector is located. We model this shielding as material with $Z=8.6$, $A=17$, and $\rho = 2.3~\g/\cm^3$.
    \item Concrete shielding is followed by \textbf{$\mathbf{89}~\m$ of the standard rock}, modeled as material with $Z=11$, $A=22$ and $\rho = 2.65~\g/\cm^3$~\cite{Tanabashi:2018oca}.
    \item The remaining section of the rock with the length of $1~\m$ and the $30$ cm long section beginning right after the rock that contains, i.a. the first front veto layers, are treated separately. In particular, any secondary production processes happening here are rejected in the analysis as they might lead to veto activation. It is useful to note that $1~\m$ of the standard PDG rock corresponds to roughly three hadronic interaction lengths, $3\,\lambda_{\textrm{had,int}}$. 
    \item Located behind the first front veto is FASER's neutrino detector, dubbed FASER$\nu$~\cite{Abreu:2019yak}. It contains $1~\m$ of tungsten out of which the first \textbf{70~cm of tungsten} are taken into account when modeling secondary production. This is modeled as material with $Z=74$, $A=183.84$, and $\rho=19.3~\g/\cm^3$. Any scatterings that occur in the remaining $30~\cm$ of tungsten, as well as within further $20~\cm$ of FASER/FASER$\nu$ equipment containing i.a. second section with front veto layers, are not taken into account when obtaining sensitivity plots. \\
    We note that the FASER$\nu$ detector consists of alternating layers of tungsten (with the total length of about $1~\m$) and emulsion films. This complicated geometry is simplified for our phenomenological study in which we model FASER$\nu$ as a single tungsten block with an average position with respect to other elements of the infrastructure around. Precise details of this modeling have a minor impact on the final results.
    \item After FASER$\nu$ and the second front veto, the FASER decay vessel begins at a distance $L\simeq 480~\m$ away from the ATLAS IP. The decay vessel has a cylindrical shape with radius $R=10~\cm$ and length $\Delta=1.5~\m$.
\end{itemize}
\item [FASER~2 (HL-LHC)] \quad 
\begin{itemize}
    \item FASER~2 is characterized by a similar geometry to FASER, but without FASER$\nu$ neutrino detector in front of the decay vessel. As a result, after the concrete shielding ($10~\m$), rock ($89~\m$+$1~\m$), and the first front veto layers ($30~\cm$), the decay vessel begins at a distance $L\simeq 480~\m$ away from the ATLAS IP. The decay vessel is a cylinder with radius $R=1~\m$ and length $\Delta=5~\m$.
\end{itemize}

\item [MATHUSLA (HL-LHC)]  (beginning with most distant places away from MATHUSLA)
\begin{itemize}
    \item The size of the CMS detector and its cavern is taken into account in the analysis in a simplified manner. In particular, we do not treat secondary production processes happening in the elements of the infrastructure there as it would be challenging to model them accurately. On the other hand, this region lies close to the CMS IP and the relevant production rate for LLPs that can reach MATHUSLA is dominated by the primary production processes. The CMS cavern has roughly a cylindrical shape with $13~\m$ radius and the length of $60~\m$~\cite{CMS:1994hea}. By taking into account the planned position of the MATHUSLA detector and the size of the CMS cavern, one can see that about $17~\m$ to $30~\m$ of the distance between the CMS IP and the bottom part of the MATHUSLA detector (details below) will be occupied by the cavern. 
    \item The rest of a distance between the CMS IP and MATHUSLA is modeled \textbf{as the standard rock} with $Z=11$, $A=22$, and $\rho = 2.65~\g/\cm^3$~\cite{Tanabashi:2018oca} that extends up until the last $1~\m$ long section of the rock in front of MATHUSLA. The last $1~\m$ long section of the rock is, instead, treated separately, similar to the aforementioned case of FASER/FASER 2. This is to make sure that secondary production processes that we take into account are shielded from veto layers by at least $3\,\lambda_{\textrm{had,int}}$.
    \item MATHUSLA decay vessel begins right after the rock. The total distance between the CMS IP and MATHUSLA, as well as the size of the detector, is dictated by geometry described by \cref{eq:MATHUSLAgeometry}.
\end{itemize}

\item [SHiP] (beginning with most distant places away from the SHiP decay vessel)
\begin{itemize}
    \item The target and hadron stopper with additional shielding correspond to about $13.2~\m$ of infrastructure elements that we neglect when modeling secondary production. This is justified by the fact that the dominant contribution to the signal yield from this region corresponds to primary production of LLPs initiated by protons hitting the target.
    \item After this front section, the remaining part of the active muon shield consists of \textbf{6 iron blocks each $\mathbf{5}~\m$ long with $\mathbf{10}~\cm$ empty gaps in between}, where iron is modeled as material with $Z=26$, $A=55.84$, and $\rho=7.874~\g/\cm^3$.
    \item This is followed by an additional section with the length of about $1~\m$ which contains other elements of the infrastructure that are not modeled in detail in our analysis.
    \item The SND detector begins at a distance about $44.7~\m$ away from the IP. In its internal $80~\cm \times 80~\cm$ part around the collision axis, it corresponds to the emulsion detector and the following downstream tracker. The emulsion detector is surrounded by a magnet with a coil made out of copper (alternative design with aluminum is also considered), which is further surrounded by an iron yoke. For the purpose of modeling, we assume that the region of the SND surrounding the emulsion detector is mostly filled with material with properties similar to iron. 
    \begin{description}
        \item [The emulsion detector and downstream tracker] The total length of the emulsion detector is $3~\m$, and it consists of alternating layers of $1$ mm thick lead plates and emulsion films that are additionally interleaved with electronic detectors. We model this as a single block of lead with the total length of about $1~\m$ that corresponds to $19$ walls that each has $57$ lead plates, as discussed in section 4.2 of Ref.~\cite{Ahdida:2654870}. The second half of this block of lead (i.e. $50~\cm$ of lead) is excluded from our analysis of secondary LLP production as it corresponds to the last $3\,\lambda_{\textrm{had,int}}$ in front of the downstream tracker and the following SND muon system acting as veto. In our simplified treatment of detector geometry, we assume that the block of lead occupies $1$ m long section in the middle of a $3$ m long emulsion detector. Hence, out of the emulsion detector with the total length of $3~\m$, only secondary production happening in the front \textbf{$\mathbf{50}$ cm long part of the lead block} is taken into account which is positioned in between $1~\m$ and $1.5~\m$ away from the beginning of the SND detector. It is modeled as material with $Z=82$, $A=207.2$, and $\rho=11.35~\g/\cm^3$. After the emulsion detector, the remaining $3~\m$ of the internal $80~\cm\times 80~\cm$ part of the SND detector is occupied by the downstream tracker. We do not study secondary production processes that occur here.
        
        Importantly, for the purpose of studying additional LLP signatures relevant for scatterings off electrons in the SND, which are discussed in \cref{sec:resultselectrons}, we model the emulsion detector as a full block of lead with $1~\m$ length.
        
        \item [Magnets surrounding the emulsion detector and the downstream tracker] As discussed above, we assume that the magnets are made out of material equivalent to iron with $Z=26$, $A=55.845$ and $\rho=7.874~\g/\cm^3$. They have a total length of about $6~\m$. We again exclude a section of the magnet toward the end with $0.5~\m$ length, as it corresponds to the last $3\,\lambda_{\textrm{had,int}}$ before the SND muon system acting as veto. On the other hand, \textbf{the first $\mathbf{5.5}$ m long section of the magnet} is taken into account when modeling secondary production.
    \end{description}
    \item The SND downstream tracker and surrounding magnet is followed by the $2$ m long muon identification system of the SND. It partially acts as the front veto before the decay vessel begins, so we do not take into account any possible secondary production processes happening in the close vicinity or inside the SND muon system.
    \item The muon identification system is followed by the decay vessel that has length of $50~\m$. We take into account its irregular shape~\cite{Ahdida:2654870} when simulating the events. 
\end{itemize}
\end{description}

%============================================================
\section{Cuts imposed on signal events\label{sec:physicscuts}}
%============================================================

When obtaining the sensitivity reach plots presented in \cref{sec:results}, we impose additional cuts that can limit the number of signal events, but, most importantly, allow one to discriminate between signal and background events more easily. We summarize these cuts briefly below for both two-body decays with all LLP energy going into visible charged tracks, as well as for three-body decays with missing energy in the final state. The latter is relevant for searches for inelastic DM. In the analysis, we also assume a $100\%$ detection efficiency after the relevant cuts are imposed.

On top of these experimental cuts, we always require the recoil momentum of the target in secondary production processes not to exceed $p_{\textrm{recoil}}=1~\gev$. Similarly, no secondary production events that occur within the last three hadronic interaction lengths, $3\,\lambda_{\textrm{had,int}}$, in front of veto are taken into account. We note that detailed detector simulations could allow one to alleviate these constraints. In particular, similar challenges have already been considered in the context of LLP searches in high-energy electron and muon beam-dump experiments, e.g. NA64-$e$~\cite{Andreas:2013lya} and NA64-$\mu$~\cite{Gninenko:2653581} employing $\sim 100$~GeV $e$/$\mu$ beams.

\begin{description}
    \item [FASER/FASER 2]
    We require the LLP decays to happen inside the decay volume and deposit at least $100~\gev$ of energy in two final state visible charged tracks. Due to the large energy threshold, no additional direction reconstruction cut is imposed since charged particles produced in the decay are highly collimated along the LLP direction in the lab frame.
    \item [MATHUSLA]
    We follow Ref.~\cite{Curtin:2018mvb} and, for LLP decays happening in the decay volume, we implement lower threshold on three-momenta of both charged particles in the final state, $|\vec{p}_{e}|>1~\gev$, that is relevant for $e^+/e^-$ charged tracks. On top of this cut, in order to suppress various possible sources of BG, additional timing and pointing information can be used to correlate the event with $pp$ collisions at the CMS IP.
    
    As for the latter, due to the small momentum exchange with the target in scattering processes, the direction of LLPs entering the decay volume is not much affected by the recoil and this effect can be neglected in the analysis (see also discussion in \cref{sec:scatteringsigma}). The direction reconstruction of decaying LLPs can also be affected in the presence of missing energy in the final state, especially for low-energy signal events. As a result of such misreconstruction, isotropic BG induced by interactions of atmospheric neutrinos might become non-negligible as it corresponds to few tens of events per year~\cite{Chou:2016lxi}. This issue has been studied in Ref.~\cite{Ibarra:2018xdl} for LLPs produced in Drell-Yan processes in $pp$ collisions at the CMS IP, as well as produced in invisible decays of the SM Higgs boson. In both cases, the sensitivity reach has been found to decrease by a factor of $3$ for new particles with mass $m_{\textrm{LLP}}\sim 10~\gev$, once neutrino-induced BG is taken into account. On the other hand, as noted in Ref.~\cite{Ibarra:2018xdl}, this effect could become even less sizable when dedicated background simulations are performed. In this study, we will then neglect possible negative impact of direction misreconstruction in searches for inelastic DM.
    \item [SHiP]
    We follow Ref.~\cite{Ahdida:2654870} and require three-momenta of both visible charged particles produced in LLP decays in the decay vessel to be large enough, $|\vec{p}_{\textrm{daughter}}|>1$~GeV. Similar to the case of MATHUSLA, we note that direction reconstruction is not much affected by the recoil in scattering processes relevant for secondary production of LLPs. 
    
    In case of 3-body decays with missing energy, loose selection efficiency could be imposed with an impact parameter with respect to target not exceeding $250$~cm~\cite{Ahdida:2654870}. While this condition allows one to more easily accept partially reconstructed signal events, it could also make BG rejection more challenging. Dedicated BG simulations that go beyond this analysis are required to thoroughly investigate this issue which we left for future studies. Instead, in our phenomenological analysis, we neglect this possible obstacle and assume that all such events can be properly reconstructed and discriminated from BG. This also allows for a better comparison to the reach of MATHUSLA experiment for which we make a similar assumption.
    
    Experimental cuts applied to additional searches based on LLP scatterings off electrons in the SND are described in \cref{sec:resultselectrons} for both considered signatures.
\end{description}

%%%%%%%%%%%%%%%%%%%

\bibliography{shortlifetime}

\end{document}